\documentclass[twocolumn,amssymb,nobibnotes,aps,prd,reprint,superscriptaddress]{revtex4-2}
\usepackage{multirow}
\usepackage{subfigure}
\usepackage{amssymb}
\usepackage{amsmath}
\usepackage{graphicx}
\usepackage{dcolumn}
\usepackage{bm}
\usepackage{wasysym}
\usepackage{appendix}
\usepackage[mathlines]{lineno}

\begin{document}
\preprint{APS/123-QED}

\title{A Local Lorentz Invariance test with LAGEOS satellites}
\author{David Lucchesi}
\email[]{david.lucchesi@inaf.it}
\affiliation{Istituto Nazionale di Astrofisica (INAF),  Istituto di Astrofisica e Planetologia Spaziali (IAPS), Via del Fosso del Cavaliere, 100, 00133 Roma, Italy}
\affiliation{Istituto Nazionale di Fisica Nucleare (INFN), Sezione di Tor Vergata, Via della Ricerca Scientifica 1, 00133  Roma, Italy}
\affiliation{Consiglio Nazionale delle Ricerche (CNR), Istituto di Scienza e Tecnologie della Informazione (ISTI),  via G. Moruzzi 1, 56124 Pisa, Italy}

\author{Massimo Visco}
\affiliation{Istituto Nazionale di Astrofisica (INAF),  Istituto di Astrofisica e Planetologia Spaziali (IAPS), Via del Fosso del Cavaliere, 100, 00133 Roma, Italy}
\affiliation{Istituto Nazionale di Fisica Nucleare (INFN), Sezione di Tor Vergata, Via della Ricerca Scientifica 1, 00133  Roma, Italy}

\author{Roberto Peron}
\affiliation{Istituto Nazionale di Astrofisica (INAF),  Istituto di Astrofisica e Planetologia Spaziali (IAPS), Via del Fosso del Cavaliere, 100, 00133 Roma, Italy}
\affiliation{Istituto Nazionale di Fisica Nucleare (INFN), Sezione di Tor Vergata, Via della Ricerca Scientifica 1, 00133  Roma, Italy}

\author{José C. Rodriguez}
\affiliation{Red de Infraestructuras Geod\'esicas, Instituto Geogr\'afico Nacional, Madrid, Spain}

\author{Giuseppe Pucacco}
\affiliation{Dipartimento di Fisica,  Universit\`a di Tor Vergata, Via della Ricerca Scientifica 1, 00133  Roma, Italy}
\affiliation{Istituto Nazionale di Fisica Nucleare (INFN), Sezione di Tor Vergata, Via della Ricerca Scientifica 1, 00133 Roma, Italy}

\author{Luciano Anselmo}
\affiliation{Consiglio Nazionale delle Ricerche (CNR), Istituto di Scienza e Tecnologie della Informazione (ISTI),  via G. Moruzzi 1, 56124 Pisa, Italy}

\author{Massimo Bassan}
\affiliation{Dipartimento di Fisica,  Universit\`a di Tor Vergata, Via della Ricerca Scientifica 1, 00133  Roma, Italy}
\affiliation{Istituto Nazionale di Fisica Nucleare (INFN), Sezione di Tor Vergata, Via della Ricerca Scientifica 1, 00133 Roma, Italy}

\author{Graham Appleby}
\affiliation{British Geological Survey (Hon Research Associate), United Kingdom}

\author{Marco Cinelli}
\affiliation{Istituto Nazionale di Astrofisica (INAF),  Istituto di Astrofisica e Planetologia Spaziali (IAPS), Via del Fosso del Cavaliere, 100, 00133 Roma, Italy}
\affiliation{Istituto Nazionale di Fisica Nucleare (INFN), Sezione di Tor Vergata, Via della Ricerca Scientifica 1, 00133  Roma, Italy}

\author{Alessandro Di Marco}
\affiliation{Istituto Nazionale di Astrofisica (INAF),  Istituto di Astrofisica e Planetologia Spaziali (IAPS), Via del Fosso del Cavaliere, 100, 00133 Roma, Italy}
\affiliation{Istituto Nazionale di Fisica Nucleare (INFN), Sezione di Tor Vergata, Via della Ricerca Scientifica 1, 00133  Roma, Italy}

\author{Marco Lucente}
\affiliation{Istituto Nazionale di Astrofisica (INAF),  Istituto di Astrofisica e Planetologia Spaziali (IAPS), Via del Fosso del Cavaliere, 100, 00133 Roma, Italy}
\affiliation{Istituto Nazionale di Fisica Nucleare (INFN), Sezione di Tor Vergata, Via della Ricerca Scientifica 1, 00133  Roma, Italy}

\author{Carmelo Magnafico}
\affiliation{Istituto Nazionale di Astrofisica (INAF),  Istituto di Astrofisica e Planetologia Spaziali (IAPS), Via del Fosso del Cavaliere, 100, 00133 Roma, Italy}
\affiliation{Istituto Nazionale di Fisica Nucleare (INFN), Sezione di Tor Vergata, Via della Ricerca Scientifica 1, 00133  Roma, Italy}

\author{Carmen Pardini}
\affiliation{Consiglio Nazionale delle Ricerche (CNR), Istituto di Scienza e Tecnologie della Informazione (ISTI),  via G. Moruzzi 1, 56124 Pisa, Italy}


\author{Feliciana Sapio}
\affiliation{Istituto Nazionale di Astrofisica (INAF),  Istituto di Astrofisica e Planetologia Spaziali (IAPS), Via del Fosso del Cavaliere, 100, 00133 Roma, Italy}
\affiliation{Istituto Nazionale di Fisica Nucleare (INFN), Sezione di Tor Vergata, Via della Ricerca Scientifica 1, 00133  Roma, Italy}


\collaboration{SaToR-G Collaboration}

\date{\today}

\begin{abstract}
Strong theoretical arguments suggest that a breakdown of Lorentz Invariance could arise under some very particular conditions. 
From an experimental point of view, it is important to  test the Local Lorentz Invariance with ever greater precision and in all contexts, regardless of the theoretical motivation for the possible violation. 
In this paper we discuss some aspects of the gravitational sector. Tests of Lorentz Invariance in the context of gravity are difficult and rare in the literature.
Possible violations could arise from quantum physics applied to gravity or the presence of vector and tensor fields mediating the gravitational interaction together with the metric tensor of General Relativity. We present our results in the latter case.
We analyzed the orbit of the LAGEOS and LAGEOS II satellites over a period of almost three decades.
The effects of the possible preferred frame represented by the cosmic microwave background radiation on the mean argument of latitude of the satellites orbit were considered. These effects would manifest themselves mainly through the post-Newtonian parameter \(\alpha_1\), a parameter that has a null value in General Relativity.
We constrain this parameterized post-Newtonian parameter down to the level of \(\sim 2\times10^{-5}\), improving a previous limit obtained through the Lunar Laser Ranging technique.
\end{abstract}

\maketitle


\section{Introduction}\label{sec:intro}
Local Lorentz Invariance (LLI) represents a cornerstone of Einstein's theory of General Relativity (GR) \cite{1916AnP...354..769E}. LLI states that the outcome of any local (in space and in time) non-gravitational experiment is independent of the velocity of the freely falling reference frame in which the experiment is carried out. LLI represents one of the pillars of Einstein Equivalence Principle (EEP) --- along with the weak equivalence principle and local position invariance --- valid in GR and in all metric theories of gravity \cite{2018tegp.book.....W}.
In GR, the only field that mediates the long-range gravitational interaction is the metric tensor \(g_{\mu\nu}\). 
However, in the context of modern unification theories, other fields --- scalar, vector or tensor in their essence --- may come to play a role in mediating the gravitational interaction in addition to the metric tensor of Einstein’s theory of gravitation.
In these unification theories, the action of a (four-) vector field \(K^{\mu}\) or of other tensor fields \(B_{\mu\nu}\), beside \(g_{\mu\nu}\), is such that the distribution of matter in the universe typically selects a preferred rest frame for local gravitational physics and, consequently, implies a violation of LLI. Conversely, if only one or more scalar fields are present, beside the metric tensor \(g_{\mu\nu}\), as in the case of tensor-multi-scalar theories of gravitation \cite{1992CQGra...9.2093D}, no violation of Lorentz Invariance is expected.

From the phenomenological point of view, and in the framework of the parameterized post-Newtonian (PPN) formalism \cite{1968PhRv..169.1017N,1971ApJ...163..611W,1972ApJ...177..775N,1972ApJ...177..757W}, valid in the Weak-Field and Slow-Motion (WFSM) limit of GR, the Preferred Frame Effects (PFE) are described by the parameters \(\alpha_1\), \(\alpha_2\) and \(\alpha_3\), that are all equal to zero in GR and in tensor-scalar theories of gravity.
LLI and, consequently, PFE, are well tested in the context of high-energy physics experiments \cite{2005LRR.....8....5M} --- see also \cite{2013CQGra..30m3001L} for a review of Lorentz Invariance tests in effective field theories --- but are much more difficult to test in the context of gravitation. 

Current best limits on LLI obtained in the case of Solar System tests, i.e. in the WFSM limit of GR, were achieved \cite{2008ASSL..349..457M} through Lunar Laser Ranging \cite{1973Sci...182..229B}: \(\alpha_1=(-7\pm9)\times10^{-5}\) and  \(\alpha_2=(+1.8\pm2.5)\times10^{-5}\); and exploiting the close alignment between the spin of the Sun and the total angular momentum of the Solar system \cite{1987ApJ...320..871N}: \(\mid{\alpha_2}\mid \lesssim 2.4 \times10^{-7}\).
In the quasi-strong-field regime of pulsar systems, current best limits are \cite{2012CQGra..29u5018S,2013CQGra..30p5019S}:  \(\hat{\alpha}_1=(-0.4\pm4)\times10^{-5}\) and  \(\mid{\hat{\alpha}_2}\mid \lesssim 1.6 \times10^{-9}\); and \cite{1996CQGra..13.3121B,2005ApJ...632.1060S}: \(\mid{\hat{\alpha}_3}\mid \lesssim 4 \times10^{-20}\).
The hat in the last results indicates the quasi-strong field regime, where possible modifications in the post-Newtonian effects may be present with respect to the standard PPN formalism \footnote{The reasons why the post-Newtonian approximation is so ``\textit{unreasonably effective}'' (quoting C. M. Will) in the quasi-strong regime of binary pulsar systems --- like the one mentioned --- but also in the strong regime of inspiraling compact systems (e.g. black holes before final fusion \cite{abbott}) are unknown \cite{2011PNAS..108.5938W}.}.
The above limits were obtained by assuming the existence of PFE with respect to the cosmic microwave background (CMB).
{This is the frame in which the CMB radiation appears nearly isotropic (after subtracting the dipole moment anisotropy due to the motion of the Sun and the foreground emission from our galaxy), with temperature inhomogeneities at the level of a few parts in \(10^5\), and an almost perfect black body spectrum \cite{2020A&A...641A...6P}.}

In this paper, we provide a new constraint on the PPN parameter \(\alpha_1\) from the analysis of the orbital residuals of the two geodetic satellites { LAGEOS and LAGEOS II} over a time interval of approximately 30 years. {LAGEOS satellites} are passive satellites very well tracked through the powerful Satellite Laser Ranging (SLR) technique \cite{2002AdSpR..30..135P,2019JGeod..93.2161P}. In particular, we consider the possible existence of PFE due to the motion of the Earth-Sun-satellite system with respect to the CMB radiation and analyze the effects on the rate of the observable \(\ell_0\) defined as the sum of the satellite argument of pericenter, \(\omega\), and its mean anomaly, \(M\), i.e.:
\begin{equation}
    \ell_0 =\omega + M.
\end{equation}
In celestial mechanics, this quantity is known as the mean argument of latitude.
This is, to our knowledge, the first constraint on the PPN parameter \(\alpha_1\) obtained in the Earth field by means of the analysis of the data from the orbit of  artificial satellites.

This work expands what {is described in \cite{2024PRL..98d4034L} }with regard to the determination {of the orbits of the two geodetic satellites and their} consequent analysis, as well as the estimation of the main systematic errors that impact the measurement of the parameter \(\alpha_1\).
This last evaluation is reported in detail in Appendix \ref{app:errors}, where the estimation of the systematic errors in the measurement and their impact on the knowledge of the parameter \(\alpha_1\) is addressed both for the main gravitational perturbations and for the more subtle non-gravitational ones.

The rest of the paper is organized as follows.
In Section ~\ref{sec:PFE}, we present the main effects on the orbital elements of an artificial satellite due to the possible existence of preferred reference frames in nature. 
In Section \ref{sec:POD}, we present our procedure to reconstruct the orbit of the {satellites} from the SLR data, {the so-called Precise Orbit Determination (POD)}. From the rate of the orbital residuals we compute the observable \(\dot{\ell}_0\), that is the key to constrain the PPN parameter \(\alpha_1\). 
In Section \ref{sec:concept}, we introduce the measurement concept we used to constrain the \(\alpha_1\) parameter. {It is made in two steps. Firstly we use the values of the observables of LAGEOS and LAGEOS II (obtained from the two PODs) to separate  the potential contribution of $\alpha_1$ to $\ell_0 $ from that due to the main systematic error source, related to the imperfect knowledge of quadrupole coefficient of the Earth's gravitational field. The second step is based on  a phase-sensitive detection at the expected frequency and phase. Appendix \ref{app:sintetici} is devoted to validation of this technique using synthetic data}
In Section \ref{sec:measure}, we present the results of the measurement and the constraints obtained on the PPN parameter and, consequently, on the possible violation of LLI.
{In Sections \ref{sec:theory} and \ref{sec:experiments}, the theoretical context of our measurement is explored in depth and compared with other results in the literature.}
Finally, in Section \ref{sec:Conclusions}, our conclusions are provided. 

\section{Orbital effects of PFE}\label{sec:PFE}
In 1994, Damour and Esposito-Farese have shown that the orbits of certain artificial satellites have the potential to improve the upper limit on the \(\alpha_1\) parameter --- down to the \(\sim10^{-6}\) level --- due to the appearance of small divisors which enhance the corresponding PFE \cite{1994PhRvD..49.1693D}.
This possibility of reaching a stringent constraint in the PPN parameter linked to the existence of possible PFE is mainly due to the precision of the SLR tracking technique. Indeed, for a few tracked satellites, SLR allows to reconstruct the orbit at cm level. 
This is the case for the two LAGEOS satellites \cite{1976anah.iafcR....J,LG1phB,LG-TN-AI-037-89}.
For these reasons their orbital reconstruction has been exploited for fundamental physics measurements for over three decades \cite{1996NCimA.109..575C,2004Natur.431..958C,2004cosp...35..232L,2010PhRvL.105w1103L,2014PhRvD..89h2002L,Lucchesietal2015,2019Univ....5..141L,2019arXiv191001941L,2019EPJC...79..872C,2020Univ....6..139L,2021Univ....7..192L}.
At the same time, it is necessary to develop a dynamic model for the orbit of these satellites that is equally reliable and accurate \cite{Lucchesietal2015,2016AdSpR..57.1928V,2018PhRvD..98d4034V,2019Univ....5..141L}.

Starting from the Lagrangian \(\mathcal{L}\) for \(\mathcal{N}\) ideal proof-masses gravitationally interacting, it can be shown  that the dependency of \(\mathcal{L}\) on the two PPN parameters \(\alpha_1\) and \(\alpha_2\) (if different from zero) will provide non-boost invariant terms depending on the velocities \(\mathbf{v}^0_a\) of the proof masses with respect to special gravitationally preferred rest frame \cite{1994PhRvD..49.1693D}.
Damour and Esposito-Farese applied their results both to the particular case of satellites in equatorial orbit (\(i=0\)) and to the more general case of satellites with a non-zero inclination \(i\) with respect to the equatorial plane.
If \(\alpha_1\ne0\), the main effects produced are on the eccentricity vector \(\mathbf{e}\) of the satellite's orbit \footnote{This is the Laplace-Runge-Lenz vector that identifies the satellite pericenter direction: a constant of motion in the ideal case of the 2-body Newtonian problem.} and on its orbital longitude \(\ell_0=\omega+M\) \footnote{The  mean argument of latitude is not a true longitude along the orbit of the satellite, as is the argument of latitude \(u = \omega + f\), with \(f\) the true anomaly.}. The former is a secular effect which is described by the sum of multiple (and independent) rotations and enhanced by small divisors in the case of particular values of the satellite's inclination.
The latter is a periodic effect with annual period.


We refer to \cite{1994PhRvD..49.1693D} for the details of the calculations that lead to the final expression of the perturbation on the eccentricity vector. In their analysis, the authors did not explain the contribution of the \(\alpha_2\) parameter in the different orbital elements because it is better constrained than \(\alpha_1\) by other measurements in the Solar System \cite{1987ApJ...320..871N}. In the present work we focus on the possible effects of the PPN \(\alpha_1\) parameter on the longitude of an artificial satellite.

Eq. (\ref{eq:Lag}) below provides the general expression for a term \(\mathcal{L}_{\alpha_1}\) depending on the \(\alpha_1\) parameter \cite{1994PhRvD..49.1693D} in the Lagrangian  of arbitrarily interacting proof masses \(m_a\) and \(m_b\) (\(G_N\), \(c\) and \(r_{ab}\) are, respectively, the Newtonian Gravitational constant, the speed of light and the distances between the masses):

\begin{equation}\label{eq:Lag}
\mathcal{L}_{\alpha_1}=-\frac{{\alpha_1}}{4c^2}\sum_{a\ne b} \frac{G_Nm_am_b}{r_{ab}}(\mathbf{v}^0_a\cdot \mathbf{v}^0_b),
\end{equation}

while Eq. (\ref{eq:Lag2}) provides the explicit expression for this Lagrangian when the two interacting masses are the Earth and the satellite:

\begin{equation}\label{eq:Lag2}
\mathcal{L}_{\alpha_1}=-\frac{{\alpha_1}}{2c^2} \frac{G_Nm_{\oplus}m_S}{r_{\oplus S}}(\mathbf{v}_{\oplus}+\mathbf{w})\cdot (\mathbf{v}_S+\mathbf{v}_{\oplus}+\mathbf{w}).
\end{equation}

In Eq.  (\ref{eq:Lag2}), $r_{\oplus S}$ represents the Earth-satellite distance,  \(\mathbf{v}_{\oplus}\) is the velocity of the Earth with respect to the Sun, \(\mathbf{w}\) is the ``absolute" (i.e. with respect to the preferred frame) 
velocity of the Sun, and \(\mathbf{v}_S\) is the orbital velocity of the satellite around the Earth. 
So,  \(\mathbf{v}_{\oplus}+\mathbf{w}\) represents the ``absolute" velocity of the Earth with respect to the preferred frame, while \(\mathbf{v}_S+\mathbf{v}_{\oplus}+\mathbf{w}\) represents the ``absolute" velocity of the satellite.


We shall use the Lagrange perturbation equations to compute the measurable effects of this  Lagrangian on two Keplerian elements, namely the argument of pericenter $\omega$ and the mean anomaly $M$:
\begin{equation}
\frac{d\omega}{dt} = -\frac{\cos i}{na^2(1-e^2)^{1/2}\sin i}\frac{\partial\mathcal{R}}{\partial i} +\frac{(1-e^2)^{1/2}}{na^2e} \frac{\partial\mathcal{R}}{\partial e},\label{eq:L_peri}
\end{equation}
\begin{equation}
\frac{d{M}}{dt} = n -\frac{1-e^2}{na^2e}\frac{\partial\mathcal{R}}{\partial e} -\frac{2}{na}\frac{\partial\mathcal{R}}{\partial a},\label{eq:L_mean}
\end{equation}
where  \(\mathcal{R}\) is the disturbing function obtained from the Lagrangian of Eq.  (\ref{eq:Lag2}). {In general, the disturbing function is the negative of the perturbing potential in the Lagrangian.}

To compute the variation of these orbital elements, a linear perturbation approach is sufficient in which the right-hand side of the above perturbing equations  are evaluated by keeping constant, to their (mean) nominal values, the semi-major axis \(a\), the eccentricity \(e\), the inclination \(i\) and the mean motion \(n=\sqrt{G_Nm_{\oplus}/a^3}\) of the satellite.

By summing up the two rates to construct the observable \(\dot{\ell}_0=\dot{\omega}+\dot{M}\), the terms involving \(\partial\mathcal{R}/\partial e\) tend to cancel out, and this cancellation is more effective the smaller the eccentricity of the orbit.
The advantage of considering the sum of the two observables of Eqs. (\ref{eq:L_peri}) and (\ref{eq:L_mean}), is that most of the perturbative effects of the main gravitational and non-gravitational perturbations on the final observable \(\dot{\ell}_0\) are reduced. {This also applies in the case of certain relativistic effects, such as Schwarzschild precession \cite{2019agr..book.....S}.}
Before explicitly calculating the variation over time of the longitude \(\ell_0\) through the two previous Lagrange perturbing equations, it is interesting to discuss the contribution of the different velocities that fall into Eq. (\ref{eq:Lag2}) and whether this contribution is significant for our goal.
By working out the dot product of Eq. (\ref{eq:Lag2}), four main terms are obtained:

\begin{equation}
 \mathcal{F} (\mathbf{v}_{\oplus}, \mathbf{v}_{S}, \mathbf{w})=  2\mathbf{w}\cdot\mathbf{v}_{\oplus} + \mathbf{v}_{\oplus} \cdot \mathbf{v}_{S}  +  \mathbf{w}\cdot\mathbf{v}_S + ({w}^2+{v}_{\oplus}^2).\label{eq:speed}
\end{equation}

The first three scalar products produce, in principle, observable effects related to the parameter \(\alpha_1\).
The first term acts in Eq. (\ref{eq:L_mean}) through the term \(\frac{\partial\mathcal{R}}{\partial a}\) and produces an effect with yearly period that will be the object of our investigation. 
The second and third scalar products generate variations of the argument of pericenter (\(\dot{\omega}\propto-\alpha_1/e\)) and of the mean anomaly  (\(\dot{M}\propto+\alpha_1/e\)) that are equal and opposite for small eccentricities, so they cancel out in the observable \(\dot{\ell}_0\). Their cancellation arises from the partial derivative of the disturbing function with respect to the eccentricity \(e\) of the orbit in Eqs. (\ref{eq:L_peri}) and (\ref{eq:L_mean}).
Finally, the last term of Eq. (\ref{eq:speed}) produces a constant, i.e. secular, effect in \(\dot{M}\), since this effect is not at the annual frequency, it is not relevant to our analysis. Moreover, it can be unobservable in principle since it can be absorbed by appropriately rescaling the universal gravitational constant \(G_N\), we refer to \cite{1971ApJ...169..141W,1994PhRvD..49.1693D,2018tegp.book.....W} for details.

Therefore, focusing on the first term of Eq. (\ref{eq:speed}) the disturbing function reduces to:

\begin{equation}
<\mathcal{R}>_{2\pi}=-\frac{\alpha_1}{c^2}\frac{G_Nm_{\oplus}}{a}(\mathbf{w}\cdot\mathbf{v}_{\oplus})\label{eq:disturbing},
\end{equation}

where the notation \(<\mathcal{R}>_{2\pi}\) implies that we take the average over the unperturbed 2-body Keplerian orbit of the satellite around the Earth.
Inserting this last expression into Eqs.  (\ref{eq:L_peri}) and (\ref{eq:L_mean}), and retaining only the terms that contain \(\frac{\partial\mathcal{R}}{\partial a}\) that are relevant to the effect we are looking for. 
By using Kepler third law we finally obtain:

\begin{equation}
<\dot{\ell}_0>_{2\pi}=\mathcal{K}+<\dot{\ell}_0>_{2\pi}^{per}-2\alpha_1n\frac{(\mathbf{w}\cdot\mathbf{v}_{\oplus})}{c^2}+ \mathcal{O}(e\alpha_1),\label{eq:long}
\end{equation}

where \(\mathcal{K}\) represents a constant contribution to the rate in longitude, while the second term takes into account possible long-term periodic perturbative effects of both gravitational and non-gravitational nature.

The constant term \(\mathcal{K}\) has been inserted for completeness, but has no role in the measurement of the \(\alpha_1\) parameter since we shall be looking, in the orbit residuals of \(\dot{\ell}_0\), for a signal that has the signature of annual periodicity, see Section \ref{sec:concept}.

As regards periodic perturbations, these are discussed in Appendix \ref{app:errors}, when the main systematic errors that can affect the measurement are introduced.

To compute the term related to the PPN parameter \(\alpha_1\) we introduce, as astronomical reference frame, the plane of the ecliptic with the Sun at its origin, the unit vector \(\hat{\mathbf{x}}\) pointing toward the Vernal Equinox \(\aries\) and the unit vector \(\hat{\mathbf{z}}\) normal to the ecliptic plane.  We can now compute the dot product in Eq. (\ref{eq:long}) between the two velocities that, in this frame, take the form (see Table \ref{tab:CMB}):

\begin{equation}
\mathbf{w} = w\left( \cos \beta_{PF} \cos \lambda_{PF} \hat{\mathbf{x}} +  \cos \beta_{PF} \sin \lambda_{PF} \hat{\mathbf{y}} + \sin \beta_{PF} \hat{\mathbf{z}} \right),\label{eq:WPFE}
\end{equation}

\begin{equation}
\mathbf{v}_{\oplus} = {v}_{\oplus}\left( \sin (\lambda_0 + \dot{\lambda}_{\oplus}t) \hat{\mathbf{x}} - \cos (\lambda_0 + \dot{\lambda}_{\oplus}t) \hat{\mathbf{y}}   \right).\label{eq:vTerra}
\end{equation}

\begin{table}[h!]
\caption{Solar-system velocity \(\mathbf{w}\) with respect to the cosmic microwave background. Coordinates are in the ecliptic reference frame. The estimated errors are \(\pm2\) km/s and \(\pm0^{\circ}.01\). Adapted from  \cite{2018tegp.book.....W}.\label{tab:CMB}}
\centering
\begin{ruledtabular} 
\begin{tabular}{lc}
 & Velocity vector \(\mathbf{w}\)  \\
\hline
Absolute value: \(w\)  & 368 km/s  \\
Latitude: \(\beta_{PF}\) & \(-11^{\circ}.13\)   \\
Longitude: \(\lambda_{PF}\) & \(171^{\circ}.55\) \\
\end{tabular}
\end{ruledtabular}
\end{table}

In Eq. (\ref{eq:WPFE}), the coordinates \((\beta_{PF}, \lambda_{PF})\) represent, respectively, the ecliptic latitude and longitude that identify the direction of the possible preferred frame represented by the CMB radiation. 

In Eq. (\ref{eq:vTerra}), \(\lambda_0\) represents the ecliptic longitude of the Earth around the Sun at a fixed epoch, while \(\dot{\lambda}_{\oplus}\) represents its angular orbital rate. This angular rate can be approximated with the Earth's mean motion \(n_{\oplus}=\sqrt{G_NM_{\odot}/a^3_{\odot\oplus}}\), assuming for simplicity a circular Earth orbit (where \(M_{\odot}\) and \(a_{\odot\oplus}\) represent, respectively, the mass of the Sun and the astronomical unit).

Substituting these expressions for the velocities \(\mathbf{w}\) and \(\mathbf{v}_{\oplus}\) into Eq. (\ref{eq:long}), we finally obtain:

\begin{equation}
<\dot{\ell}_0({\alpha_1})>_{2\pi}=-2\alpha_1n\frac{wv_{\oplus}}{c^2}\cos \beta_{PF}\sin(\lambda_0 +n_{\oplus}t-\lambda_{PF}) + \quad \dots.
\label{eq:long2}
\end{equation}

Eq. (\ref{eq:long2}) shows that, when PPN effects are considered,
$\ell_0$ acquires a rate of change  proportional, through well known quantities, to the sought PPN parameter $\alpha_1$, and modulated with a yearly period.
The dots in Eq. (\ref{eq:long2}) represent minor contributions and we have eliminated the contribution of the constant \(\mathcal{K}\) and temporarily removed the unmodeled or poorly modeled periodic effects. These effects will be considered separately in the analysis of the main sources of systematic error.

The result obtained for the longitude rate \(<\dot{\ell}_0>_{2\pi}\) coincides with that obtained in \cite{1994PhRvD..49.1693D} and reported in their Eq. (56), where the effect is expressed in equatorial coordinates rather than in ecliptic coordinates. It is in fact sufficient to set the obliquity of the ecliptic equal to zero in their derivation to obtain the result of Eq. (\ref{eq:long2}).

\section{Orbit Determination and Residuals}\label{sec:POD}
Precise Orbit Determination (POD) is the process of accurately determining the position and velocity of a satellite (its state-vector) at any given time as it orbits the Earth. The process involves calculating the satellite's orbit with a very high level of precision and accuracy, using a combination of tracking measurements and modeling techniques.
{This analysis focuses on the POD of the two geodetic satellites LAGEOS and LAGEOS II. Using two satellites allows for the reduction of the errors in the measurement of \(\alpha_1\), particularly those arising from the Earth's gravitational field.}
The analysis for the data reduction of the {satellites} normal points  was carried on using two different software packages: GEODYN II, developed at NASA-GSFC \cite{1998pavlis}, and SATAN package (SATellite ANalysis), developed at the Royal Greenwich Observatory in the UK \cite{SinclairAppleby1988}.  This helps to avoid possible biases from the dynamical model implemented.
Both codes are routinely used by some ILRS analysis centers for the POD of geodetic satellites and, in particular, of the two LAGEOS.

{Laser ranging data are available in two formats: full-rate and normal points. Full-rate are the observations collected directly from the ILRS stations, while normal points are obtained by averaging the original observations collected over a given time interval, 2 minutes { for LAGEOS and LAGEOS II}.}
In the case of the analysis carried on by GEODYN II, the POD integration step size is 60 s and the period of the analysis covers a timespan of about { 28.3 years, starting from October 30, 1992 (i.e. MJD 48925), about one week after the LAGEOS II launch.}  Conversely, in the case of the analysis performed with SATAN, the analysis covers a timespan of about { 31.4 years, starting from February 7, 1993 (i.e. MJD 49025) and with the same step size of 60 s.}

The overall period of the two analyses was  divided into 7-day arcs not causally connected to each other, and the initial state-vector for each arc was adjusted in the POD procedure to best fit the normal points. Table \ref{tab:elementi} shows the mean orbital elements of LAGEOS and LAGEOS II \footnote{These mean elements were calculated from the Keplerian osculating elements estimated during a previous analysis of the satellite orbit.}.

\begin{table}[h]
  \caption[]{Mean orbital elements { of LAGEOS and }LAGEOS II satellite.}
  \centering
  \begin{tabular}{@{}cccrr@{}}
  \hline
     \hline
\rm{Element}  & Unit & Symbol& \rm{LAGEOS } & \rm{LAGEOS II} \\
\hline
 semi-major axis \, & [{\rm km}] & {\it a}   & 12 270.01   & 12 162.07 		     \\
 eccentricity  & [--] & {\it e}                                & 0.0044		          & 0.0138       	 \\
 inclination \, & [{\rm deg}] & \it {i}                                  & 109.84     & 52.66       		  \\
\hline
\hline
\end{tabular}
\label{tab:elementi}
\end{table}

The modeling in GEODYN accounts for: i){ satellites} dynamics, ii) measurement procedure and iii) reference frames transformations. In this context, our models comply, wherever possible, with the International Astronomical Union (IAU) 2000 Resolutions~\cite{2003AJ....126.2687S} and the International Earth Rotation Service (IERS) Conventions 2010~\cite{2010IERS-Conv-2010}.
In Table \ref{tab:modelli2}  we have summarized the main models included in the orbital determinations made with GEODYN II.
{In Appendix \ref{app:errors},  we described the main sources of systematic errors and the models used to account for key perturbative effects related to the gravitational field, tides and non-gravitational forces.}

\begin{table*}[t] 
\scriptsize
\caption{Models currently used for the POD of GEODYN II and SATAN\footnote{{Reproducibility Note: The headline results for the PPN parameter $\alpha_1$ presented in Section V were obtained using GEODYN II (NASA-GSFC \cite{1998pavlis}) and the SATAN (RGO \cite{SinclairAppleby1988}) packages. The final dynamical configuration for both solvers is defined by the models summarized in this Table \ref{tab:modelli2}, employing a 60 s integration step size and 7-day orbital arcs. Specifically, the GEODYN II analysis covers the MJD 48925–59250 span, while the SATAN analysis covers the MJD 49025–60500 span, both utilizing the IERS Conventions 2010 and IAU 2000 standards as the baseline references.}}. The models are grouped in gravitational perturbations, non-gravitational perturbations and reference frames realizations.\label{tab:modelli2}}
\begin{ruledtabular} 
\begin{tabular}{llll}
Model for &  GEODYN II &  SATAN  & Reference \\
\hline
Geopotential (static) & EIGEN-GRACE02S/GGM05S & EIGEN-GRGS RL04 & \cite{2005JGeo...39....1R,Tapley2013,JGRB:JGRB50058,Lemoine2019} \\
Geopotential (time-varying: even zonal harmonics) & GRACE/GRACE FO &  EIGEN-GRGS RL04 & \cite{Tapley2013,JGRB:JGRB50058,Lemoine2019,PETER20224155} \\
Geopotential (time-varying: ocean tides) & Ray GOT99.2 &  FES2014 & \cite{1999Ray,2021OcSci..17..615L} \\
Geopotential (time-varying: Earth tides) & IERS2010 &  IERS2010 & \cite{2010IERS-Conv-2010} \\
Geopotential (time-varying: non tidal) & IERS Conventions (2010) & FES2014& \cite{2010IERS-Conv-2010,2021OcSci..17..615L} \\
Third--body & JPL DE-403 & JPL DE-200 & \cite{1995Standish} \\
Relativistic corrections & Parameterized post-Newtonian & Parameterized post-Newtonian & \cite{2003AJ....126.2687S,1990CeMDA..48..167H} \\
\hline
Direct solar radiation pressure & Cannonball & Cannonball & \cite{1998pavlis} \\
Earth albedo & Knocke-Rubincam & McCarthy-Martin& \cite{1987JGR....9211662R,1977repMcCarthyMartin} \\
Earth-Yarkovsky & Rubincam  & -- & \cite{1987JGR....92.1287R,1988JGR....9313805R,1990JGR....95.4881R} \\
Solar-Yarkovsky & Farinella-Vokrouhlick\'y &  -- & \cite{1996PSS...44.1551F} \\
Neutral drag & JR-71/MSIS-86 & NRLMSISE-00 & \cite{1976STIN...7624291C,1987JGR....92.4649H,2002JGRA..107.1468P} \\
Spin & LASSOS  &  -- & \cite{2018PhRvD..98d4034V} \\
\hline
Stations position & ITRF2008/2014 &  ITRF2020 & \cite{2011JGeod..85..457A,https://doi.org/10.1002/2016JB013098,2023JGeod..97...47A} \\
Centre of mass corrections & Rodriguez 2019 & Rodriguez 2019 & \cite{2019JGeod..93.2553R}\footnote{$https://datos-geodesia.ign.es/SLR/centre\_of\_mass_models/$.}\\
Ocean loading & Schernek and GOT99.2 tides & FES2014 & \cite{1998pavlis,1999Ray,1999Ray,2021OcSci..17..615L} \\
Earth Rotation Parameters & IERS EOP C04 &  IERS EOP C04\_05 & \cite{IERS-EOP_C04} \\
Nutation & IAU 2000 &  IAU 2000 & \cite{2002JGRB..107.2068M} \\
Precession & IAU 2000 &  IAU 2000 & \cite{Capitaine2003}  \\
\end{tabular}
\end{ruledtabular}
\end{table*}

{ The relativistic effects in GEODYN II} can be included according to the model described in \cite{1990CeMDA..48..167H}. In particular, the model takes into account the accelerations linked to Schwarzschild (or Einstein), Lense-Thirring and de Sitter precessions \cite{1916AnP...354..769E,1918PhyZ...19..156L,1916MNRAS..77..155D}.
However, in the orbital analysis, we did not  model GR in order to obtain, \textit{a posteriori}, residuals in the orbital parameters carrying information on all possible relativistic effects: both the ``classical'' ones due to GR and those (possibly) due to alternative theories of gravitation. Subsequently, from these residuals, we removed the nominal values of the relativistic effects predicted by GR, see Table \ref{tab:rel}. 

{The main effect is due to the Schwarzschild precession, which affects the two Keplerian elements to nearly the same extent (but in opposite direction), due to the small eccentricity of the satellite's orbit.
The Lense-Thirring precession affects only the argument of pericenter. The de Sitter effect does not cause any precession in either the argument of the pericenter or in the mean anomaly of the satellite.
For completeness, we also included the relativistic effect of the Earth's quadrupole moment \(J_2\) on the two orbital elements. 
There are two different contributions: the first is a {\it direct} perturbative effect generated by the static quadrupole moment (\(J_2/c^2\)) on the satellite orbit, and is derived by \(1^{st}\)-order perturbation theory. The second {\it indirect} contribution represents a mixed term, which derives from the interaction of the classical Newtonian \(J_2\) coefficient with the Schwarzschild (i.e. post-Newtonian) metric.
This term comes from a \(2^{nd}\)-order perturbation theory, see \cite{2018CeMDA.130...40S} and references therein for details.
 

\begin{table} [h!]
\caption{Main relativistic precessions on the argument of pericenter and on the mean anomaly of{ LAGEOS and LAGEOS} II. Units are in mas/yr (milli-arc-sec per year). 
\label{tab:rel}}
\begin{ruledtabular} 
\begin{tabular}{lcc}
 \rm{{Rate}} (mas/yr)     &LAGEOS & LAGEOS II\\
 \hline
\(\dot{\omega}_{Schw}\)&+3278.77   & +3352.58  \\
\(\dot{\omega}_{LT}\)  & +31.23 & \(-57.33\)  \\
\(\dot{\omega}_{J_2}^{rel}\) & \(-3.62\)  & +3.01  \\
\hline
Total \(\dot{\omega}_{rel} \)  & +3306.38 &+3298.26  \\
\hline \hline
\(\dot{M}_{Schw}\)    & \(-3278.74\) & \(-3352.26\)  \\
\(\dot{M}_{J_2}^{rel}\) & \(-0.92\)  & \(+0.15\)  \\
\hline
Total  \(\dot{M}_{rel} \)& \(-3279.66\)  & \(-3352.11\)  \\
\end{tabular}
\end{ruledtabular}
\end{table}

In the data reduction, we do not estimate parameters that are not (in general) well known a priori, such as the radiation coefficient \(C_R\) \footnote{This is not the case for LAGEOS II, whose radiation coefficient (\(C_R\simeq1.12\)) has a relative error of 0.5\(\%\), and is therefore sufficiently known. However, we must take into account that this coefficient, if estimated in the POD, acts as a sort of scaling factor for the entire surface of the satellite and is therefore capable of absorbing the imperfections of the dynamical model.} --- this is a measure of the mean reflectivity properties of the satellite surface \cite{1987nongrav.book.....M} --- or particular coefficients of the Earth's gravitational field.
In this way we avoid to unintentionally absorb relativistic effects not included in the dynamic model, as previously highlighted, and gives us confidence in the presence of the relativistic effects in the residuals of the different orbital elements.
In particular, empirical accelerations have not been estimated, as is usually done in the POD. Indeed, to overcome deficiencies in the dynamic model, especially concerning non-gravitational forces, it is customary in a data reduction to introduce and adjust empirical accelerations  (constant and once-per-revolution) in the POD to absorb modeling errors.


After the POD, the residuals are obtained computing the difference between the satellite state-vector estimated at the beginning of each arc, within the POD data reduction, and the estimated state-vector of the previous arc propagated at the same epoch \cite{2006P&SS...54..581L}.
Therefore, the residuals represent the variation of orbital elements along the arc length (7 days) and contain the impact of unmodeled perturbing effects on the satellite's orbit.
Figure \ref{fig:residui1} shows, in the same plot, the residuals in the rate of both  the argument of pericenter \({\omega}\) and of the mean anomaly \({M}\){ for LAGEOS and LAGEOS II in the case of the analysis performed with GEODYN.}

 \begin{figure}[h!]
 	\centering
 	\begin{subfigure}
 		\centering
 		\includegraphics[width=0.4\textwidth]{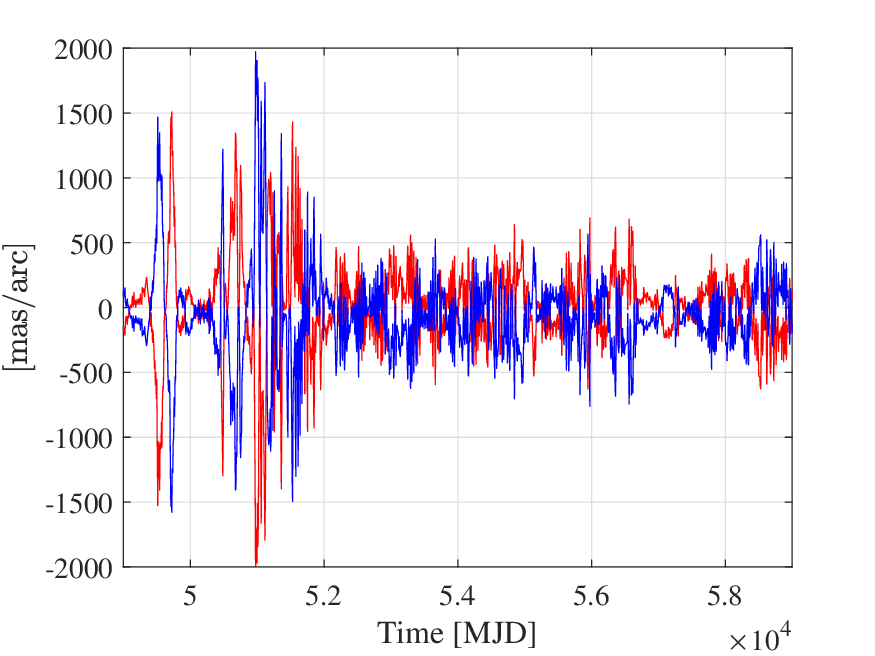}
 	
 	\end{subfigure}
 \begin{subfigure}
 	\centering
 	\includegraphics[width=0.4\textwidth]{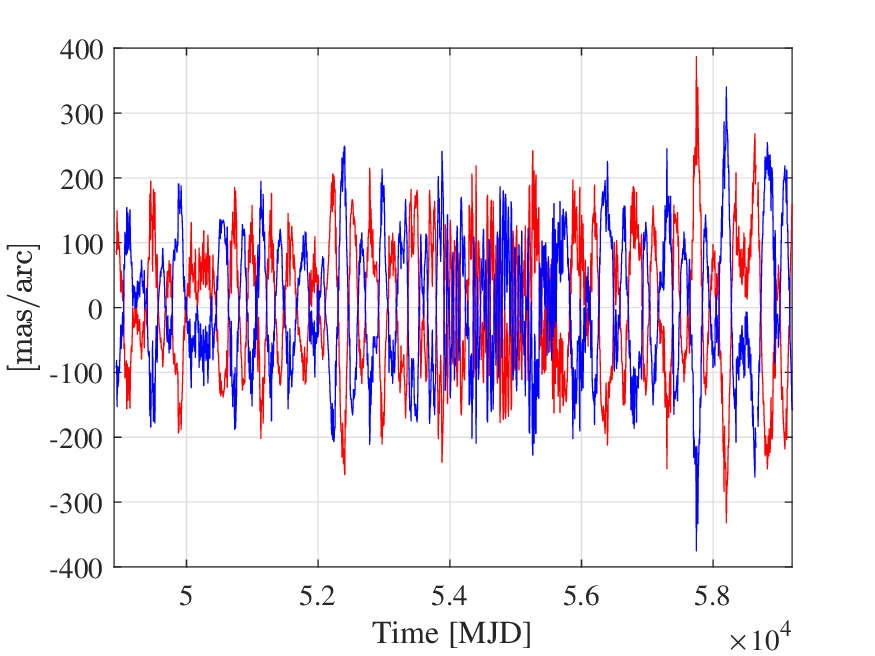}
 	
 \end{subfigure}
 \caption{{GEODYN II analysis: LAGEOS (top) and LAGEOS II (bottom)} 7-day residuals in the rate of the argument of pericenter (red) and in the rate of the mean anomaly (blue) as a function of time. \label{fig:residui1}}
 \end{figure}
 

As can be seen  by comparing the residuals versus time with their sum, the effects tend to almost cancel out, as highlighted in Section \ref{sec:PFE}.
In this way we achieve a significant reduction of the residuals for \(\dot{\ell}_0=\dot{\omega}+\dot{M}\), as shown in Figure \ref{fig:residui2}.
The amplitudes of the oscillations shown in Figure \ref{fig:residui1} are in fact significantly reduced: from a factor of 10 to a factor of 40 {in the case of LAGEOS II} over the time interval of the analysis, based on a simple visual inspection, {and by even a greater factor for the older LAGEOS.}

%

 \begin{figure}[h!]
 \includegraphics[width=0.4\textwidth]{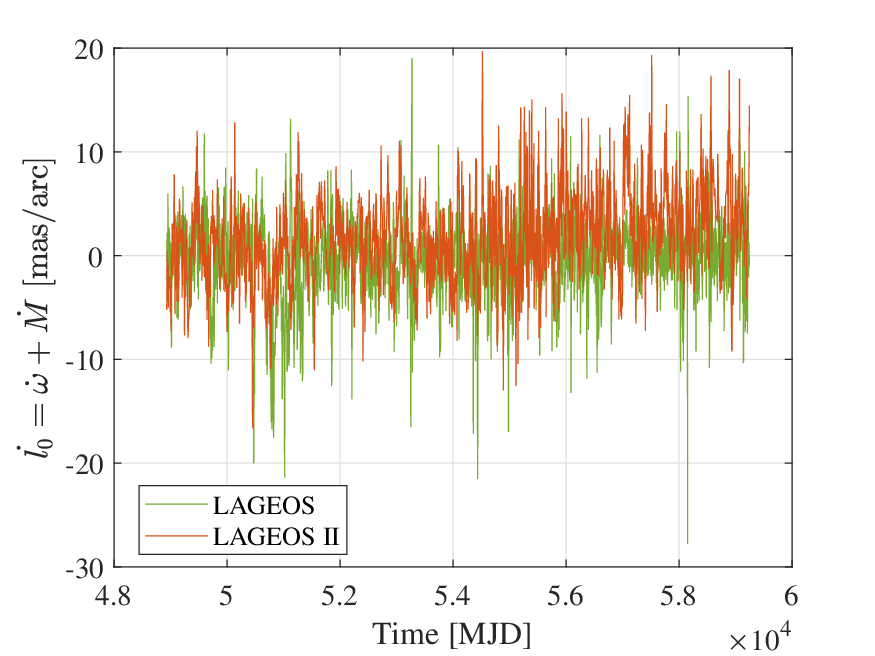}
 \caption{{GEODYN II analysis: LAGEOS and LAGEOS II} residuals in the rate of the longitude \({\ell}_0={\omega}+{M}\) versus time.  Note the different scale along the vertical axis with respect to Figure \ref{fig:residui1}.
\label{fig:residui2}}
 \end{figure}
 
These residuals were obtained by adding the residuals of Figure \ref{fig:residui1} after removing, on the basis of the rates reported in Table \ref{tab:rel}, the total relativistic precession predicted by GR on the rate of the argument of pericenter and on the rate of the mean anomaly. 

{As highlighted above, to cross-validate our results and identify possible biases in GEODYN II, we repeated the POD of the two LAGEOS satellites with SATAN.}
{The SATAN dynamical model (see Table \ref{tab:modelli2}) differs from the one implemented in GEODYN II primarily for the adoption of the following models \footnote{Specifically, we mean the differences in the models used by the two software in our analyses.}: the ITRF-2020 terrestrial reference frame, the EIGEN-GRGS RL04 mean-field model for the background gravity field \cite{Lemoine2019}, and the FES2014 model for ocean tides and ocean loading \cite{2021OcSci..17..615L}. We also note that, unlike the analysis conducted with GEODYN II, for the POD results obtained with SATAN we estimated the range biases for each station and satellite. This is now a common practice at ILRS Analysis Centers, which consists of accounting for systematic errors in the data or their modeling. The disadvantage of this approach is greater noise in the estimated parameters. The advantage is that it accounts for the possible presence of systematic errors, which would result in constant range errors over an orbital arc. 
The long-term performance comparison yielded positive results. For instance, both packages successfully identified signatures in the residuals arising from unmodeled thermal thrust effects in the POD. Furthermore, we observed consistent agreement in the predicted secular precession induced by relativistic effects, such as the Schwarzschild precession of the argument of pericenter for the LAGEOS satellites.

Figure \ref{fig:satan1} shows the residuals in the rate of the argument of pericenter and in the rate of the mean anomaly obtained with SATAN in the case of LAGEOS and LAGEOS II over the time interval of  the  analysis. The trend of the residuals for the two rates obtained by SATAN is fairly consistent with the residuals obtained by GEODYN II, see Figure \ref{fig:residui1}.

 \begin{figure}[h!]
 	\centering
 	\begin{subfigure}
 		\centering
 		\includegraphics[width=0.4\textwidth]{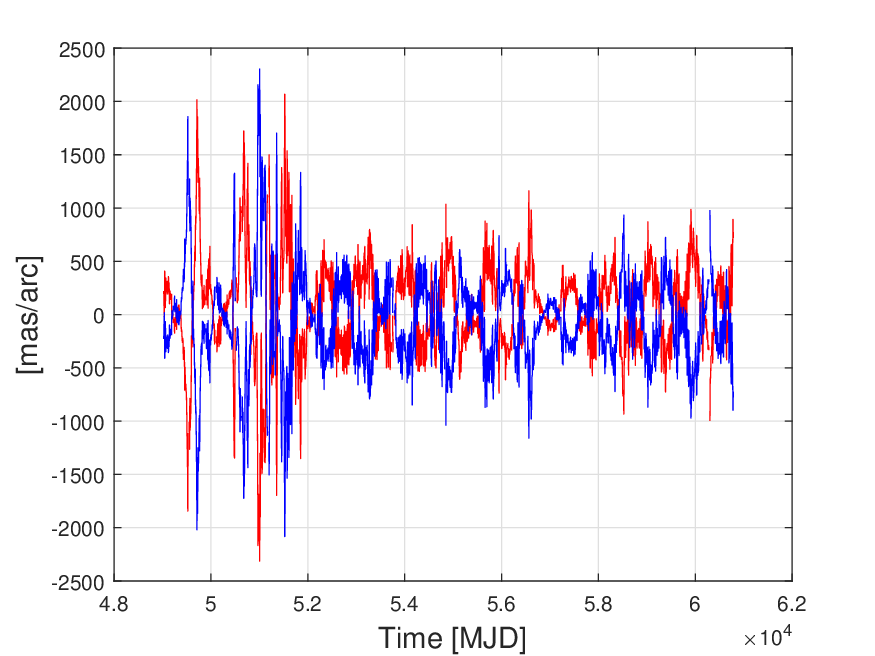}
 	
 	\end{subfigure}
 \begin{subfigure}
 	\centering
 	\includegraphics[width=0.4\textwidth]{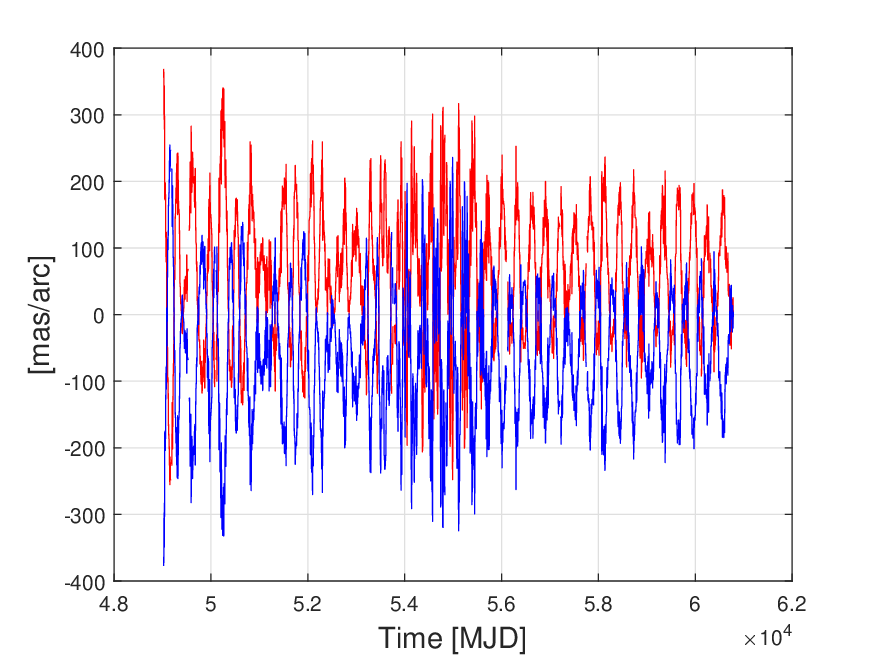}
 	
 \end{subfigure}
 \caption{{LAGEOS (top) and LAGEOS II (bottom)} 7-day residuals in the rate of the argument of pericenter (red) and in the rate of the mean anomaly (blue) as a function of time from the analysis performed with SATAN. \label{fig:satan1}}
 \end{figure}
 

{Figure \ref{fig:satan2} shows the mean argument of latitude for LAGEOS, as determined by GEODYN II (blue line) and SATAN (red line). The agreement between the two independent analyses is good, with in-phase oscillations observed across the entire time span.}

 \begin{figure}[h!]
 \includegraphics[width=0.4\textwidth]{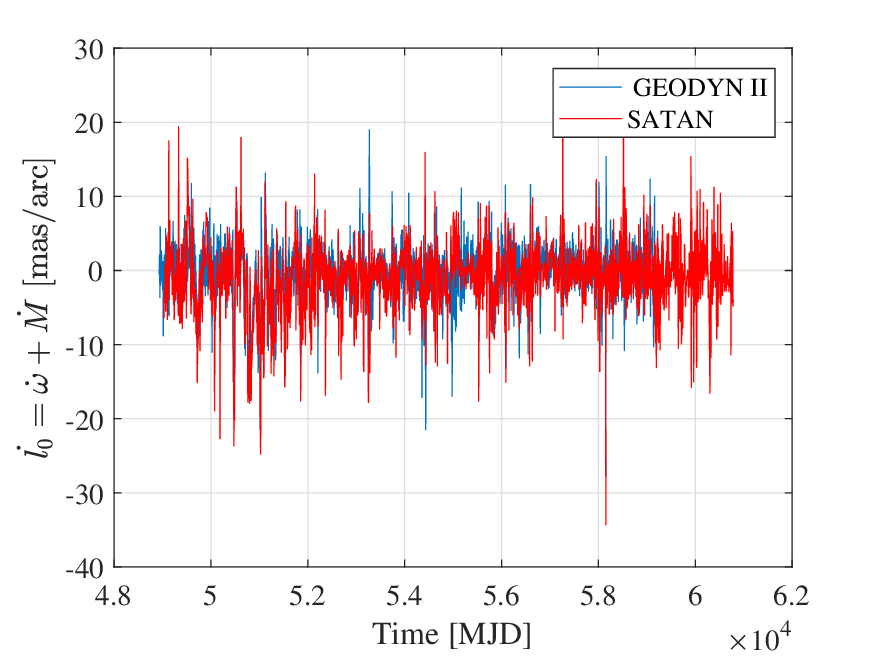}
 \caption{LAGEOS 7-day residuals in the rate of the mean argument of latitude (blue) obtained from the GEODYN II analysis compared with the same residuals obtained from the analyis performed with SATAN (red). 
\label{fig:satan2}}
 \end{figure}
 

{We obtained comparable results for the analogous observable constructed from LAGEOS II orbital residuals.}

\section{Measurement concept}\label{sec:concept}
The primary effect on the longitude \(\ell_0\) of a satellite that may depend on the existence of a preferred frame is an oscillation with annual periodicity and with an amplitude proportional to the PPN parameter \(\alpha_1\), as shown in Eq. (\ref{eq:long2}).
However, numerous periodic and secular effects --- both gravitational and non-gravitational --- mask this possible gravitational effect.
This was clearly confirmed by the orbital residuals of Figure \ref{fig:residui1}, where a number of unmodeled signals are present over a wide spectrum.
These signals combine in \(\dot{\ell}_0\) in a rather complex oscillation,  as is illustrated in Figure \ref{fig:residui2}. 

{We can take advantage of  the two independent measurements of the longitudes $\ell_0$ of the LAGEOS satellites. Indeed, as briefly highlighted in the previous section, we can exploit the two independent measurements of the longitude $\ell_0$ of the LAGEOS satellites to reduce the error in the measurement of the parameter \(\alpha_1\). As will be widely discussed in Appendix \ref{app:grav}, the main systematic error on these observable is due to the imperfect knowledge of the Earth's quadrupole coefficient $\bar{C}_{2,0}$  used in the POD.  Following Kozai \cite{1959AJ.....64..367K},  and neglecting the contribution of the terms of  \(\mathcal{O}(e^2)\), we can relate the error $\delta\bar{C}_{2,0}$ with that of the longitude rate $\delta(\dot{\ell}_0)$:}	
\begin{equation}
		\delta(\dot{\ell}_0)|_{\delta\bar{C}_{2,0}} \simeq \frac{3}{2}\sqrt{5}\left(\frac{R_{\oplus}}{a}\right)^2n(3-4\sin^2 i)\delta\bar{C}_{2,0}\label{eq:errC20}.
\end{equation}  
{By using equations (\ref{eq:errC20}) and (\ref{eq:long2}) together, we can treat the longitudes  $\dot{\ell}_0^{LG1}$ and $\dot{\ell}_0^{LG2}$ obtained by the PODs of LAGEOS and LAGEOS II as the two known terms in a linear system of two equations in two unknowns.  The two unknowns are the variables $\delta\bar{C}_{2,0}$ and $A_1(t)$, the latter containing the possible contribution of $\alpha_1$ at the annual frequency, but net of its main systematic error $\delta\bar{C}_{2,0}$}. The system to solve is:
\begin{equation}
\left\{\begin{array}{@{}l@{}}
\dot{\ell}_0^{LG1}=K_1^{LG1} \cdot A_1(t) + K_2^{LG1}\cdot \delta\bar{C}_{2,0} \\
\dot{\ell}_0^{LG2}=K_1^{LG2} \cdot A_1(t) + K_2^{LG2}\cdot \delta\bar{C}_{2,0} 
\end{array}\right.\, 
\label{equ_comb}
\end{equation}
with
\begin{eqnarray*}
K_1^{LG}&=&-2n^{LG}\frac{wv_{\oplus}}{c^2}\cos \beta_{PF}\\
K_2^{LG}&=&\frac{3}{2}\sqrt{5}\left(\frac{R_{\oplus}}{a^{LG}}\right)^2n^{LG}(3-4\sin^2 i^{LG})	
\label{system}
\end{eqnarray*}
where $n^{LG}$, $i^{LG}$ and $a^{LG}$ are respectively the mean motion, the semi-major axis and the inclination of the satellite LG (see Table \ref{tab:elementi}).   

{Figure \ref{fig:alpha1_t} shows the results for the variable $A_1(t)$, while  Figure \ref{fig:FFT_alpha1_t} shows its Fast Fourier Transform (FFT).  The FFT reveals several spectral lines, including one with an annual period and an amplitude of about $3\times 10^{-5}$ in the GEODYN analysis and not greater than $5\times 10^{-6}$ in the case of SATAN.}


%

 \begin{figure}[h!]
 	\centering
 	\begin{subfigure}
 		\centering
 		\includegraphics[width=0.4\textwidth]{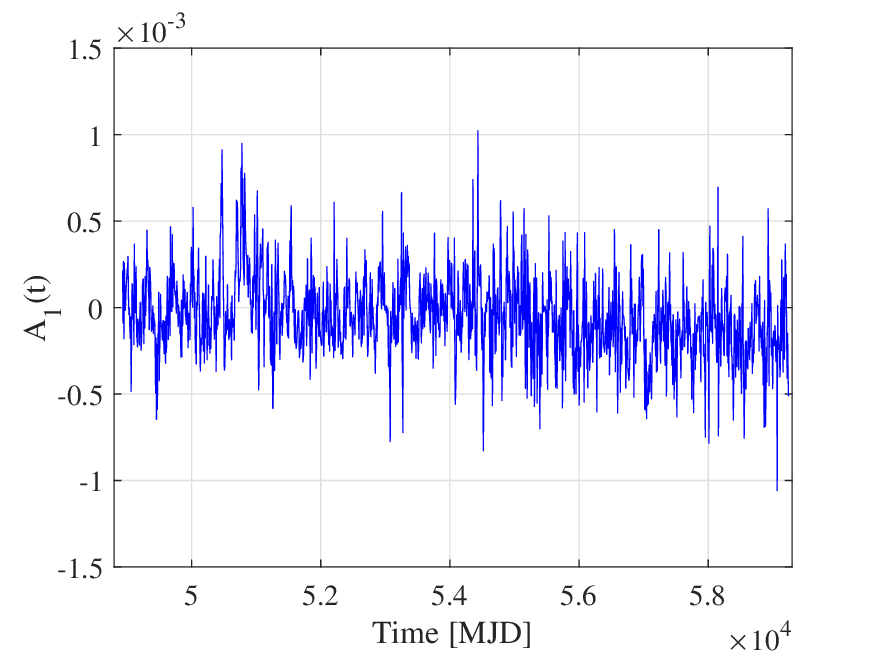}
 	
 	\end{subfigure}
 \begin{subfigure}
 	\centering
 	\includegraphics[width=0.4\textwidth]{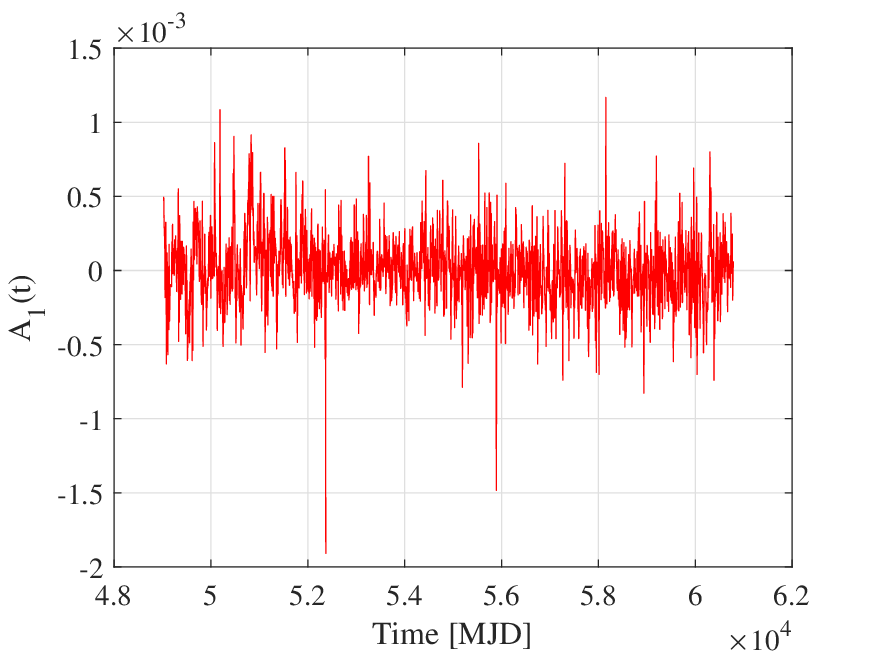}
 	
 \end{subfigure}
 \caption{GEODYN II (top) and SATAN (bottom). Time behavior for the variable $A_1(t)$  from the solutions of the system shown in Eq. (\ref{equ_comb}). \label{fig:alpha1_t}}
 \end{figure}
 

 \begin{figure}[h!]
 	\centering
 	\begin{subfigure}
 		\centering
 		\includegraphics[width=0.4\textwidth]{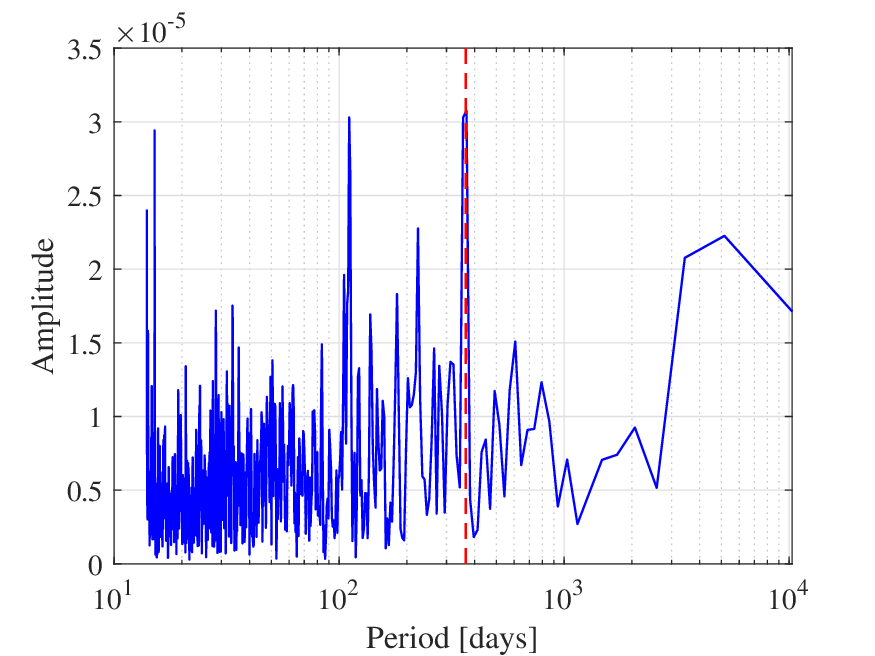}
 	
 	\end{subfigure}
 \begin{subfigure}
 	\centering
 	\includegraphics[width=0.4\textwidth]{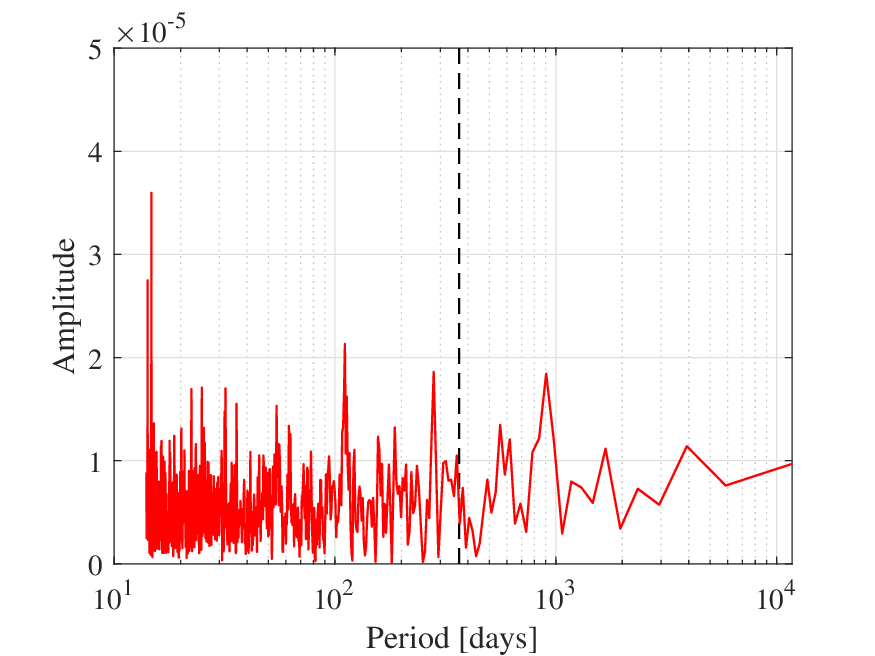}
 	
 \end{subfigure}
 \caption{GEODYN II (top) and SATAN (bottom). FFT of the variable $A_1(t)$ of Figure \ref{fig:alpha1_t}. \label{fig:FFT_alpha1_t}}
 \end{figure}
 

{We are interested in this signal, and therefore we treat all the other periodic effects in Figure \ref{fig:FFT_alpha1_t} as a ``noise'' that masks the ``physical'' signal we aim to measure.}

To isolate our signal, we adopt a Phase-Sensitive Detection (PSD) scheme (also called a Lock-In scheme \cite{Cosens_1934,1941RScI...12..444M}): a homodyne demodulation at the target frequency (the annual one) and phase, followed by a low-pass filter. 
The PSD is achieved by multiplying the input data stream by a sine wave (the reference)  with the target frequency $f_{ref}=f_0$ and target phase $\phi_0$. This process shifts all the frequencies in the input "down" (and also up) by an amount $f_{ref}$. The demodulated signal is then low-pass filtered to remove components at \(f \ne f_0\) \footnote{Indeed, thanks to the relation $\sin^2(x) = \frac{1- \cos2x}2$, the signal at $f_0$ splits into a d.c. component and one oscillating at twice the target frequency.}.
In this way we obtain the amplitude of the target signal as a function of time at zero frequency.

In conclusion, applying the phase-sensitive detection to Eq. (\ref{eq:long2}), we obtain:

\begin{equation}
\alpha_1 =  2<\sin(\lambda_0 +n_{\oplus}t-\lambda_{PF}) {A_1(t)}>_{2\pi}.\label{alpha}
\end{equation}

This DC component represents the possible violation signal to be extracted from the final data.
As for the other solution of the system reported in Eq. (\ref{equ_comb}), \(\delta\bar{C}_{2,0} \), this will be discussed within the analysis of systematic errors in Appendix \ref{app:errors}.

\section{Measurement and constraints}\label{sec:measure}
{From the previously described analyses of the LAGEOS and LAGEOS II orbits --- a 28.3-year period with GEODYN II and a 31.4-year period with SATAN --- we derived the time-varying quantity $A_1(t)$, which is sensitive to the possible annual contribution from the PPN parameter $\alpha_1$.}


The PSD technique described in Section \ref{sec:concept} was then applied to this new observable using a reference signal with annual frequency \(f_{0}\simeq 2.738\times 10^{-3}\) days\(^{-1}\) and phase \(\phi_0=\lambda_0-\lambda_{PF}\). 
The initial phase $\phi_0$ was fixed for each analysis based on the Earth's ecliptic longitude, $\lambda_0$, at the respective start dates. For the GEODYN II analysis (starting MJD 48925), $\lambda_0 \simeq 223^{\circ}.83$ required $\phi_0 = 52^{\circ}.28$, while for the SATAN analysis (starting MJD 49025), $\lambda_0 \simeq 318^{\circ}.23$ required $\phi_0 = 146^{\circ}.68$.


We tuned the parameters for our PSD analysis --- specifically the integration time and the order of the Butterworth low-pass filter --- by simulating the expected signal for a range of hypothetical $\alpha_1$ values. These same simulations were also used to define a measurement interval that avoids the filter's boundary effects, where a pure sinusoidal input would not yield a constant amplitude output. Consequently, we restricted our final analysis to the interval between day 2000 (MJD 50925 for GEODYN II and MJD 51025 for SATAN) and day 9000 (MJD 57925 for GEODYN II and MJD 58025 for SATAN) of the full orbital dataset.



Figures \ref{fig:residui3a} and \ref{fig:residui3b} show the results for $A_1(t)$ --- respectively for  the analyses performed with GEODYN II and SATAN ---  at the output of the detection chain. 
The DC component estimated from this plot represents the value of \(\alpha_1\) and, therefore, of a possible violation of the LLI: $\alpha_1 = <A_1(t)^{PSD}>$, see  Eq. (\ref{alpha}).
We used a 3-order low-pass filter, corresponding to an attenuation of about 60 dB per decade of the amplitude of the signal above the cutoff frequency. The low-pass filter removes the higher frequencies and leaves the long-term trend we are looking for. A value of 3000 days was assumed for the integration time of the low-pass filter.
 \begin{figure}[h!]
 \includegraphics[width=0.5\textwidth]{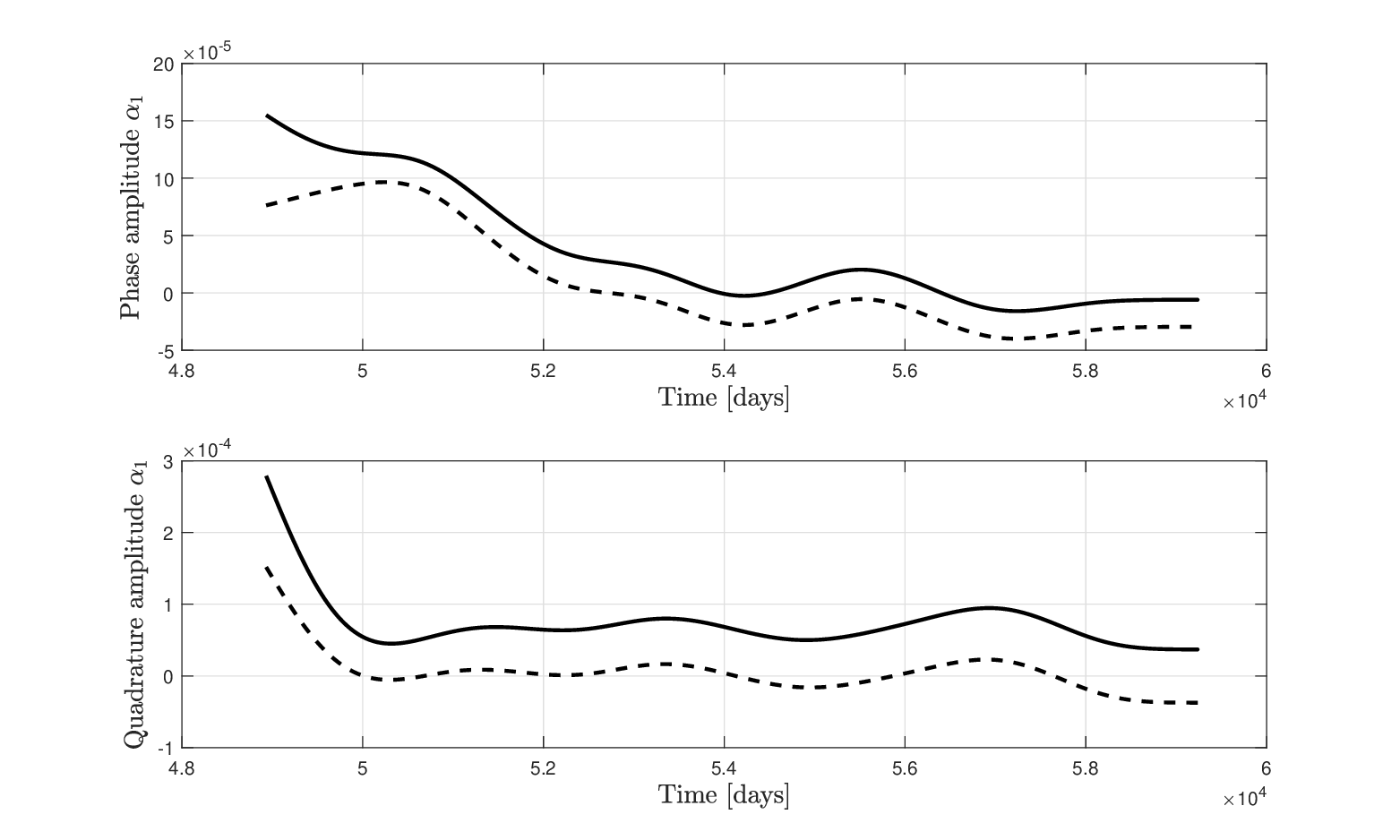}
 \caption{GEODYN II analysis. Time behavior of \(A_1(t)\) after the PSD for the in-phase component (top plot) and the quadrature component (bottom plot). The dashed lines give the PSD (or Lock-In) response to a modified signal from which the part of the signal around the annual periodicity has been significantly reduced. \label{fig:residui3a}}
 \end{figure}
 \begin{figure}[h!]
 \includegraphics[width=0.5\textwidth]{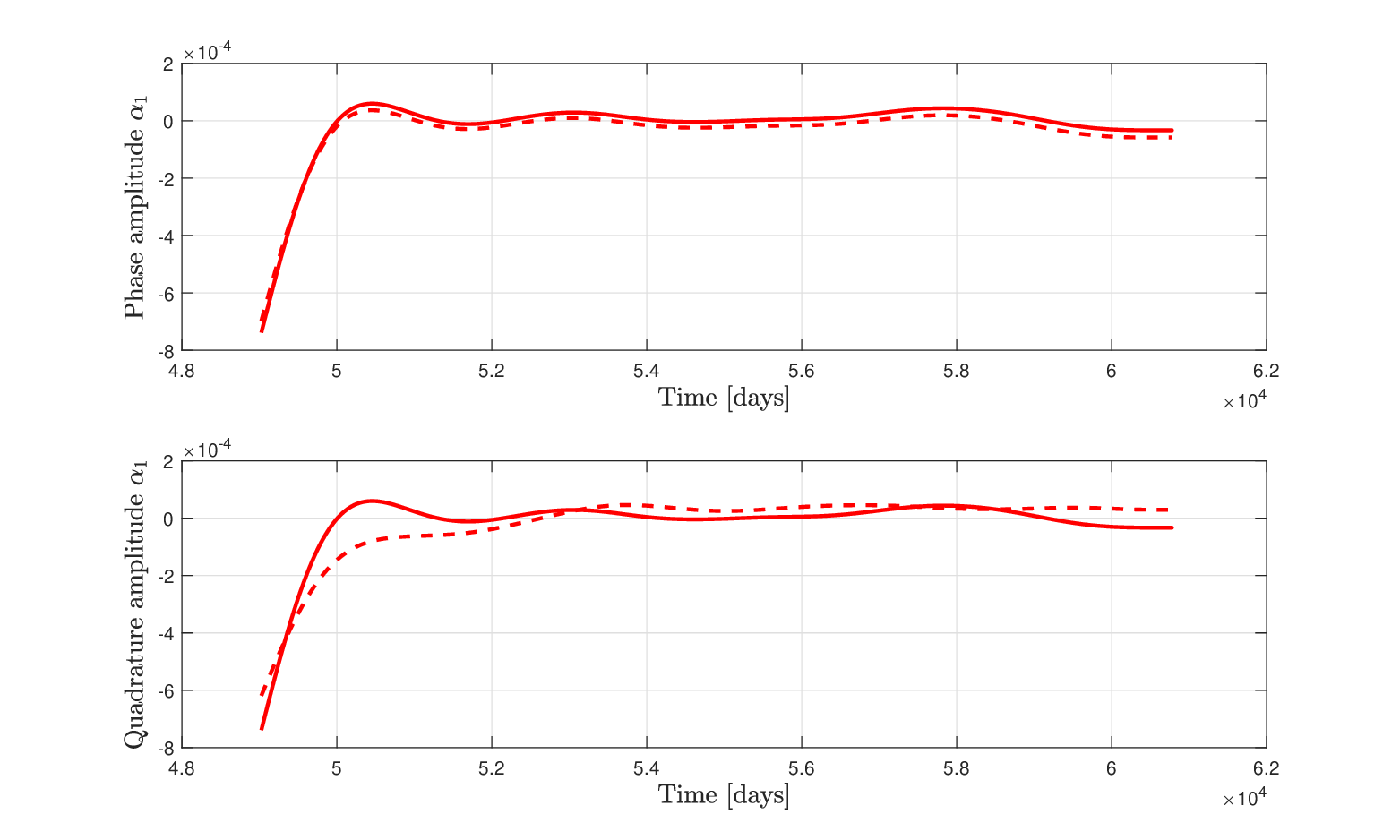}
 \caption{SATAN analysis. Time behavior of \(A_1(t)\) after the PSD for the in-phase analysis (top plot) and the quadrature analysis (bottom plot).  The dashed lines give the PSD (or Lock-In) response to a modified signal from which the part of the signal around the annual periodicity has been significantly reduced. \label{fig:residui3b}}
 \end{figure}

From the PSD output in the time interval between day 2000 and day 9000, within which we do not expect {that the  signal} is altered by the low-pass filter (ringing effects), we compute the mean and standard deviation of the constrained parameter and obtain:

\begin{equation}
\alpha_1 = \left(+2\pm3\right)\times 10^{-5},\label{risultato}
\end{equation}

in the case of the analysis performed with GEODYN II and

\begin{equation}
\alpha_1 = \left(+1\pm2\right)\times 10^{-5},\label{risultato2}
\end{equation}

in the case of the analysis performed with SATAN.
We verified that the obtained constraints did not depend on the filter characteristics, by changing its order between 3 and 5 and the integration time between 2000 and 5000 days.
These figures also shows the Lock-In response (dashed line) to a 
modified signal obtained by reducing  by a factor of 1000 the amplitude of the component at the annual frequency \(f_0\), through a notch filter with a width of \(\pm  0.25\) d\(^{-1}\), thus eliminating from our output every component oscillating at the sought signal frequency.

We must highlight the source of the standard deviation we derived from the PSD, i.e., the origin of the statistical error in our measurements. The main contribution to the standard deviation does not come from amplitude variations of the target signal, but rather from broadband "noise," specifically the signal components around the annual peak. 
This is evident by observing the PSD response to the notched signal. 
As can be seen, the modified signal (dashed lines) maintains, after the PSD, the same time behavior of the original signal (continuous lines).
This suggests that the amplitude variation at the PSD output is not mainly due to the annual component.
We further verified this conclusion using several synthetic random signals with a Gaussian amplitude distribution and a power spectral density equivalent to that of the physical signal after the annual peak has been removed. These simulations also confirmed that the signal's standard deviation depends primarily on spectral components outside the annual peak. See Appendix \ref{app:sintetici} for details.

{The same time evolution, without and with the annual notch, is shown in Figures \ref{fig:residui3a} and \ref{fig:residui3b} for the quadrature component of the PSD (where a phase $\pi/2 $ is added to the phase of the reference signal). This component does not contain any signal at the sought frequency and phase. We observe a behavior quite similar to that of the in-phase component, and this suggests again that the time dependence of the PSD output does not depend on a physical signal.}

In the case of the analysis performed with GEODYN II, Figure \ref{fig:scatter} provides a synoptic view of the result obtainable for the constraint on the PPN parameter \(\alpha_1\) by varying the PSD frequency around the annual frequency (the period of the reference sinusoid is indicated in the plot) and the phase over a complete cycle. 
The black circle indicates the result obtained for the constraint on the parameter \(\alpha_1\) at the expected frequency and phase for the possible violation.
The figure shows the expected correlation existing between the period and the phase of the reference sinusoid.
A similar figure was obtained with the analysis conducted with SATAN.

 \begin{figure}[h!]
 \includegraphics[width=0.45\textwidth]{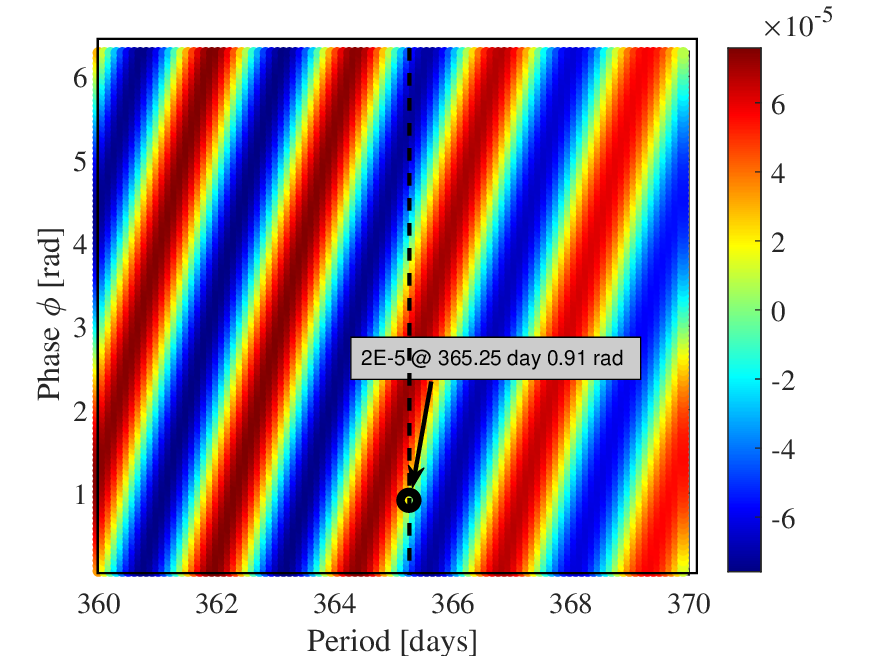}
 \caption{Scatter plot for \(\alpha_1\) in the case of the in-phase analysis of Figure \ref{fig:residui3a}. The horizontal axis shows the period of the reference sinusoid used for homodyne detection, while the vertical axis represents its phase variation.\label{fig:scatter}}
 \end{figure}
%

Figure \ref{fig:alpha1} represents  the estimate obtained for the parameter \(\alpha_1\) by further varying the phase of the (ideal) demodulation sinusoid: with
\(\phi=[0, 2\pi]\) while keeping the reference period at 365.25 days. 
The results of the analyses obtained with GEODYN II and SATAN, respectively, are shown in blue and red.
The  solid line shows how the estimate of the PPN parameter (mean value over the analyzed interval) varies  while the dashed line shows the variation of the standard deviation.
The blue solid line corresponds to the black dashed line shown in Figure \ref{fig:scatter}.


 \begin{figure}[h!]
 \includegraphics[width=0.5\textwidth]{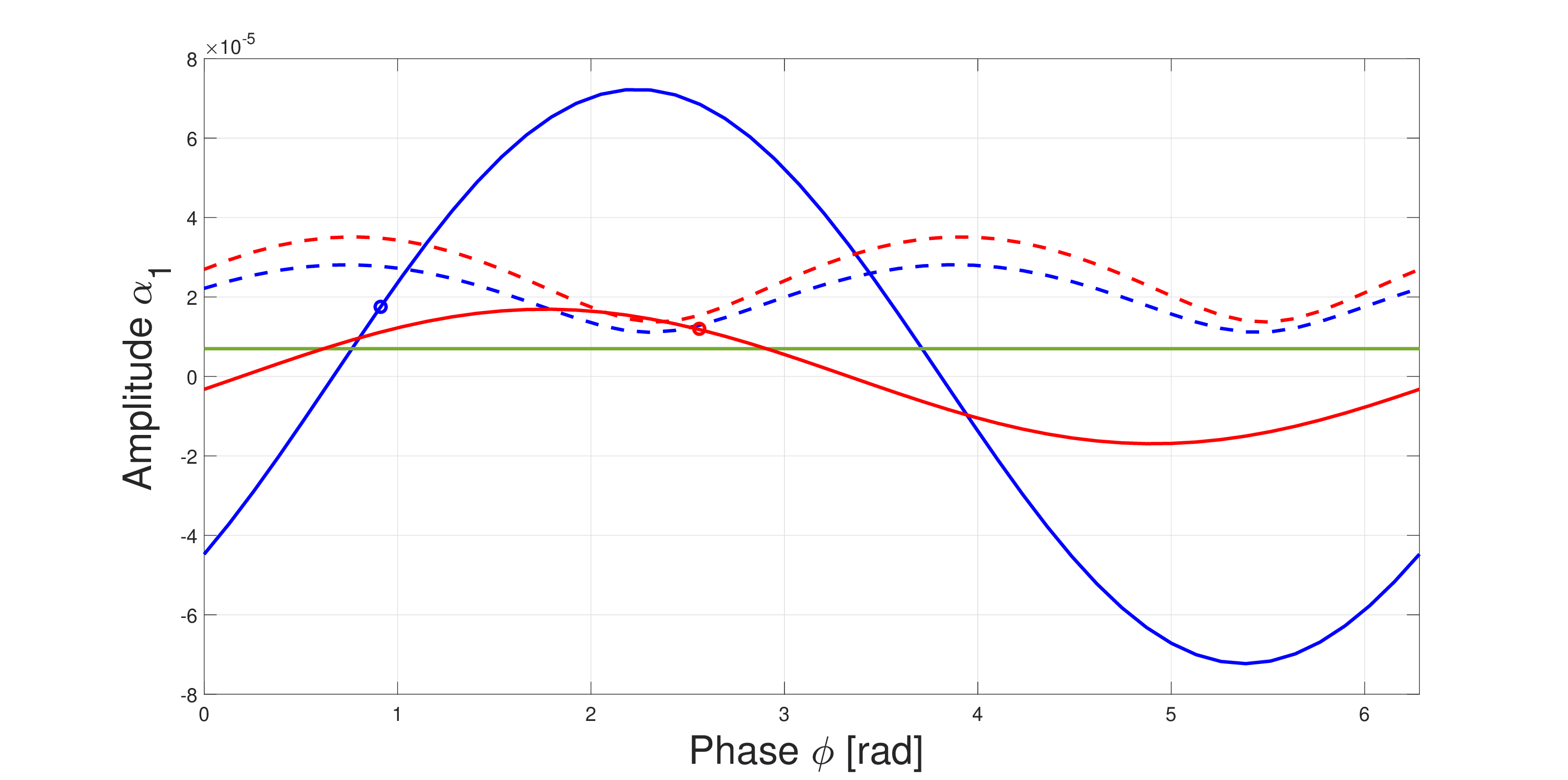}
 \caption{Behavior of \(\alpha_1\) (solid line: blue for GEODYN II and red for SATAN) as the phase of the demodulation sinusoid varies. The frequency of the sinusoid is fixed at the annual value. The blue and red dashed lines represent the standard deviation. 
The two circles, blue and red, correspond to the values of $\alpha_1$ obtained for GEODYN II and SATAN respectively, see Eqs. (\ref{risultato}) and (\ref{risultato2}).
The green continuous line shows the overall estimated systematic error to \(\alpha_1\)  due to both gravitational and non-gravitational perturbations, about \(8\times10^{-6}\). 
The filter parameters are the same as those used for Figures \ref{fig:residui3a} and \ref{fig:residui3b},  the measurement interval is [2000, 7000] days.\label{fig:alpha1}}
 \end{figure}

The value of \(\alpha_1\) at the  phase \(\phi = \phi_0 = 0.9125\) rad (blue circle) corresponds to the specific case of Figure \ref{fig:residui3a} (top).
{It is worth emphasizing that the maximum value of \(\alpha_1\) is not obtained at the phase \(\phi = \phi_0\), but at a higher value equal to approximately 2.18 rad. This reinforces our conclusion that what we obtained at the expected frequency and phase value for the violation, actually represents a null measurement for the PPN parameter.}

{In Figure \ref{fig:alpha1}, the horizontal green line represents the overall systematic error \(\lesssim8\times10^{-6}\) in the measurement of \(\alpha_1\) estimated in Appendix \ref{app:errors}, which we consider to be independent of the phase of the possible violation signal.}
{The indicated value refers to the resulting error on the joint measurement provided by the analysis of the two orbits according to Eq. (\ref{eq:alpha1_sys}) in Appendix \ref{app:errors}, obtained from the solution of the system (\ref{equ_comb}). This overall error is dominated by the gravitational field error related to the hexadecapole coefficient \(\bar{C}_{4,0}\).}

We also performed a sensitivity analysis using a Monte Carlo simulation, quantifying how the uncertainties in the preferred-frame velocity and direction (that of CMB dipole) propagate into the final uncertainty in $\alpha_1$.
Specifically, in the Monte Carlo we repeated the entire analysis pipeline leading to the measurement of $\alpha_1$. In each run, the three parameters defining the CMB reference frame (absolute velocity w, ecliptic latitude $\beta_{PF}$ and ecliptic longitude $\lambda_{PF}$) were randomly varied within (reasonable) ranges, but above their estimated uncertainties (see Table \ref{tab:CMB}). We found that the uncertainties in these parameters propagate on the final uncertainty in $\alpha_1$ down to $\sim10^{-7}$, well below the quoted statistical errors of $3\times10^{-5}$ and $2\times10^{-5}$. 


{The analysis that yielded  the upper limit on the $\alpha_1$ value in Eqs. (\ref{risultato}) and (\ref{risultato2}) was aimed at reducing  noise at the annual frequency, therefore  we used a Lock-In integration time of 3000 days. However, this yielded few independent samples over the 7000-day period, precluding any consideration about  variations in $\alpha_1$ estimate. To address this, we reduced the low-pass filter time constant to 500 days in the case of SATAN analysis, see Figure \ref{fig:time_ev}. This produced less correlated data (blue line). Averaging these points over 500 days yielded 21 nearly uncorrelated  $\alpha_1$ estimates (red points with error bars). The higher cutoff frequency allowed us to extend the observation time to the 600–11000 day span.
In this case, the estimated value of the PPN parameter is  $\alpha_1=(0.7\pm4)\cdot10^{-5}$.  The noise increased by a factor of approximately 2 compared to previous measurements due to wider bandwidth associated with the shorter integration time. 	
We tested for possible trends in these less correlated data. A linear regression of the current data resulted in a slope of $2\cdot10^{-9}$~d$^{-1}$, with a standard deviation of $3\cdot10^{-9}$~d$^{-1}$ and a p-value of 0.39. Therefore, we conclude that there is no statistically significant evidence that the parameter $\alpha_1$ changes over time. The observed variations are attributable to random noise. 
\begin{figure}[h!]
	\centering
	\includegraphics[width=0.45\textwidth]{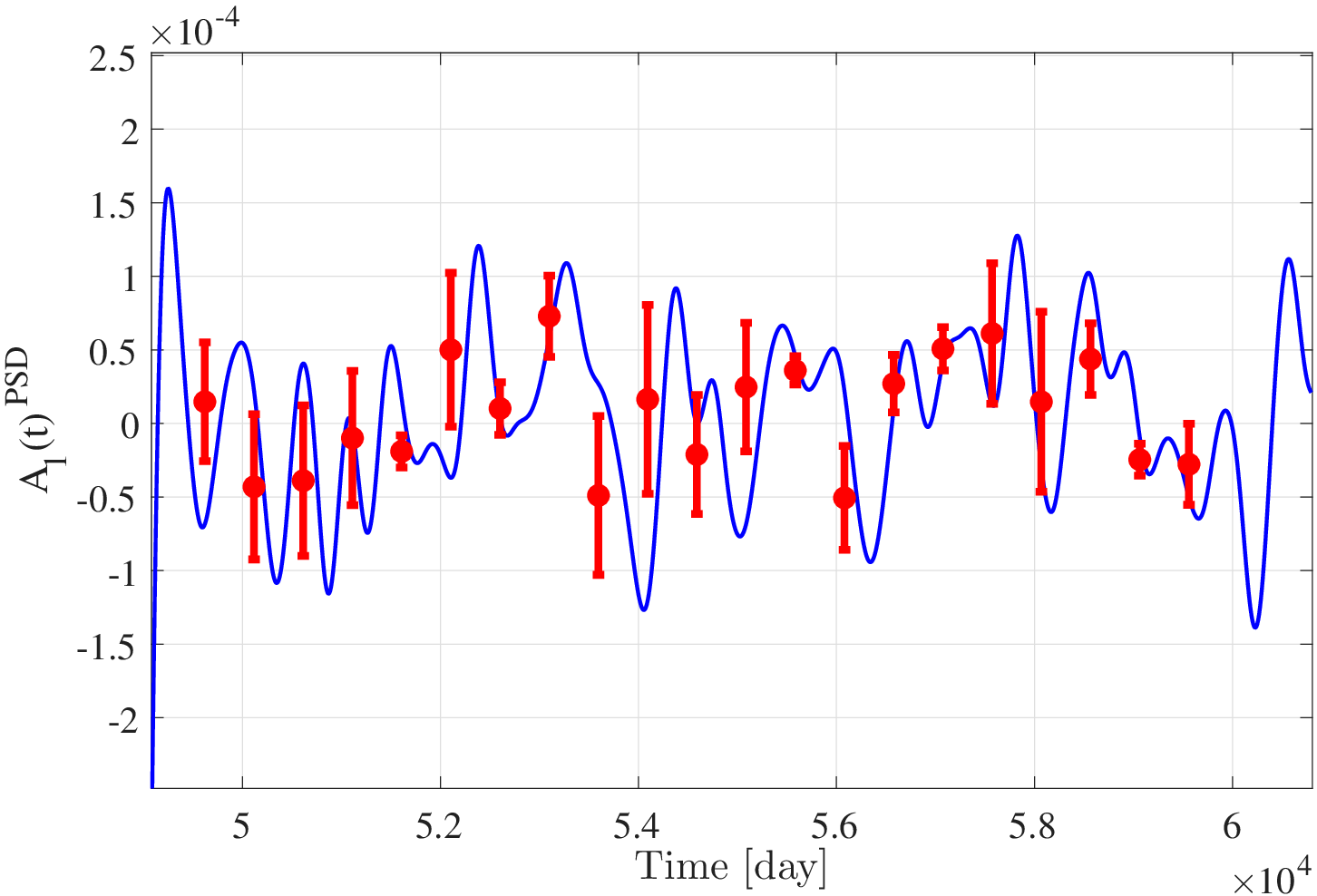}
	\caption{SATAN analysis. Lock-In output (blue line) if the integration time of the low-pass filter is reduced to 500 days. The red points represent the uncorrelated estimates of $\alpha_1$ obtined averaging the output over the reduced integration time.}
	\label{fig:time_ev}
\end{figure}}

\section{Theoretical context}\label{sec:theory}

It is important to reiterate that Lorentz violation is considered a possible residual effect of quantum gravity at low energies, for instance from  Ho{\v{r}}ava-Lifshitz quantum theory \cite{2009PhRvL.102p1301H}. Therefore, identifying possible evidence of Lorentz violations will provide insights into the essence of quantum gravity.
Hence, the tiny upper limit we obtained for $\alpha_1$ ($|\alpha_1|<3\times10^{-5}$ and $|\alpha_1|<2\times10^{-5}$) has  profound implications for alternative theories of gravity.

Vector-tensor theories are the most natural class of models that predict a non-zero  $\alpha_1$, in particular vector theories. These theories introduce a fundamental, dynamical vector field, $K_{\mu}$, in addition to the metric tensor $g_{\mu\nu}$ of GR.
In a cosmological context, this vector field can acquire a non-zero vacuum expectation value ($<K_{\mu}>=u_{\mu}$, where $u_{\mu}$ is a timelike four-velocity) that is aligned with the cosmic time direction. This vacuum expectation value  singles out a specific frame of rest --- the cosmological rest frame --- which we can identify with the frame in which the CMB is isotropic.
This background vector field $u_{\mu}$ acts as a gravitational "aether." The motion of matter (Earth, planets, Sun, $\dots$) relative to this aether leads to new physical effects that are not present in GR. In the weak-field limit, these effects manifest precisely as a non-zero $\alpha_1$, and the magnitude of this parameter is directly determined by the way the vector field couples to the metric and to itself. 

In the following we mainly focus on Einstein-aether theory of gravity \cite{2005LRR.....8....5M}. In Section \ref{sec:AE}, we introduce the constraints that arise on the coupling constants of the theory from the result obtained on the parameter $\alpha_1$. In Section \ref{sec:NZ}, we comment on the theoretical implications of a small but non-zero value of the PPN parameter, in particular the differences between a possible positive and a negative value.

\subsection{Einstein-aether theory}\label{sec:AE}
{A non-zero value of $\alpha_1$ (and also of $\alpha_2$) has  specific implications for each particular gravitational theory, in particular with respect to coupling constants and parameter space. While a  dedicated paper is in preparation, we give here, as  an example, some implications for the Einstein-aether theory.}

Einstein-aether theory was specifically developed to explore possible violations of Lorentz invariance in gravity 
\cite{2001PhRvD..64b4028J,2002cls..conf..331M,2004PhRvD..69f4005E,2005LRR.....8....5M,PhysRevD.70.024003}, similarly to what was done in the case of the Standard Model Extension (SME) \cite{1989PhRvL..63..224K,1997PhRvD..55.6760C,1998PhRvD..58k6002C,2002PhRvD..66e6005K}.  In this theory the vector field is constrained to have unit norm ($K_{\mu}K^{\mu}=-1$) and the following values are expected for the PPN parameters: $\beta=\gamma=1$, $\xi=\alpha_3=\zeta_1=\zeta_2=\zeta_3=\zeta_4=0$, and

\begin{equation}
\alpha_1 =  \frac{8(c_3^2+c_1c_4)}{(c_1^2-2c_1-c_3^2)},\label{eq:EA1}
\end{equation}
\begin{equation}
\alpha_2 =  \frac{\alpha_1}{2}-\frac{(2c_{13}-c_{14})(c_{13}+c_{14}+3c_2)}{c_{123}(2-c_{14})}.\label{eq:EA2}
\end{equation}

In the last equation $c_{13}=c_1+c_3$, $c_{14}=c_1+c_4$ and $c_{123}=c_1+c_2+c_3$. We refer to \cite{2018tegp.book.....W} for a comprehensive description of the PPN parameters in the framework of metric theories of gravity.
The parameters $c_i$ (with $i=1,2,3,4$) represent four of the five arbitrary coupling constants that characterize this vector-tensor theory of gravity. These parameters enter in the action $\mathcal{S}$ of the theory. 

Moreover, in Einstein-aether theory, the gravitational coupling constant can be affected by Lorentz violation:
\begin{equation}
G = G_N \big(1-  \frac{c_{14}}{2}\big),\label{eq:EA3}
\end{equation}

where $G_N$ represents the current Newtonian constant, while $G$ is the coupling constant that enters in the action $\mathcal{S}$.
From Eqs. (\ref{eq:EA1})-(\ref{eq:EA3}) we can derive some general considerations on the coupling constants. 
The denominators of Eqs.  (\ref{eq:EA1}) and (\ref{eq:EA2}) provide the three conditions:
\begin{equation}
\begin{cases}
c_1^2 -2c_1 -c_3^2  \ne 0 \\
c_{123} = c_1 + c_2 + c_3 \ne 0, \\
c_{14} = c_1 + c_4 \ne 2 
\end{cases}
\end{equation}

while Eq. (\ref{eq:EA3}) provides the condition $c_{14}<2$ to ensure gravity is attractive and has the correct Newtonian limit. 
Other general considerations, for instance on the positivity of the energy, implies that $c_1>0$, $c_{14}>0$ and $c_{123}>0$.
By imposing $\alpha_1=0$, Eq. (\ref{eq:EA1}) returns the constraint

\begin{equation}
c_1c_4 =-c_3^2,\label{eq:V1}
\end{equation}

while imposing $\alpha_1=\alpha_2=0$ Eq. (\ref{eq:EA2}) returns two constraints:

\begin{equation}
c_{13}=c_1+c_3 =0\label{eq:V2}
\end{equation}

and

\begin{equation}
c_{2}=\frac{c_{13}}{3c_1}(c_3-2c_1). \label{eq:V3}
\end{equation}

These solutions are already well known in the literature. To strengthen the theoretical context of the result we obtained, we now discuss the implications of a very small but non-zero $\alpha_1$ result,  particularly for the parameter space and coupling constants for the Einstein-aether theory; we assume $\alpha_1=3\times10^{-5}$.

From Eq. (\ref{eq:EA1}) we obtain a second degree equation for $c_1$ whose solutions are obtained by solving the equation:
\begin{equation}
c_{1} = \frac{\alpha_1+4c_4 \pm \sqrt{(\alpha_1+4c_4)^2+\alpha_1c_3^2(\alpha_1+8)}}{\alpha_1},\label{eq:c1}
\end{equation}

To obtain two real solutions, we must require that the discriminant of this equation be greater than or equal to zero. This condition provides us with another quadratic equation to solve for $c_4$, whose solutions are
\begin{equation}
\begin{cases}
c_{4}^+ = -\frac{\alpha_1}{4} \Big(1-\sqrt{1-\frac{8c_3^2}{\alpha_1}} \Big) \\
c_{4}^- = -\frac{\alpha_1}{4} \Big(1+\sqrt{1-\frac{8c_3^2}{\alpha_1}} \Big),\label{eq:c4+-}
\end{cases}
\end{equation}

with $c_4\le c_4^-$ and $c_4\ge c_4^+$ and the constraint $c_3^2\le \alpha_1/8$, see Figure \ref{fig:vincolo1}.

\begin{figure} 
	\centering
	\includegraphics[width=0.7\linewidth]{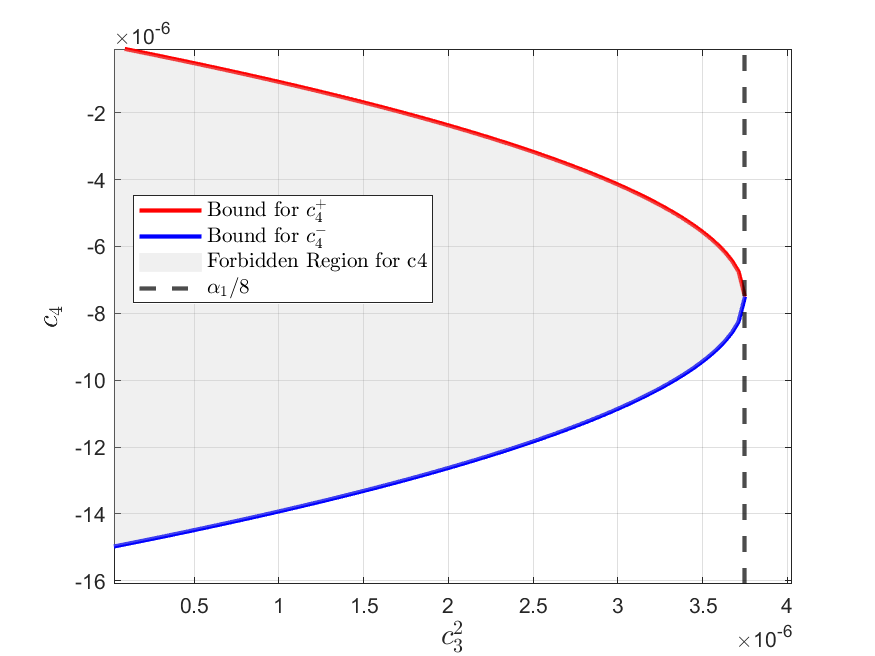}
	\caption{Forbidden (gray) and allowed (white) regions for the coupling constant $c_4$ when $c_3^2$ varies from zero to its maximum value $\alpha_1/8$ (black dashed line).}
	\label{fig:vincolo1}
\end{figure}

Substituting these two solutions into Eq. (\ref{eq:c1}) we obtain four solutions for $c_1$.
By imposing the previously introduced energy positivity conditions, $c_1>0$ and $c_1+c_4>0$, we obtain constraints on the sum $c_1+c_4$ as a function of $c_3^2$. The resulting allowed region in the parameter space is shown in Figure  \ref{fig:vincolo2}. 

\begin{figure} 
	\centering
	\includegraphics[width=0.7\linewidth]{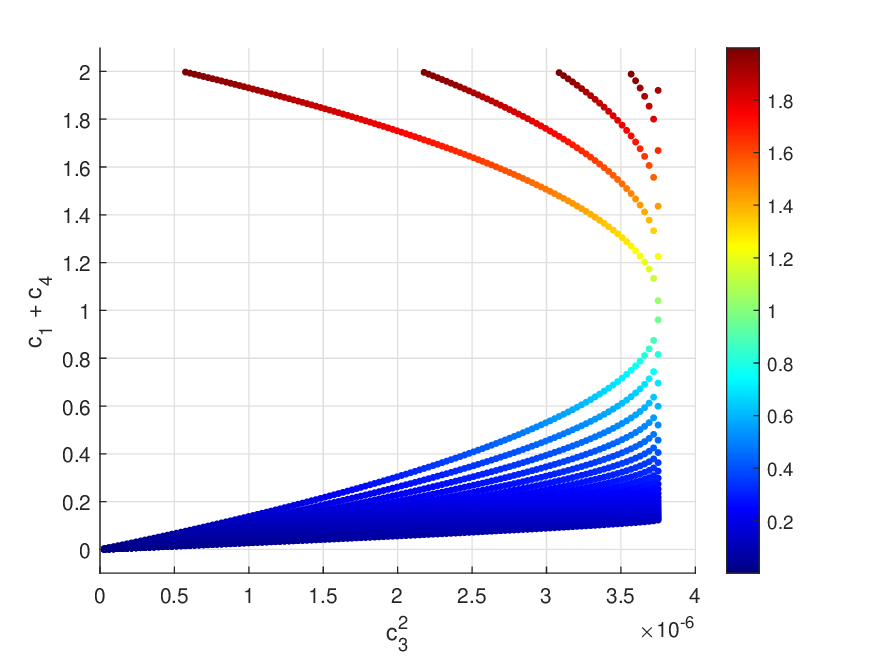}
	\caption{Scheme of some allowed values for $c_1+c_4$ as a function of $c_3^2$ when considering the conditions on the positivity of the energy.}
	\label{fig:vincolo2}
\end{figure}

The region inside the innermost two curves is forbidden, and so are the region defined by $c_1+c_4\ge2$ and the region $c_3^2> \alpha_1/8$.
We finally note that $c_3$ is confined to very small values: $|c_3|\le \sqrt{\alpha_1/8}\simeq1.94\times10^{-3}$.


\subsection{Non-zero results}\label{sec:NZ}
To further strengthen the context of our results, we explored the hypothetical implications of our central measured values for $\alpha_1$, even though they are statistically consistent with zero. What we can say if the value were positive or negative? 
If we consider a positive central value, as $\alpha_1 = +1 \times 10^{-5}$, the analysis is qualitatively identical to the one presented above. The parameter space constraints shown in Figures \ref{fig:vincolo1} and \ref{fig:vincolo2} retain the same structure, with the constraint on the coupling constant $c_3$ becoming slightly tighter: $|c_3| \leq \sqrt{\alpha_1/8} \approx 1.12 \times 10^{-3}$.

More interestingly, considering a negative central value, such as $\alpha_1 = -1 \times 10^{-5}$, leads to a significant change in the theoretical constraints. The discriminant in Eq. (\ref{eq:c1}) becomes always positive, as the term under the square root in Eqs. (\ref{eq:c4+-}) transforms into $1 + 8c_3^2/|\alpha_1|$. This has the profound physical implication that for a negative $\alpha_1$, this formalism would no longer place any upper bound on the value of $|c_3|$, allowing for a much larger region of the theory's parameter space to be viable.

This highlights that the sign of $\alpha_1$ is not merely a mathematical curiosity; it carries distinct physical meaning. A positive versus a negative $\alpha_1$ would correspond to opposite-in-direction physical effects on the orbits, such as anomalous argument of pericenter precessions. More fundamentally, the sign would imply different relationships between the theory's coupling constants, thus influencing other observable phenomena like the effective gravitational constant $G$ and the Nordtvedt effect (a potential violation of the Strong Equivalence Principle). Ultimately, measuring the sign of $\alpha_1$ would provide a crucial clue to the intrinsic nature of the underlying vector field in theories like Einstein-aether, distinguishing between different classes of models and their interaction with matter and metric.

While this remains a hypothetical exercise, it demonstrates how the sign, not just the magnitude, of a future, more precise measurement of $\alpha_1$ could be used to powerfully constrain --- or even rule out --- specific classes of alternative gravity models.

\section{Comparison with Other Tests of Lorentz Invariance}\label{sec:experiments}
We now place our results in the context of other experimental tests of Lorentz invariance. A crucial aspect of our analysis is that it probes the weak-field, quasi-static limit of gravity. This stands in contrast to many of the most stringent existing constraints, which are derived from highly-dynamical, strong-field environments such as binary pulsars and gravitational wave events. 
In this section, we compare our findings with results from Solar System tests, binary pulsar timing, and the gravitational sector of the Standard-Model Extension (SME).
 




\subsection{Solar System}

For the constraints obtained from "experiments" performed in the Solar System, therefore in the WFSM limit of GR, we begin by recalling the constraints previously cited in the Introduction:
\begin{equation}
\begin{cases}
\alpha_1=(-7\pm9)\times10^{-5} \\
\alpha_2=(+1.8\pm2.5)\times10^{-5},\label{eq:VLLR}
\end{cases}
\end{equation}

obtained with the LLR technique \cite{2008ASSL..349..457M} and:
\begin{equation}
\mid{\alpha_2}\mid \lesssim 2.4 \times10^{-7},\label{eq:VSole}
\end{equation}

obtained by exploiting the close alignment between the Sun's spin and the total angular momentum of the Solar System \cite{1987ApJ...320..871N}.
This last result is not sufficiently robust, being based on some approximations and not supported by an error analysis.

As a further example of constraints obtained for $\alpha_1$ and $\alpha_2$ in the solar system, we report the results obtained in 2014 by \cite{2014IJMPD..2350006I}:
\begin{equation}
\begin{cases}
\alpha_1=(-1\pm6)\times10^{-6} \\
\alpha_2=(-0.9\pm3.5)\times10^{-5}.\label{eq:VIorio}
\end{cases}
\end{equation}

These results were obtained from the analysis of the orbits of the planets of the solar system, in particular from the analysis of the longitude of the perihelia of the terrestrial planets: Mars, Earth, Venus and Mercury, according to the INPOP10a model of the ephemerides of these planets \cite{2011CeMDA.111..363F}.
The study described in \cite{2014IJMPD..2350006I}, while interesting,  has an issue that prevents us from considering it robust: it consists of a secondary analysis, meaning that the work took into account the "anomalous" precessions (the differences between observations and GR predictions) and attributed the resulting values entirely to the possible effects of $\alpha_1$ and $\alpha_2$.

This approach is susceptible to two main sources of uncertainty. First, the formal errors from global ephemeris fits may not fully capture unmodeled systematic effects (e.g., from asteroid masses). Second, without access to the full covariance matrix of the original fit, it is challenging to properly account for strong correlations between model parameters, such as the solar quadrupole moment $J_2$, the Lense-Thirring effect, and the PPN parameters themselves. While the analysis attempted to mitigate some of these issues by constructing a linear combination of planetary data, the resulting constraints may be optimistic.

\subsection{Binary Systems and GW emission}

The orbital motion of a binary pulsar system through the galaxy provides a varying velocity vector relative to any cosmic preferred frame.
 A non-zero $\alpha_1$ would induce periodic variations in the orbital parameters, particularly the orbital eccentricity and the longitude of periastron. The effect is strongest for low-eccentricity binaries. By timing the pulses over many years and seeing no such anomalous variations, tight constraints can be placed. 

The best constraints are those already presented in the Introduction \cite{2012CQGra..29u5018S,2013CQGra..30p5019S,1996CQGra..13.3121B,2005ApJ...632.1060S}:
\begin{equation}
\begin{cases}
\hat{\alpha}_1=(-0.4\pm4)\times10^{-5} \\
\mid{\hat{\alpha}_2}\mid \lesssim 1.6 \times10^{-9} \\ \label{eq:VPulsar}
\mid{\hat{\alpha}_3}\mid \lesssim 4 \times10^{-20}.
\end{cases}
\end{equation}

We remind that this is a regime of strong or nearly-strong gravitational fields, and that the meaning of the PPN parameters may differ from that of weak fields.

Conversely, a constraint obtained in a strong-field regime is the one obtained by \cite{2016PhRvD..94h4002Y}. 
Specifically, this constraint was obtained exploiting the Fermi GBM (Gamma-ray Burst Monitor) signal in the delayed emission scenario case, i.e. assuming that the signal measured by Fermi was a counterpart to GW150914. 
This is a constraint on the coupling constant $\hat{c}_{13}=\hat{c}_1+\hat{c}_3$ of Einstein-aether theory obtained from the joint analysis of GW150914 \cite{abbott} and Fermi \cite{2016ApJ...826L...6C}:
\begin{equation}
\hat{c}_{13}\le10^{-17}.
\end{equation}

This represents a severe constraint on the departure of speed of gravity from that of light and on Lorentz-violating mechanism.

\subsection{Standard-Model Extension}\label{sec:sme}
Standard Model Extension provides a comprehensive effective field theory framework for studying violations of Lorentz and CPT symmetry \cite{1997PhRvD..55.6760C,1998PhRvD..58k6002C,1999PhRvD..59k6008C}. 
SME tests have the characteristic of being transversal to multiple fields of physics: from field and particle physics to gravitation, from weak to strong fields to cosmology.

In its gravitational sector \cite{2004PhRvD..69j5009K}, the minimal SME introduces Lorentz-violating terms into the action. Within the Riemannian formalism, the leading-order correction to the Einstein-Hilbert action is given by \cite{2006PhRvD..74d5001B,2011PhRvD..83a6013K}:
\begin{equation}
\mathcal{S}_{LV} = \frac{1}{16\pi G}\int d^4x \emph{e} \Big(-uR + s^{ab}R_{ab}^T +t^{abcd}C_{abcd}\Big),
\end{equation}
where  $u$, $s^{ab}$ and $t^{abcd}$ are background fields that break Lorentz invariance by coupling to the scalar curvature $R$, the trace-free Ricci tensor $R^T_{ab}$ 
and the Weyl tensor $C_{abcd}$, respectively, while \emph{e} represents the determinant of the vierbein $e_{\mu}^a$.
This framework links the PPN parameters to the vacuum expectation values of the SME coefficients. Of particular relevance to our work is the connection between 
 $\alpha_1$ and the time-time component of the background tensor $\bar{s}^{TT}$. Numerous experiments constrain the components of the SME coefficients \cite{2016Univ....2...30H,2011RvMP...83...11K}. A global analysis of direct bounds on $\bar{s}^{TT}$  from cosmic-ray Cherenkov radiation, pulsars, VLBI, and GRBs yielded \cite{2016Univ....2...30H}:
\begin{equation}
\bar{s}^{TT} = (-4.6\pm7.7)\times10^{-5}.
\end{equation}
Separately, combined spectroscopy of white dwarfs and radio timing of pulsars provided the limit \cite{2014PhRvD..90l2009S}:
\begin{equation}
\mid\bar{s}^{TT}\mid <1.6\times10^{-5}.
\end{equation}
Our constraint on the PPN parameter  $\alpha_1$
 can be translated into a bound on this SME coefficient. In the isotropic limit, the two are related by \cite{2006PhRvD..74d5001B}:
\begin{equation}
\bar{s}^{TT} =-\frac{3}{16}\alpha_1.
\end{equation}
Applying our results, this yields new (more stringent) constraints:  
\begin{equation}
\bar{s}^{TT}\lesssim6\times10^{-6}
\end{equation}
and
\begin{equation}
\bar{s}^{TT}\lesssim4\times10^{-6}, 
\end{equation}
respectively in the case of the analyses performed with GEODYN II and SATAN.

\section{Conclusions}\label{sec:Conclusions}

Local Lorentz invariance (LLI) is a cornerstone of modern physics, underpinning both the Standard Model of particle physics and General Relativity. Testing this fundamental principle is crucial, as any observed violation would provide compelling evidence for \emph{new physics}. Our work is motivated by this search, specifically probing for gravitational theories that extend General Relativity with additional fields (e.g., $K^{\mu}$ and/or $B_{\mu\nu}$) that could mediate the gravitational interaction and couple to matter in a way that violates LLI.
In this paper, we presented a new test of LLI in the gravitational sector by searching for preferred-frame effects, with the Cosmic Microwave Background (CMB) chosen as the natural candidate for the preferred frame. Our analysis, based on a long time series of precise orbital data from the LAGEOS and LAGEOS II geodetic satellites, provides a new constraint on the Parameterized Post-Newtonian Parameter (PPN) parameter $\alpha_1$. The satellites were tracked using the Satellite Laser Ranging (SLR) technique, and the data were processed using two independent software packages: GEODYN II (over a time interval of 28.3 years) and SATAN (over a time interval of 31.4 years), for precise satellite orbit determination. 
The use of two distinct data reduction strategies, each employing a different set of dynamical and measurement models, see Table \ref{tab:modelli2}, provides a crucial cross-validation and enhances the robustness of our final result against potential software-specific systematic errors.
Our final results are (respectively for GEODYN II and SATAN):
\begin{equation}
\alpha_1 = (+2 \pm 3) \times 10^{-5},
\end{equation}
and
\begin{equation}
\alpha_1 = (+1 \pm 2) \times 10^{-5}.
\end{equation}
The measurement relies on the phase-sensitive detection of an annual signal in a specific linear combination of the rates (of the residuals) of the argument of perigee $\omega$ and the mean anomaly $M$ for the two satellites: $\dot{\ell}_0=\dot{\omega}+\dot{M}$. This technique isolates a signal with a well-defined period, the annual one in our case, significantly simplifying the analysis of systematic errors, as only perturbations with the same period need to be considered.
An advantage of using $\dot{\ell}_0$ as observable, the rate of the mean argument of latitude of a satellite, is that some of the perturbative effects (both gravitational and non-gravitational) acting individually in the rates of the two keplerian elements ${\omega}$ and ${M}$ are significantly reduced by their sum (Section \ref{sec:PFE}).
Furthermore, using this observable in the case of two satellites, as detailed in Section \ref{sec:concept}, allows us to cancel  the error from the uncertainty in the Earth's dominant zonal gravitational coefficient, $\bar{C}_{2,0}$. This uncertainty represents the leading source of error in a single-satellite analysis, and its removal is crucial for achieving high precision. 
With the dominant error from $\bar{C}_{2,0}$ eliminated, our systematic error budget is dominated by the uncertainty in the Earth's hexadecapole coefficient, $\bar{C}_{4,0}$. As shown in Table \ref{tab:error-budget}, the contributions from tides and non-conservative perturbations are negligible. The total systematic error is estimated to be \(7.7\times10^{-6}\), which is smaller than the estimated statistical uncertainties of $3\times10^{-5}$ and $2\times10^{-5}$.
Note that the main contribution to the standard deviation of the results comes from the noise that characterizes our precise orbit determination, in particular broadband noise.
 A detailed discussion of the error analysis is provided in Appendix \ref{app:errors}.
Our results are consistent with zero, in full agreement with the predictions of General Relativity. This represents the first constraint on the PPN parameter $\alpha_1$ derived from the orbital dynamics of Earth-orbiting satellites. Furthermore, they improve by a factor of 3 to 4.5 the previous best bound, obtained by Lunar Laser Ranging in the gravitational field of the Sun \cite{2008ASSL..349..457M}.
This work places new, tighter constraints on the existence of a gravitational preferred frame and, consequently, on theories of gravity that extend GR with additional vector or tensor fields. A prominent example is Einstein-aether theory \cite{PhysRevD.70.024003}, which is particularly relevant as it is considered the low-energy limit of Hořava-Lifshitz quantum gravity \cite{2009PhRvL.102p1301H}. As detailed in Section \ref{sec:AE}, our measurement of $\alpha_1$ constrains a linear combination ($c_1+c_4$) of the theory's coupling constants as a function of $c_3^2$.
 Furthermore, in the context of the Standard-Model Extension (SME) in the pure-gravity sector (Section \ref{sec:sme}), our results implies a limit on the time-time component of the Lorentz-breaking coefficient:  $\bar{s}^{TT}\lesssim6\times10^{-6}$ and $\bar{s}^{TT}\lesssim4\times10^{-6}$, respectively for the analyses performed with GEODYN II and SATAN. These constraints are significantly more stringent than those obtained from strong-field regimes.
We conclude by highlighting that our measurement constitutes a "direct" test of LLI within the Dicke framework \cite{2021Univ....7..192L}, as it directly probes the imprint of a non-zero $\alpha_1$ on the orbital motion of the LAGEOS satellites.
A natural continuation of this work, that we plan to pursue, consists of extending this analysis to other orbital elements. Specifically, analyzing the eccentricity vector holds the potential to simultaneously constrain both PPN parameters $\alpha_1$ and $\alpha_2$ in the Earth's gravitational field, further refining these fundamental tests of gravity.

\begin{acknowledgments}
This work has been performed by the SaToR-G (Satellite Test of Relativistic Gravity) collaboration  funded by the Commissione Scientifica Nazionale (CSN2) for Astroparticle Physics of the Istituto Nazionale di Fisica Nucleare (INFN), to which we are very grateful. 
The authors acknowledge the ILRS for providing high-quality laser ranging data of the {LAGEOS and} LAGEOS II satellites, {and an anonymous Referee for helpful comments and remarks that have considerably improved the original manuscript}.
This work is partially supported by ICSC – Centro Nazionale di Ricerca in High Performance Computing, Big Data and Quantum Computing, funded by European Union – NextGenerationEU.
\end{acknowledgments}
\section{DATA AVAILABILITY}
The data that support the findings of this article are openly available at \footnote{The supporting data for this article are openly available from the Zenodo repository: https://zenodo.org/records/18441938.}.
\appendix
\section{Validation of the Phase Sensitive Detection using synthetic data}
\label{app:sintetici} %
The  Phase Sensitive Detection (PSD, or Lock-In), is the final stage and a crucial step of our analysis, enabling the isolation of a small signal of known frequency and phase buried in background signals and noise. 

To validate its use, it is appropriate to characterize the PSD, with respect  to both  its detecting capability and  the noise produced at its output.  We recall that the PSD consists of two modules in cascade: in the first,  the input time series is multiplied by a sine wave of given frequency (in our case $f_0$, the inverse of a sidereal year) and phase, shifting  the spectral component of interest to DC. Then, a low-pass filter narrows the band around zero frequency,   in order to further reject signals and noise at $ f \ne 0$ ($ f \ne f_0$ in the original signal). We used a third-order Butterworth filter with a time constant of $\Delta t = 3000$ d.  The low-pass, like any filter, produces some {\it ringing},  
(distortion of the signal shape)
at the edges of the processed time series, visible in Figures \ref{fig:residui3a} and \ref{fig:residui3b}: that is why we discarded, in our analysis, the transients at the beginning and end of the time series, limiting our analysis to  7000 days. 
The length of the data stream and the filter parameters were chosen to optimize the Signal-to-Noise Ratio, and they were kept fixed, both in the analysis of Section \ref{sec:measure} and in the tests described in this Appendix, unless explicitly varied.

{\bf Noise Characterization --}
Characterization of the noise at the output of the PSD requires a statistical approach, which is not viable when only one time series is available.  
Splitting the data stream into shorter segments, a typical workaround, is not feasible here due to the limited length of the dataset.
We therefore  generated random time series, each having the same average, standard deviation and amplitude distribution (Gaussian) as the real data stream. 
To remove any potential physical signal at the yearly frequency  $f_0$, the spectral component at that frequency was replaced by interpolating the amplitudes of neighboring bins.
The Power Spectral Density was further smoothed using a moving average over $3.5\times10^{-4}$ ~d$^{-1}$. \\
The synthetic noise time series were fed as input to the Lock-In filter: Figure \ref{fig:B2} displays the envelope of 30 synthetic noise trials alongside the physical signal. The time domain behavior of the physical signal is very similar in shape and amplitude to the synthetic signals, This suggests that many of its characteristics can be attributed to wideband noise or, at least, that wideband noise represents the main contribution, well over narrowband systematic effects.
\begin{figure}[h!]
\begin{center}
\includegraphics[width=0.45\textwidth]{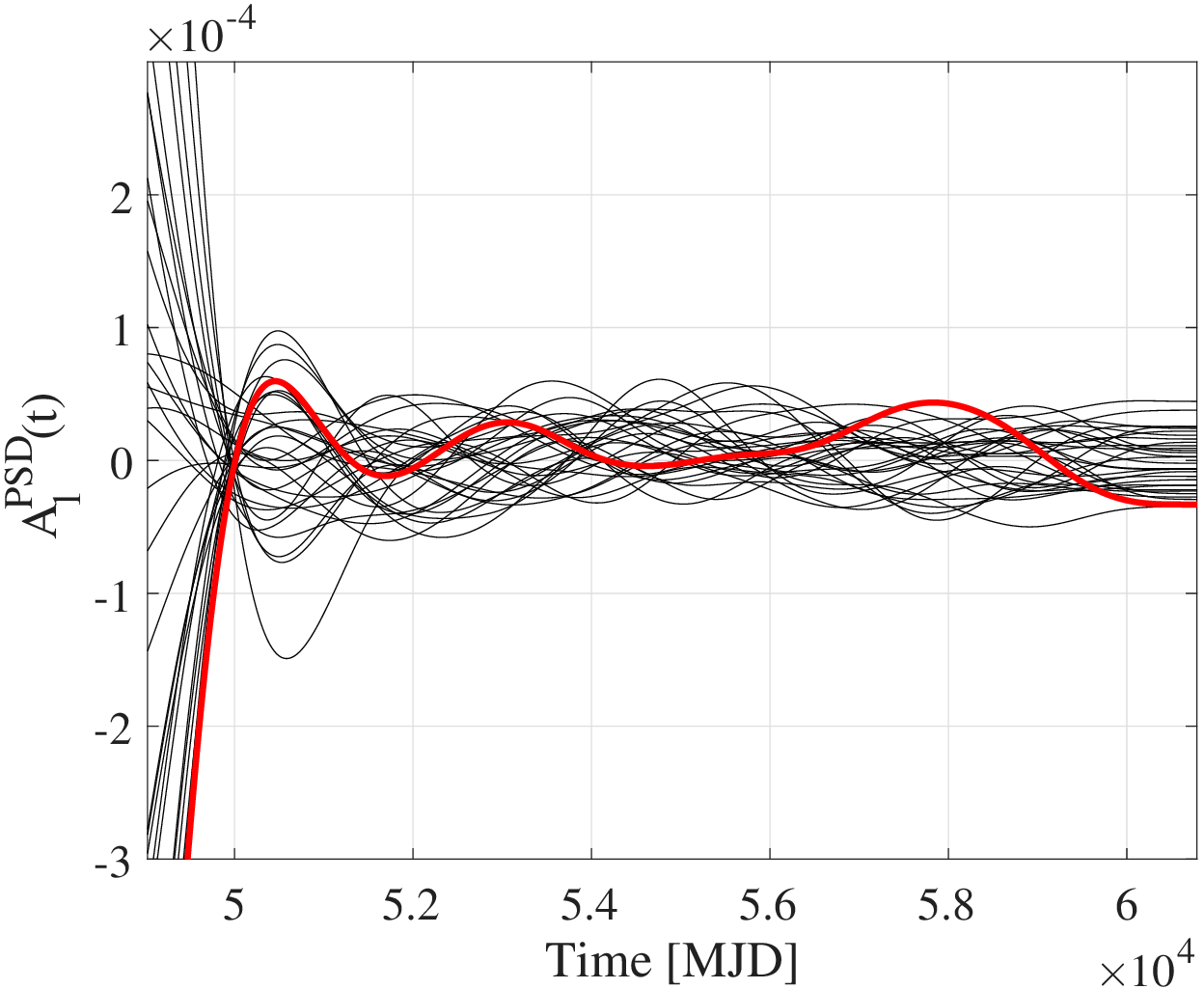}
\caption{Gray: 30 data streams of synthetic noise, at the
output of the Lock-In.  Red: the physical data stream.}
\label{fig:B2}
\end{center}
\end{figure}

{\bf Dispersion of response to a signal due to noise --}
We evaluated the sensitivity to two periodic signals with annual frequency and fixed phase, corresponding to $\alpha_1= 2\times10^{-5}$ and  $\alpha_1= 5\times10^{-5}$, respectively.
{We added to each of this signals $10^5$ synthetic data-sets of noise. These simulations were then processed in the same configuration used with real data.} 
Figure \ref{fig:B3} illustrates the histogram obtained for the {time} average of the Lock-In output. 
The solid lines in the figure represent the expected distributions. These remain Gaussian, as with the input, but with a standard deviation modified by the correlation introduced by the low-pass filter.
From {these} simulations, we obtained values of $\alpha_1= (2.0\pm1.1)\times10^{-5}$ and $\alpha_1=(5.0\pm1.1)\times10^{-5}$ respectively. 
{The two averages are identical to the amplitude of the injected signals. The two standard deviations ($\sigma=1.1\times10^{-5}$) are both equal, within error, to the expected value derived from an analytical model of the low-pass filter.}
\begin{figure}[h!]
\begin{center}
\includegraphics[width=0.45\textwidth]{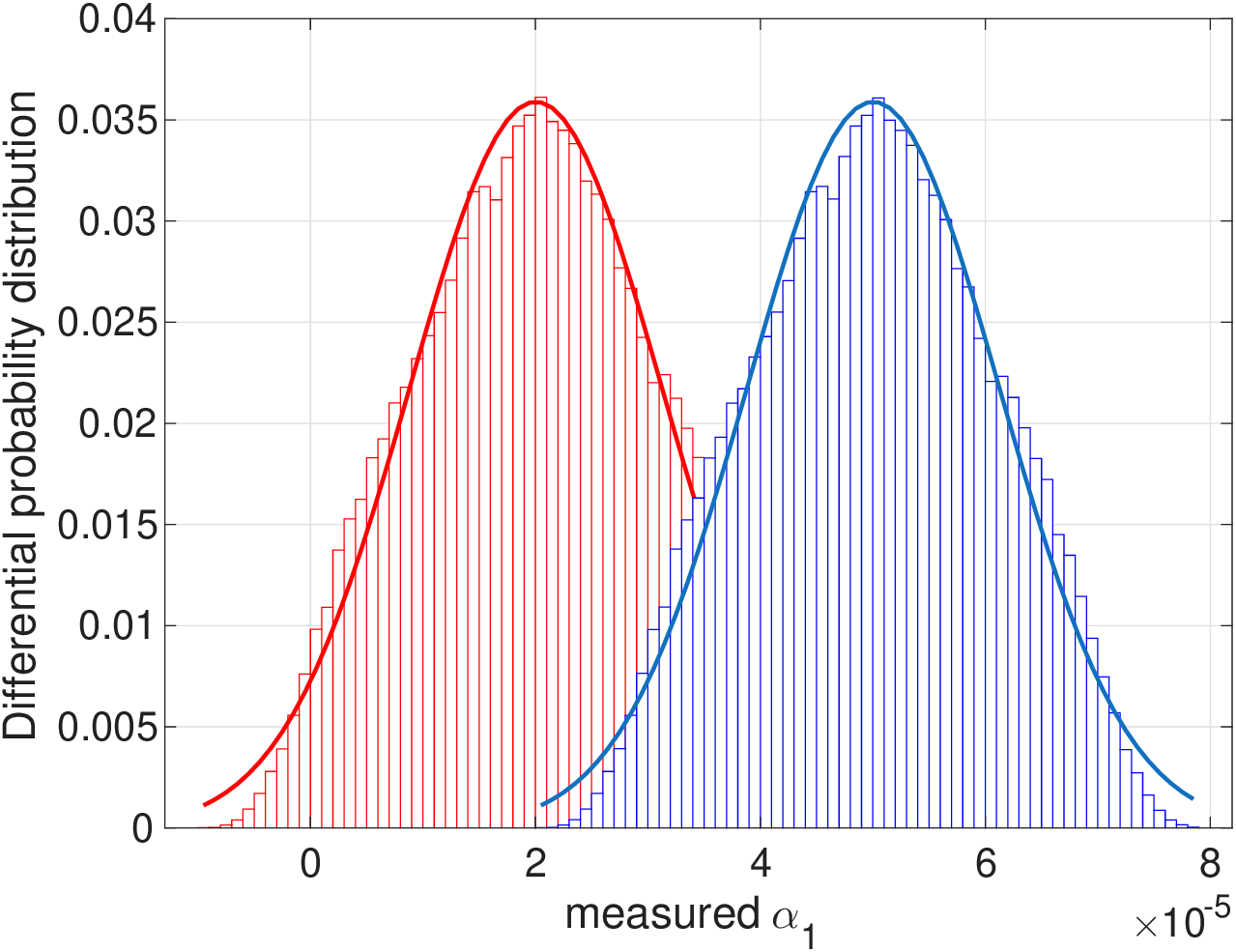 }
\caption{
Histogram of the detection pipeline output for simulated signals with two different amplitudes  added to synthetic noise ($2.0\times10^{-5}$ in red and $5\times10^{-5}$ in blue). The solid lines represent the expected distributions,}
\label{fig:B3}
\end{center}
\end{figure}

{\bf Response to simulated detuned signals --}
Using synthetic signals, we computed the response of our pipeline to signals not perfectly tuned to our target (i.e., signals deviating from the annual frequency and the expected phase).
To test the pipeline, we added signals with an amplitude equivalent to an  $\alpha_1=2\times10^{-5}$, to different streams of simulated noise. We varied the phase by $\pm \pi/4$  and the period by $\pm1$ d, around the target.
Specifically, we tested a grid of 21x11 phase and frequency combinations. For each combination, we performed 1000 noise trials. 
Figure \ref{fig:var_phas_fre} reports the results, showing that the detection pipeline can measure the amplitude of sinusoids deviating by up to 3 hours in period and 40 degrees in phase from the predicted values, remaining within the measurement error.

\begin{figure}[h!]
	\begin{center}
		\includegraphics[width=0.45\textwidth]{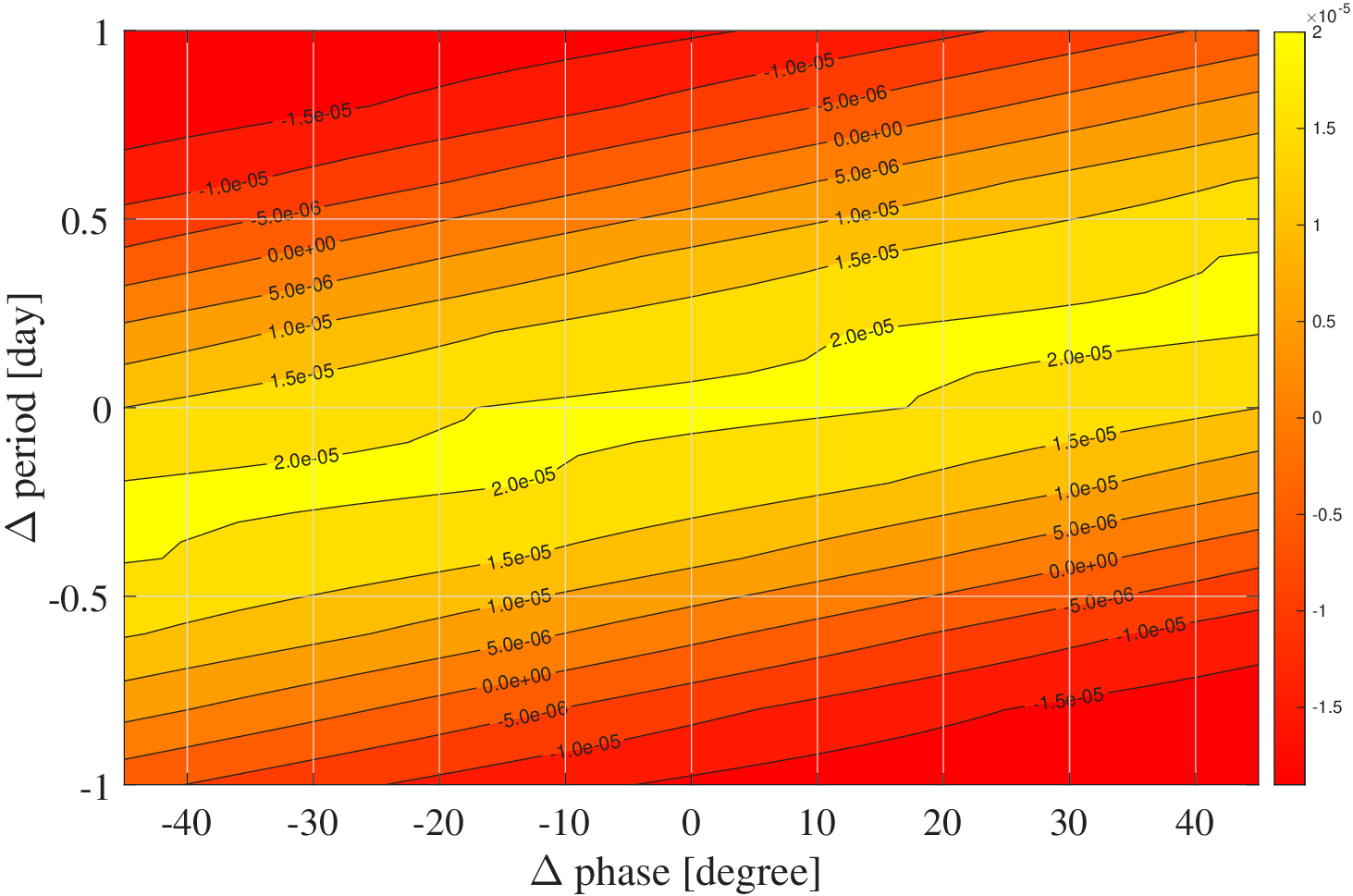}
		\caption{Surface plot of detected $\alpha_1$, for injected signals with a fixed amplitude of $2\times10^{-5}$, over synthetic noise, plotted versus period and phase detuning.}
		\label{fig:var_phas_fre}
	\end{center}
\end{figure}
\section{Systematic Errors}
\label{app:errors} %
{The sources of error in the measurement of \(\alpha_1\) are to be found in disturbance effects in the motion of the satellites related to both gravitational and non-gravitational perturbations (NGPs).
The first group includes imperfect modeling of the Earth's gravitational field and tides, both solid and oceanic; these errors are discussed in Sections \ref{app:grav} and \ref{app:maree}.
For NGPs, the main factors to consider are effects related to solar and terrestrial radiation pressure and thermal forces; these errors are discussed in Sections \ref{app:sole-terra} and \ref{app:termici}.
Clearly, we are primarily interested in detecting a signal related to our observable that varies with the annual period of the possible LLI violation.}
{The analyses described in this Appendix follow three different approaches.} 

{In the first approach, we estimated the error induced on the observable \(\dot{\ell}_0\) by a given annual perturbation effect and computed its effect on the parameter \(\alpha_1\) from the combined analysis of the orbits of the two LAGEOS, Eq. (\ref{equ_comb}), and from the application of the filter produced by the PSD based on Eq. (\ref{eq:alpha1_sys}) below.
Therefore, in this specific case, in the analysis of the different errors, we focused on the perturbation effects characterized by an annual periodicity.}

{In the second approach Section \ref{app:ngp_end_to_end}), the error estimate was performed by applying the entire pipeline used to estimate \(\alpha_1\) to a single perturbation model. In this case, therefore, the possible annual periodicity effect of the model on our observable was not considered, but rather the entire model with all its temporal characteristics. This approach was particularly relevant for assessing the systematic bias from NGP.}
{Finally, in the third approach, we performed a sensitivity analysis of our measurement to some characteristic parameters, assuming extremely pessimistic maximum errors (Section \ref{app:sensitivity}).}

Section \ref{app:errori-discussione} finally summarizes our general considerations on the error budget in measuring the PPN parameter \(\alpha_1\).


\subsection{Gravitational Field}\label{app:grav}

The long-term perturbative effects of the Earth’s gravitational field are linked to the deviation from spherical symmetry in the Earth’s mass distribution.  
In particular, we are interested in the long-term effects of the zonal harmonics \(J_{\ell}\) \footnote{With \(J_{\ell}=-\sqrt{2\ell+1}\bar{C}_{\ell,0}\) and \(\bar{C}_{\ell,0}\) the normalized zonal harmonics.}, whose disturbing function \(\mathcal{R} \) is:

\begin{equation}
\mathcal{R} =-\frac{Gm_{\oplus}}{r}\sum_{\ell=2}^{\infty}\left(\frac{R_{\oplus}}{r}\right)^{\ell}J_{\ell}P_{\ell}(\cos\theta),
\end{equation}

where \(\ell\) is the degree of the expansion and \(P_{\ell}(\cos\theta)\) are the Legendre Polynomials with \(\theta\) the colatitude on Earth \footnote{Where \(\cos\theta=\sin i\sin(\omega+f)\) and \(f\) represents the true anomaly. By expressing the true anomaly as a function of the mean anomaly and averaging over this variable, the long-term effects are obtained.}.
If the gravity field were static, these effects would not be characterized by any periodicity, neither in the rate of the argument of pericentre, nor in the rate of the mean anomaly of the satellite.
At the same time, it is well known that the time-dependent solutions for the low-degree harmonics of the gravitational field are characterized not only by secular trends, but also by periodic variations with annual and semi-annual periodicities \footnote{See ICGEM Temporal Models at https://icgem.gfz-potsdam.de/sl/temporal.}. This is the case of the quadrupole coefficient \(\bar{C}_{2,0}\), which provides the largest perturbation due to the Earth's oblateness, as can be seen from the measurements provided by the GRACE and GRACE-FO missions \cite{Tapley2013,JGRB:JGRB50058}, as well as from SLR data \cite{1997JGR...10222377C,JGRB:JGRB50058,2018GeoJI.212.1218C,https://doi.org/10.1029/2022JB025459}. 

Figure \ref{fig:C20}(a) shows the time evolution of the quadrupole coefficient from the monthly solutions estimated at the Center for Space Research (CSR -- University of Texas at Austin) exploiting the SLR data of (up to eight) geodetic satellites with the UTOPIA software \cite{JGRB:JGRB50058}\footnote{The values reported for the quadrupole were obtained from two different analyses performed by the CSR, the first and of greater temporal extension is plotted in black while the second is plotted in gray. In the time interval common to the two analyses, the values for the coefficient are in good agreement.}. This is known as the historical solution for the terrestrial quadrupole coefficient and is a zero-tide solution \footnote{UTOPIA is based on a zero-tide potential, i.e., based on the removal of the permanent part of the external potential from the mean-tide potential.}. The main periodic component that characterizes the oscillation of the coefficient is the one with annual period, as clearly shown in Figure \ref{fig:C20}(b).
This implies that the secular trends on the pericenter and on the mean anomaly are also characterized by an annual oscillation which overlaps with their linear behavior.

\begin{figure*}
\centering
\subfigure[About five decades of the Earth's quadrupole coefficient \(\bar{C}_{2,0}\) from SLR data. The behavior can be fitted, in appropriate sub-intervals, with a main oscillation with annual frequency superimposed to a linear trend.]{\includegraphics[width=0.45\textwidth]{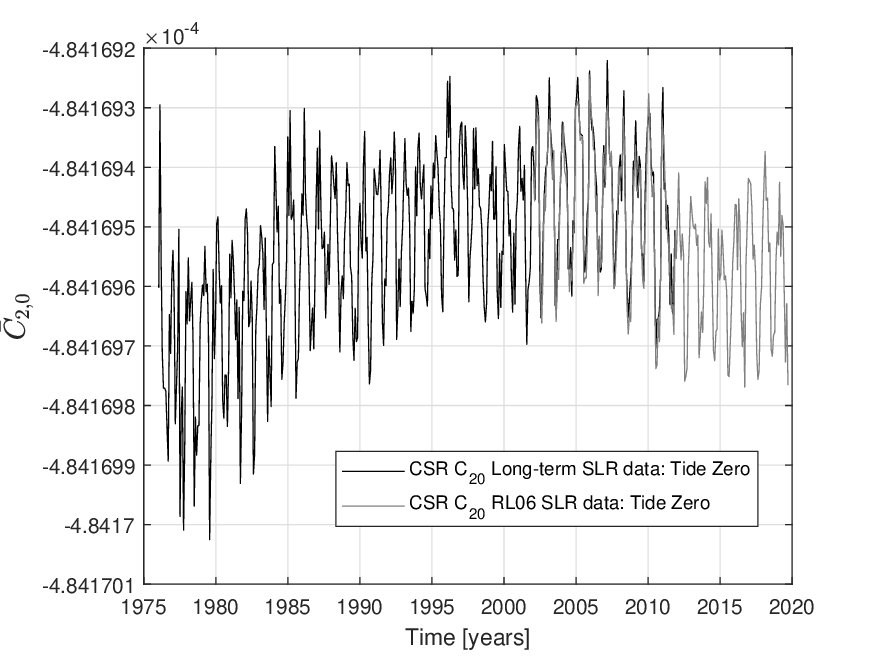}}
\hfill
\subfigure[Detail of the behavior shown in (a) which highlights the annual periodic oscillation of the quadrupole coefficient and centered on the beginning of our analysis (Oct 31, 1992). It highlights the annual periodicity and allows us to estimate the initial phase.]{\includegraphics[width=0.45\textwidth]{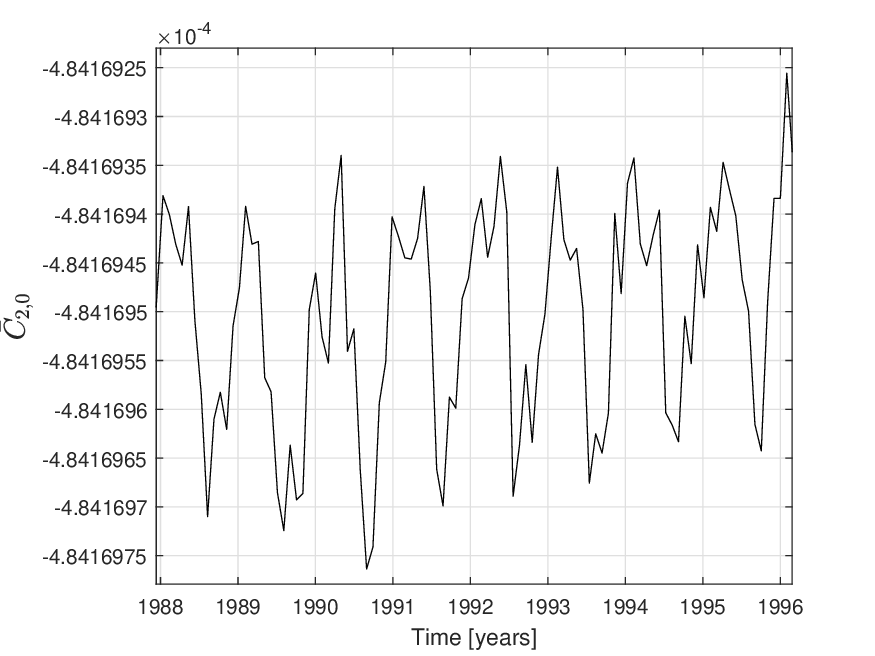}}
\caption{Earth quadrupole coefficient estimated on a monthly basis by CSR using SLR data from up to 8 geodetic satellites.}
\label{fig:C20}
\end{figure*}

Indeed, in our analysis with GEODYN II, the low-degree harmonics of the terrestrial geopotential have been modeled to take into account their long-term drifts. In particular, we considered the long-term behavior of the even zonal harmonics (those of order \(m\)=0) of degree \(\ell\)=2 and \(\ell\)=4 on the basis of the available knowledge for these coefficients --- obtained from SLR data of geodetic satellites and those from the GRACE and GRACE-FO missions --- provided by different Analysis Centers \footnote{The GRACE and GRACE-FO data were considered to compare whether the trends measured for these coefficients using the SLR technique were still consistent, including their periodic behavior, with those obtained with a different technique and with non-geodetic satellites.}. This analysis was also applied to the octupole coefficient \(\bar{C}_{3,0}\), i.e., the first odd zonal harmonic coefficient (\(\ell\)=3), which is a measure of the North-South asymmetry in the distribution of mass within our planet.

Figure \ref{fig:C20b} shows part of the historical solution for the terrestrial quadrupole coefficient obtained from the CSR that we have used to fit the two linear trends shown (on which the annual oscillations are superimposed) to better capture the long-term behavior of the coefficient over the time interval of our analysis with GEODYN II.

\begin{figure} 
	\centering
	\includegraphics[width=0.7\linewidth]{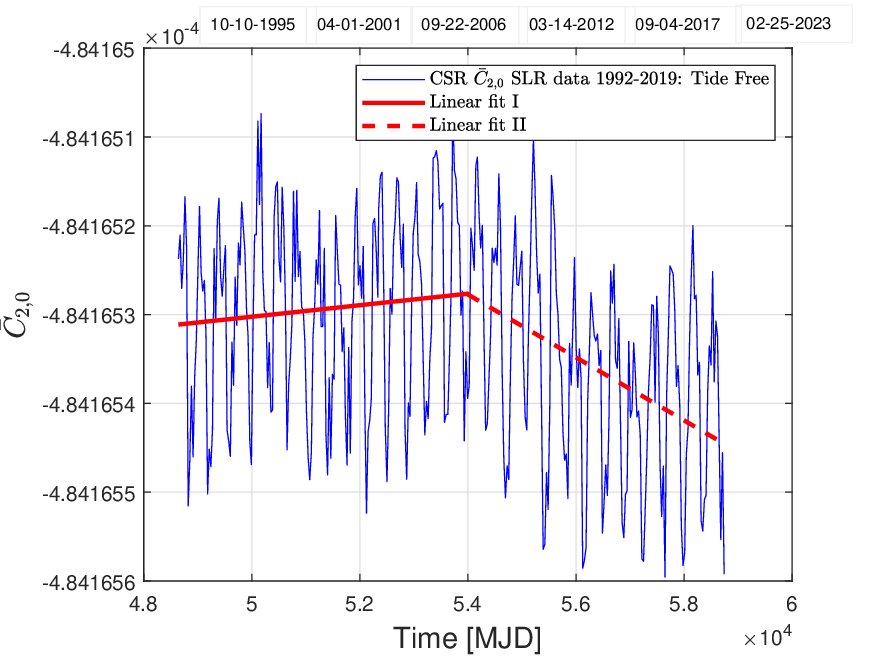}
	\caption{Long-term behavior for quadrupole coefficient $\bar{C}_{2,0}$ of the Earth gravitational field obtained by CSR from SLR data. The original data of Figure \ref{fig:C20} were converted to a tide-free potential. The two red lines represent the best linear fit, in two successive periods, to the long-term oscillatory trend of the coefficient. To estimate the phase of the annual frequency component, the \(\bar{C}_{2,0}\) data were fitted with a linear trend and a sinusoid with annual periodicity. We obtained \(\phi_{\bar{C}_{2,0}}=0.7\pm0.2\) rad.}
	\label{fig:C20b}
\end{figure}

To achieve this, we converted the zero-tide quadrupole coefficient of previous Figure \ref{fig:C20} into a tide-free quadrupole coefficient, as shown in Figure \ref{fig:C20b}. This is because GEODYN is based on a tide-free potential and not on a tide-zero potential like UTOPIA. The difference \(\Delta \bar{C}_{2,0}\) between zero-tide and tide-free is of the order of \(4.2\times10^{-9}\). This value can be tuned to the specific solution obtained for the gravitational field. We refer to IERS Conventions [2010] for more details~\cite{2010IERS-Conv-2010}.

In Table \ref{tab:camp-grav}, are shown the results of our linear fit for the low degree zonal harmonics previously cited. The intercept is given at the epoch \(t_0\)=51544 MJD (J2000.0), the first period covers the time interval 1992 – 2006, while the second period covers the time interval 2006 – 2019. The transition epoch is August 30, 2006 (MJD 53977) \footnote{In SATAN, the secular trends of the gravity field coefficients were modeled following \cite{Lemoine2019}.}.

\begin{table}[h!]
  \caption{Temporal variation of the gravitational field that we implemented in our analysis with GEODYN II for the first two even zonal harmonics and the first odd zonal harmonic. The first column indicates the degree of the zonal harmonic, the second column gives the value of the coefficient at epoch J2000.0, finally the third column indicates the rate of linear variation of the coefficient over time in yr\(^{-1}\), see Eq. (\ref{eq:trend}). \label{tab:camp-grav}}
  \begin{ruledtabular} 
  \begin{tabular}{lcc}
Zonal harmonic & Intercept J2000.0  & Rate [1/yr]  \\
\hline
$\bar{C}_{2,0}$ \(1^{st}\) period & \(-4.8416529  \times 10^{-4}\) & \(+2.37 \times 10^{-12}\)  \\
\(\bar{C}_{2,0}\) \(2^{nd}\) period & \(-4.8416519  \times 10^{-4}\)  & \(-1.295 \times 10^{-11}\)  \\
\(\bar{C}_{3,0}\)  entire period & \(+9.57235    \times 10^{-7}\)  & \(+0.490 \times 10^{-11}\)  \\
\(\bar{C}_{4,0}\)  entire period & \(+5.40007    \times 10^{-7}\)  & \(+0.470 \times 10^{-11}\)  \\
\end{tabular}
\end{ruledtabular}
\end{table}

As regards the values of the rates of the octupole and hexadecapole coefficients, they coincide with those suggested by current IERS Conventions, while the values for the coefficients set at the epoch J2000.0 differ slightly from those envisaged by the Conventions.
These trends have been modelled in GEODYN II by means of the following equation:

\begin{equation}
\bar{C}_{\ell,0}(t) = \bar{C}_{\ell,0}(t_0) + \dot{\bar{C}}_{\ell,0}\cdot (t-t_0). \label{eq:trend}
\end{equation}

The value of the formal error for the amplitude of the quadrupole coefficient is found in the literature, but care must be taken in adopting it: 
the current best value is provided by the latest releases of the Earth gravitational model from GRACE and GRACE-FO missions, and is of the order of few times \(10^{-13}\).
However, it is important to note that, due to the low altitude, around 500 km, of the twin satellites of both GRACE and GRACE-FO missions, the low-degree coefficients are not estimated with the same precision as the medium and high-degree coefficients, and their values, in the provided solutions, are usually replaced with values derived from the SLR.
This is especially true in the case of the quadrupole coefficient \(\bar{C}_{2,0}\), as already highlighted above (see Figure \ref{fig:C20}).  
We therefore conservatively assume for this error the value \(\delta\bar{C}_{2,0}=9\times10^{-12}\) obtained through the SLR technique \cite{https://doi.org/10.1029/2022JB025459}.
With this value, Eq. (\ref{eq:errC20}) yields a systematic contribution to the error of \(\alpha_1\) equal to approximately \(1.8\times10^{-5}\) for LAGEOS and \(1.6\times10^{-5}\) in the case of LAGEOS II.

However, the contribution to the error on $\alpha_1$ from this coefficient has been removed by using the longitudes (i.e., the mean argument of latitude $\ell_0=\omega+M$) of both LAGEOS satellites, as explained in Section \ref{sec:concept}. Indeed, from the system (\ref{equ_comb}), and after the application of the PSD, we obtained the following solution for the PPN parameter \(\alpha_1\) which is free from the error due to the knowledge of the terrestrial quadrupole:

\begin{equation}
\alpha_1 = -3.396\times10^{-5}\delta\dot{\ell}_0 |_{LG1} -3.765\times10^{-5}\delta\dot{\ell}_0|_{LG2}, \label{eq:alpha1_sys}
\end{equation}

where \(\delta\dot{\ell}_0\) represents the error, in mas/arc, in the mean argument of latitude rate of the satellites.  
Note that this same equation represents the point values for the quantity \(A_1(t)\) plotted in Figure \ref{fig:alpha1_t}, the other solution of system (\ref{equ_comb}), when the errors in mas/arc are replaced by the (point) values of the observables \(\dot{\ell}_0^{LG1}\) and \(\dot{\ell}_0^{LG2}\) appearing in the left-hand side of system (\ref{equ_comb}).

Figure \ref{fig:delta_C20}(a) shows the solution for the corrections $\delta \bar{C}_{2,0}$ to the quadrupole coefficient obtained from the system (\ref{equ_comb}), while Figure \ref{fig:delta_C20}(b) shows its FFT.
The correlation between the two solutions, $A_1(t)$ and $\delta \bar{C}_{2,0}$, of the system (\ref{equ_comb}) is (as expected) relatively high, about 0.09.
In the FFT of $\delta \bar{C}_{2,0}$ it is interesting to note the presence of a line with annual periodicity, and whose amplitude \(\sim8.4\times10^{-12}\) is very close to the error we estimated for the annual component of the terrestrial quadrupole, about \(9\times10^{-12}\).

\begin{figure*}
\centering
\subfigure[Corrections $\delta \bar{C}_{2,0}$ to the quadrupole coefficient of the Earth gravitational field. These corrections have a mean of about \(4\times10^{-11}\) and a standard deviation of about \(1.1\times10^{-10}\).]{\includegraphics[width=0.45\textwidth]{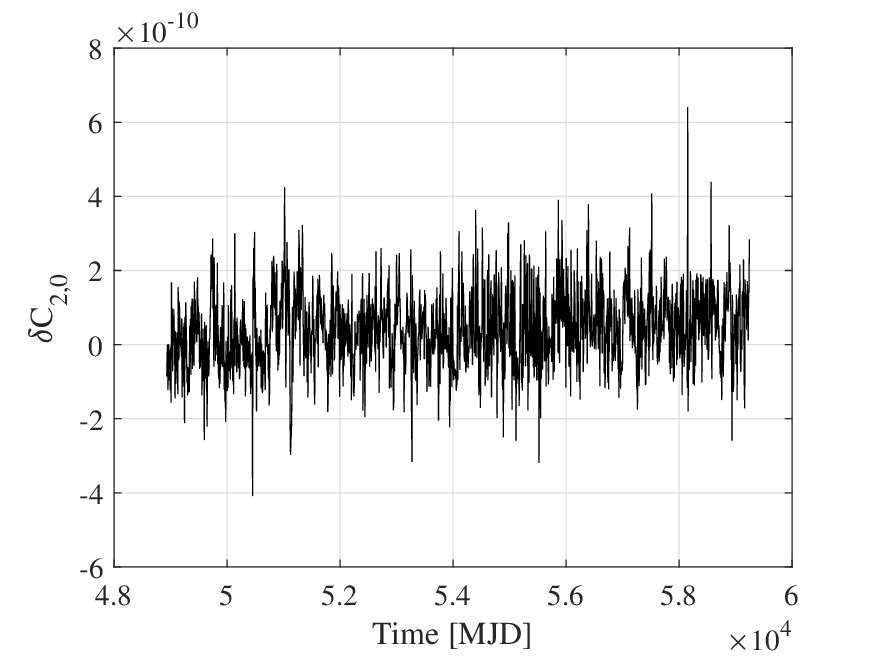}}
\hfill
\subfigure[FFT of the corrections $\delta \bar{C}_{2,0}$ to the quadrupole coefficient of the Earth gravitational field. A line with annual period is still present.]{\includegraphics[width=0.45\textwidth]{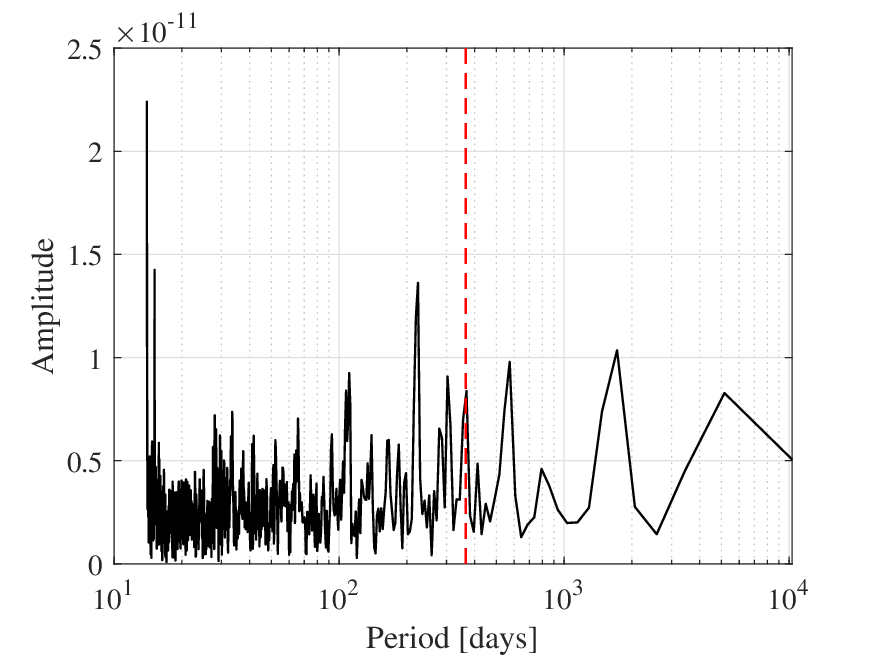}}
\caption{Earth quadrupole coefficient corrections estimated in the POD of the two LAGEOS satellites on a timespan of 28.3 years.}
\label{fig:delta_C20}
\end{figure*}

Thus, the main contribution to the systematic error in \(\alpha_1\) due to the gravitational field is related to the hexadecapole coefficient $\bar{C}_{4,0}$ and not to the quadrupole one. Eq. (\ref{eq:esadecapolo}) provides the error in the mean argument of latitude due to our lack of knowledge of the second even zonal harmonic:

\begin{widetext}
\begin{eqnarray}
 \delta\dot{\ell}_0|_{\delta \bar{C}_{4,0}}  & \simeq +\frac{15}{1024}\sqrt{9}\left(\frac{R_{\oplus}}{a}\right)^4 n\Big[
\frac{3e^2}{(1-e^2)^{7/2}}\big(9+20\cos(2i)+35\cos(4i)\big)- \frac{1}{(1-e^2)^4}\big(-27(4+5e^2) + \nonumber\\
 &4\cos(2i)(-52-63e^2) +7\cos(4i)(-28-27e^2)\big) \Big]\delta\bar{C}_{4,0}.\label{eq:esadecapolo}
\end{eqnarray}
\end{widetext}

We assumed an error of about \(6.8\times10^{-12}\) for this coefficient, in agreement with the calibrated error provided by the GGM05S model of the Earth's gravitational field \footnote{This estimate is comparable with the error provided by \cite{Lemoine2019}.}.
Therefore, by calculating separately from Eq. (\ref{eq:esadecapolo}) the errors on the mean argument of latitude of the two LAGEOS satellites produced by our knowledge of the hexadecapole coefficient, using Eq. (\ref{eq:alpha1_sys}) we obtain for \(\alpha_1\) an estimate of the systematic error due to the gravitational field equal to \(\sim6.5\times10^{-6}\).

The error $\delta\bar{C}_{6,0}$ of the third even zonal harmonic coefficient, the hexacontatetrapole, is about \(3.2\times10^{-12}\) (in agreement with the calibrated error provided by the GGM05S model) and gives a systematic error in \(\alpha_1\) of about \(6\times10^{-7}\), fully negligible.

Finally, we note that the analysis of the orbits of the two satellites was repeated by modeling, in addition to the secular trend of the quadrupole coefficient, also its oscillation with an annual period in a second case and by modeling, in a third case, both the oscillation with an annual period and that with a semiannual period. No significant differences were found with respect to the analysis described. In particular, an oscillation with an annual periodicity and amplitude comparable to that of Figure \ref{fig:FFT_alpha1_t} is always present: the amplitude is slightly reduced in the case of LAGEOS II and remains substantially unchanged in the case of LAGEOS.


\vspace{1em} 
{Furthermore, as detailed in Section \ref{app:sensitivity} (Sensitivity Analysis and Robustness Checks), the robustness of our results against potential mismodeling in key gravitational parameters has been further investigated through extreme sensitivity tests.} 

\subsection{Solid and Ocean Tides}\label{app:maree}

The effects of solid and ocean tides have been deeply investigated and computed on the orbit of the two LAGEOS satellites and on that of LARES in \cite{2018CeMDA.130...66P}. That analysis was mainly devoted to the estimation of tidal errors for the measurement of the Lense-Thirring effect \cite{2019arXiv191001941L,2020Univ....6..139L}, consequently the analysis was focused on the right ascension of the ascending node of the satellites and on the argument of pericenter of LAGEOS II.
In this work, we extended that analysis to the mean anomaly of LAGEOS and LAGEOS II.
We underline that among the perturbative effects that we are analyzing on our observable, the tides alone are characterized by a periodicity, among the many that distinguish them, exactly annual, like that of the possible violation of Eq. (\ref{eq:long2}). 
Furthermore, as shown in \cite{2018CeMDA.130...66P}, the phase \(\phi_{tide}\) of the solid and ocean tides are also well identifiable.

In Table \ref{tab:maree} the tide contributions obtained for the argument of pericenter $\Delta\dot{\omega}$ and for the mean anomaly $\Delta\dot{M}$ of the two LAGEOS satellites are shown.
As can be seen, the amplitudes are very small and the effects on the two elements tend to partially compensate, as already mentioned. 

\begin{table}[h!]
  \caption{Effect of solid and ocean tides on the argument of pericenter and on the mean anomaly of LAGEOS and LAGEOS II.  The second column provides the Doodson number of the tide with annual period. The third and fourth columns provide, respectively, the amplitudes in mas/d (milli-arc-seconds per day) of the solid and ocean tides on the rate of the argument of pericenter and on the rate of the mean anomaly of the satellite. Finally, the last column provides the phase in degrees of the two tides computed at the start date of the  satellites' orbit analysis. \label{tab:maree}}
  \begin{ruledtabular} 
  \begin{tabular}{ccccc}
Satellite & Tide 056.554 & \(\Delta\dot{\omega}\)  & \(\Delta\dot{M}\)  & \(\phi_{tide}\) \\
\hline
\multirow{2}*{LAGEOS } & Solid & \(-0.0169\) & \(+0.0784\) & \(+352^{\circ}.38\) \\
& Ocean & \(+0.0365\)  & \(-0.1688\) &  \(+312^{\circ}.38\) \\
\hline
\multirow{2}*{LAGEOS II} & Solid & \(+0.0346\) & \(-0.0128\) & \(+216^{\circ}.17\) \\
& Ocean & \(-0.0745\)  & \(+0.0276\) &  \(+176^{\circ}.82\) \\
\end{tabular}
\end{ruledtabular}
\end{table}
To estimate the impact of tides on the measurement of the parameter \(\alpha_1\) we need to add the effects in the two elements and take into account the typical errors associated with solid and oceanic tides \cite{2018CeMDA.130...66P}.
Typically, the relative error \(\delta k/k\) for the amplitude \(k\) of solid tidal coefficients is of the order of \(10^{-4}\) or less.
In contrast, ocean tides are characterized by larger uncertainties, with a typical relative error of the order of \(10^{-2}\).

We take the maximum relative error \(\delta k/k\) as well as the phase \(\phi_{tide}\) of the  tides  into account to estimate the tides error by means of the following equation:

\begin{equation}
\delta\dot{\ell}_0|_{tide} \simeq (\Delta\dot{\omega}+\Delta\dot{M})\cos \phi_{tide} \frac{\delta k}{k}. \label{eq:err-maree}
\end{equation}

By computing Eq. (\ref{eq:err-maree}) for each satellite --- and separately for both the solid and the oceanic tides --- and including the results in Eq. (\ref{eq:alpha1_sys}) we obtain a contribution to the error on \(\alpha_1\) of the order of  \(10^{-9}\) in the case of solid tides, and of the order of  \(10^{-7}\) in the case of ocean tides.
These errors are negligible compared to that estimated for the Earth's hexadecapole coefficient.


\subsection{Solar and Terrestrial Radiation Pressure}\label{app:sole-terra}

Direct solar radiation pressure (SRP) and the Earth's albedo are responsible for several long-term effects on the argument of perigee and the mean anomaly of spherical satellites like LAGEOS and LAGEOS II. However, they do not produce a direct periodic effect with a precise annual frequency \(\dot{\lambda}_{\oplus}\). The main frequency components are combinations of \(\dot{\lambda}_{\oplus}\), \(\dot{\omega}\) and \(\dot{\Omega}\) (the right ascension of the ascending node), in the form (\(\dot{\lambda}_{\oplus}\pm \dot{\omega}\)) and (\(\dot{\Omega}\pm\dot{\lambda}_{\oplus}\pm \dot{\omega}\)) \cite{2001P&SS...49..447L}.
This has been explicitly verified by extending the analytical calculations developed in \cite{2001P&SS...49..447L} to the satellite mean anomaly and analyzing its overall effect on the mean argument of latitude of a LAGEOS-type satellite.

Using the Gauss equations for the rate of the argument of perigee and for the rate of the mean anomaly (see for instance \cite{1987nongrav.book.....M}), we finally obtained --- at the first order in the orbital eccentricity \(e\) --- for the observable:

\begin{widetext}
\begin{eqnarray}
\dot{\ell}_0=\dot{\omega}+\dot{M} \simeq \frac{R}{na}\Big(-2 + \frac{3}{2} e\cos f \Big) +
\frac{T}{na}\frac{1}{2} e\sin f -\frac{W}{na}\cot i\Big(1 - e\cos f \Big)\sin (\omega +f) +\mathcal{O}(e) ,\label{eq:sole}
\end{eqnarray}
\end{widetext}

where the terms proportional to \(1/e\) that characterize both rates cancel each other out in their sum. In the last expression \(R\), $T$ and $W$ represent the components of the SRP acceleration in the Gauss co-moving frame, respectively along the radial, transverse and out-of-plane directions.

Averaging Eq. (\ref{eq:sole}) over the mean anomaly, we obtain long-term effects characterized by the above-mentioned frequencies, but we do not obtain a periodic term at the annual frequency \(\dot{\lambda}_{\oplus}\) (see for instance Eq. (11) in \cite{2001P&SS...49..447L}).
If we consider the effect of eclipses (see again \cite{2001P&SS...49..447L}), these do not introduce a periodicity with annual frequency either in the argument of the perigee or in the mean anomaly of the satellite. In fact, the elements most sensitive to the effect of eclipses are the inclination and the node of the satellite.
These last elements will instead be characterized by several new frequencies in the presence of eclipses, among these, in addition to the frequency of eclipses 2(\(\dot{\Omega}-\dot{\lambda}_{\oplus}\)), by the annual \(\dot{\lambda}_{\oplus}\) and twice-annual 2\(\dot{\lambda}_{\oplus}\) frequency.

The non-dependence on the annual periodicity of the observable \(\dot{\ell}_0\) is also obtained in the more general case in which the analytical computation that leads to Eq. (\ref{eq:sole}) is not restricted to the first order in eccentricity. The exercise has been repeated in the general case using the symbolic calculation of Matlab: indeed, no direct dependence on the annual frequency alone has been obtained.

The above considerations and results are valid in the approximation of interacting point bodies or bodies with spherical mass symmetry. When considering the effects of direct solar radiation pressure on a satellite around a non-spherical Earth, new effects emerge of the order of the direct solar acceleration multiplied by the Earth's quadrupole coefficient \(J_2\), i.e., \(\sim10^3\) times smaller in magnitude. In the case of the mean argument of latitude, the effects are further multiplied by the eccentricity of the satellite and still do not give rise to a purely annual periodicity. This has been verified in three different cases: i) in the case of an explicit first-order analytical calculation in eccentricity, ii) in the case of an extended symbolic calculation and iii) in the case of a numerical propagation of the orbit over a long time interval. The latter case is described explicitly in the next subsection.

Finally, we note that in the unlikely hypothesis that the different long-term effects arising from Eq. (\ref{eq:sole}) combine with each other to give an oscillating component with an annual period, this would give a contribution to the error in \(\alpha_1\) that is negligible compared to the effect previously estimated for the gravitational field.
An upper bound for the amplitude of such an effect is:
\begin{equation}
\dot{\ell}_0\simeq \frac{e}{na}C_R\Big(\frac{A}{m}\frac{\Phi_{\odot}}{c}\Big),
\end{equation}
where $C_R$ is the radiation pressure coefficient already introduced, $A/m$ represents the area-to-mass ratio of the satellite and $\Phi_{\odot}$ the solar irradiance at 1 astronomical unit.
In this equation we have maximized the effect in the observable by taking as acceleration the absolute value of the direct SRP, and not a fraction of it as Eq. (\ref{eq:sole}) would suggest.
Solving this equation for the two satellites, we obtain a value of about 0.9 mas/arc for LAGEOS II and about 0.3 mas/arc for LAGEOS. Even if we assume an annual periodicity for these long-term effects, the impact on the \(\alpha_1\) parameter of a 1\% error in the radiation coefficient would produce an error of the order of \(7\times10^{-7}\) and \(2\times10^{-7}\), respectively for LAGEOS II and LAGEOS \footnote{In \cite{2014PhRvD..89h2002L}, we conservatively assumed a 1\% error in knowledge of the radiation coefficient based on the \(C_R\) estimate performed in 2004 by \cite{2004P&SS...52..699L} in a dedicated POD. A 1\% error was independently confirmed by \cite{2014PhDT........14S} and by \cite{2019AdSpR..63...63H}.}. 
The overall impact on $\alpha_1$ by means of Eq. (\ref{eq:alpha1_sys}) would give an error $\sim4\times10^{-7}$.

If we now consider the long-term effects in the observable related to the albedo, these have the same periodicities as the SRP but with a characteristic amplitude given by that of the previous formulae rescaled by the Earth’s mean albedo coefficient: \(A_{\oplus}\simeq0.3\) \cite{1986CeMec..38..233R,1987JGR....9211662R,1992JGR....97.7121L}.
As regards eclipses and albedo, the previous considerations remain valid. In this case, however, the shadow function of the albedo must also be taken into account \cite{1972CeMec...5...80F,1997JGR...102.2711M}. As demonstrated in Eq. (36) of \cite{2001P&SS...49..447L}, this introduces frequencies given by the combinations of \(\dot{\Omega}\) and \(\dot{\lambda}\), but not a long-term effect with annual periodicity.

\subsubsection{Numerical propagation}\label{sub:SRP_numerical}

We numerically propagated the orbits of the LAGEOS and LAGEOS II satellites over a 10-year time interval, modeling both gravitational perturbations and NGPs due to solar and terrestrial radiation pressure. The model used also includes the gravitational field of the Earth up to degree and order 30, the third body perturbations of the Sun and Moon and the main GR corrections. The starting epoch was November 6, 1992. 

The propagation was performed under two distinct scenarios. In the first scenario (hereafter, the nominal \(C_R\) scenario), the radiation pressure coefficients, \(C_R\), for LAGEOS and LAGEOS II were held constant at their nominal values of 1.13 and 1.12, respectively. In the second scenario (hereafter, the perturbed \(C_R\) scenario), \(C_R\) was intentionally increased by 10\% relative to the nominal values, resulting in \(C_R\) values of 1.24 for LAGEOS and 1.23 for LAGEOS II. The radiation pressure coefficient encapsulates the average reflectivity properties of the satellite's surface and serves as a scaling parameter in the dynamical models of NGPs arising from solar and terrestrial radiation. This parameter is typically estimated during data reduction procedures when necessary.

To quantify the impact of solar and terrestrial radiation pressure on the satellite orbits, we calculated the difference between the Keplerian elements obtained from the perturbed \(C_R\) scenario and those obtained from the nominal \(C_R\) scenario. These differences represent the orbital perturbations induced by the considered NGP model.
Figure \ref{fig:gmat_L1} illustrates the temporal evolution of the argument of perigee (blue), the mean anomaly (red), and their sum—the mean argument of latitude (black)—for LAGEOS.

\begin{figure} 
	\centering
	\includegraphics[width=0.7\linewidth]{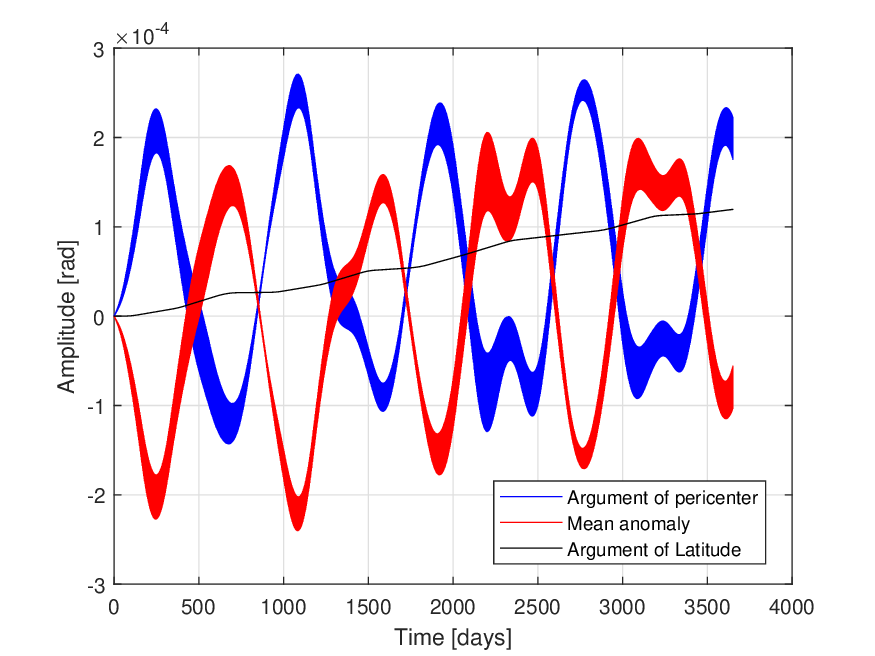}
	\caption{Evolution of the argument of perigee and of the mean anomaly of LAGEOS over a 10-year period produced by the 10\% variation of the satellite radiation coefficient. The black line represents the evolution of the mean argument of latitude of the satellite.}
	\label{fig:gmat_L1}
\end{figure}

Figure \ref{fig:fft_gmat_L1} presents the spectral analysis of these three orbital elements. Notably, Figure \ref{fig:fft_gmat_L1_part}, which provides a magnified view around the annual periodicity, reveals the absence of a contribution at the annual component in the evolution of the argument of perigee and the mean anomaly, and consequently, in the mean argument of latitude.

\begin{figure} 
	\centering
	\includegraphics[width=0.7\linewidth]{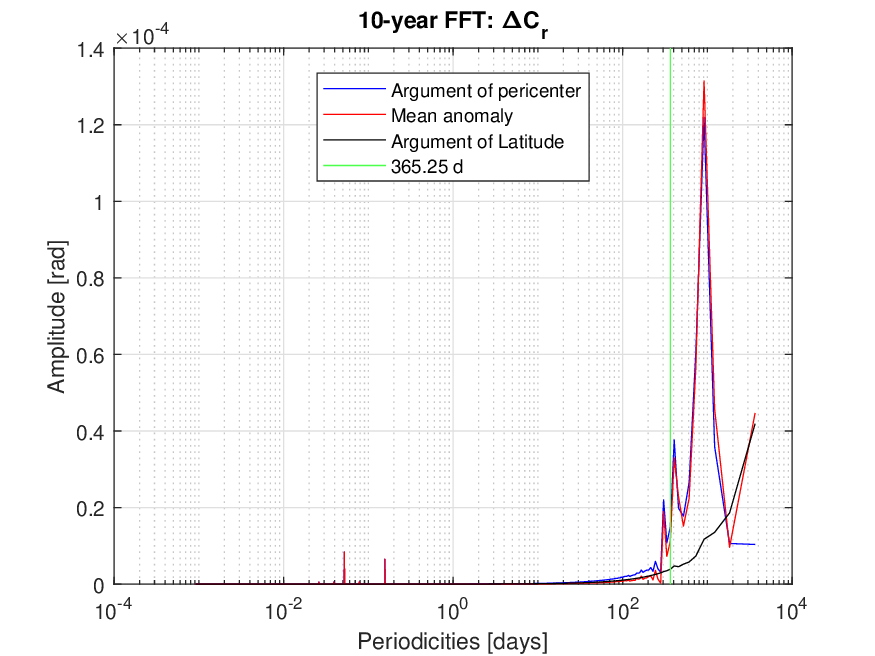}
	\caption{FFT of the variables shown in Figure \ref{fig:gmat_L1}.}
	\label{fig:fft_gmat_L1}
\end{figure}
\begin{figure} 
	\centering
	\includegraphics[width=0.7\linewidth]{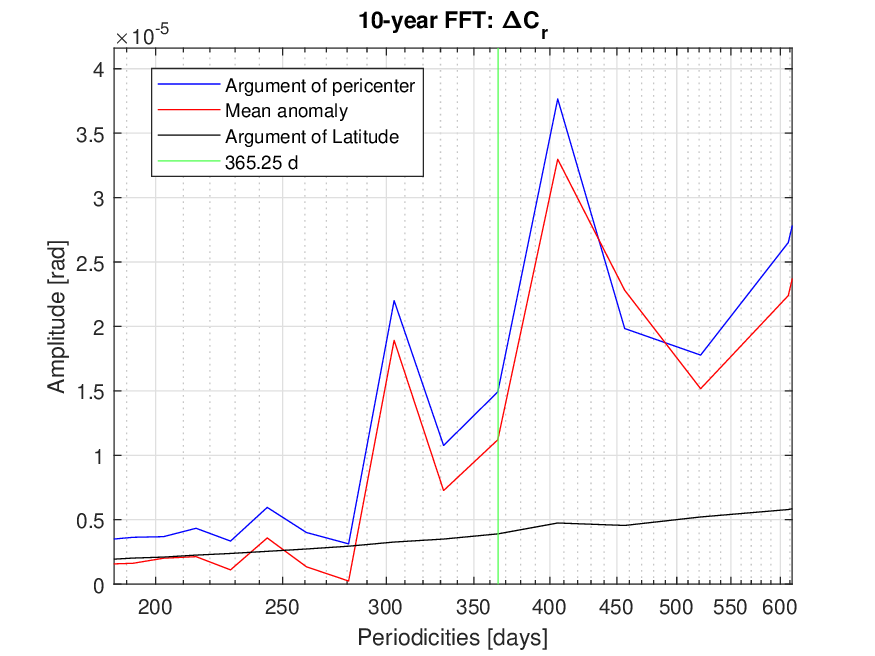}
	\caption{Particular of Figure \ref{fig:fft_gmat_L1} around the annual period.}
	\label{fig:fft_gmat_L1_part}
\end{figure}

The periods of the spectral components appearing in these last two figures are typical of the perturbative effects on these elements related to solar radiation and the Earth's albedo. For example, the line (\(\dot{\Omega}-\dot{\lambda}_{\oplus}+ \dot{\omega}\)) with a period of 406 d, the line (\(\dot{\lambda}_{\oplus}- \dot{\omega}\)) with a period of 304 d, the line (\(\dot{\Omega}-\dot{\lambda}_{\oplus}- \dot{\omega}\)) with a period of 900 d and the line (\(\dot{\Omega}+\dot{\lambda}_{\oplus}- \dot{\omega}\)) with a period of 243 d.

Similar considerations apply to LAGEOS II. Figure \ref{fig:fft_gmat_L2_part} shows the spectral analysis obtained for this satellite in the simulation time interval.

\begin{figure} 
	\centering
	\includegraphics[width=0.7\linewidth]{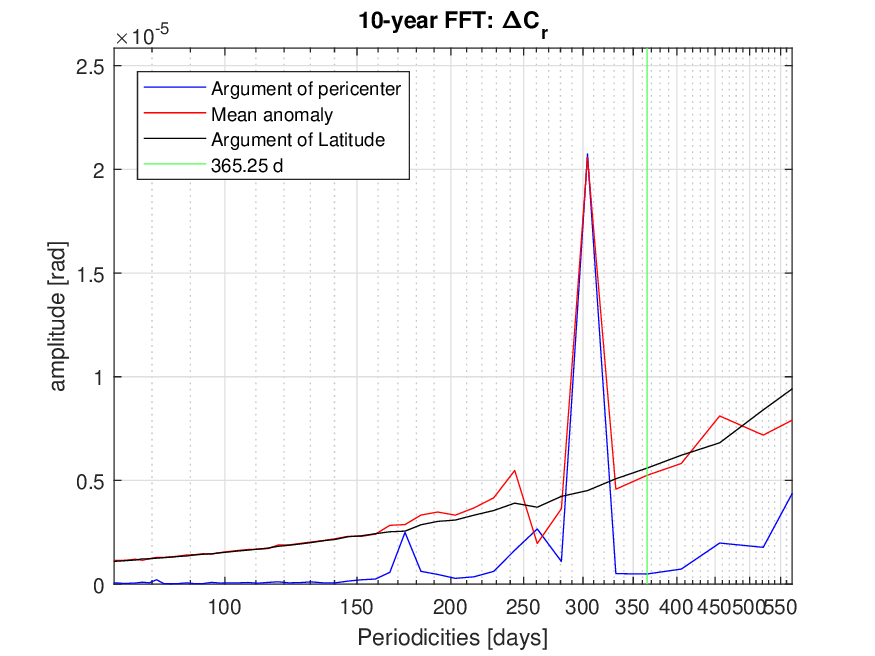}
	\caption{FFT of LAGEOS II argument of perigee (blue), mean anomaly (red) and mean argument of latitude around the annual period.}
	\label{fig:fft_gmat_L2_part}
\end{figure}

{While these analyses provide valuable insights into the spectral content of SRP and albedo perturbations and their lack of a direct annual component, a more rigorous end-to-end evaluation of their systematic bias on \(\alpha_1\) is presented in Section \ref{app:ngp_end_to_end}, following the "second approach" outlined in the introduction of this Appendix.}

\subsection{Thermal Thrust effects}\label{app:termici}

Perturbations arising from thermal thrust—namely the terrestrial Yarkovsky (Rubincam) effect and the solar Yarkovsky-Schach effect—were not modeled in our primary analysis. We adopted this approach because the models we previously developed for these phenomena \cite{2019Univ....5..141L,2018PhRvD..98d4034V,2001P&SS...49..447L,2002P&SS...50.1067L,2004P&SS...52..699L} are more accurate than those currently implemented in the GEODYN II software. Consequently, these effects contribute to our orbital residuals and will be accounted for a posteriori.

Following our first error analysis approach, we investigate here whether these thermal effects could produce a spurious signal with a purely annual periodicity in our observable, $\dot{\ell}_0$. Similar to the other NGPs discussed, neither of these two thermal effects is characterized by a dominant annual period \cite{2002P&SS...50.1067L}. For the terrestrial Yarkovsky effect, the principal periodicities are driven by the precession rates of the satellite's node  (\(\dot{\Omega}\) and \(2\dot{\Omega}\)) and of the pericenter (\(\dot{\omega}\) and \(2\dot{\omega}\)), and of their combinations (\(2\dot{\omega}\pm\dot{\Omega}\), \(2\dot{\omega}\pm2\dot{\Omega}\), \(4\dot{\omega}\pm\dot{\Omega}\) and \(4\dot{\omega}\pm2\dot{\Omega}\)) \cite{1987JGR....92.1287R,1988JGR....9313805R}. 
For the solar Yarkovsky-Schach effect, the characteristic frequencies include those from direct solar radiation pressure, along with additional terms involving the Sun's mean longitude \(\dot{\lambda}_{\oplus}\), such as (\(\dot{\Omega}\pm2\dot{\lambda}_{\oplus}\pm \dot{\omega}\),  \(\dot{\Omega}\pm \dot{\omega}\), \(2\dot{\lambda}_{\oplus}\pm \dot{\omega}\) and \(\dot{\omega}\)) \cite{1996CeMDA..66..131S,2004P&SS...52..699L,2007Andres}.
Furthermore, the spectral content of these thermal perturbations is made more complex by the satellite's spin vector. The evolution of the spin axis is itself primarily governed by the nodal precession rate \(\dot{\Omega}\) of the satellite's orbit \cite{2002P&SS...50.1067L,2018PhRvD..98d4034V}.
Again, the analytical treatment of these effects does not provide a periodic component with a strictly annual period.

Following the same methodology used for solar and terrestrial radiation pressure, we numerically simulated the long-term perturbations from thermal thrusts to investigate their potential contribution to our results. These simulations span a 28-year period, matching the time interval of our main analysis. We focused on the effects on the argument of perigee and the mean anomaly of both LAGEOS satellites. As previously stated, these non-gravitational perturbations were not explicitly modeled in our primary data analysis and they are therefore a potential source of unmodeled signals in the orbital residuals.

The simulations address the solar Yarkovsky-Schach effect and the terrestrial Yarkovsky (Rubincam) effect, based on the models described in \cite{2002P&SS...50.1067L,2004P&SS...52..699L,2019Univ....5..141L}. In both scenarios, the evolution of the satellite's spin vector was modeled over the entire time frame according to \cite{2016AdSpR..57.1928V,2018PhRvD..98d4034V}. In the numerical analysis a sampling step of 1 day was used and the effects on the orbital elements were calculated using the Gauss equations.
The primary goal of these simulations is to determine whether thermal thrusts produce a significant signal with an annual periodicity in our observables. While the results presented here are for LAGEOS II, our conclusions are equally applicable to LAGEOS.

Figure \ref{fig:oss_YS_L2} shows the long-term effects of the perturbation on our observable, the mean argument of the latitude, while Figure  \ref{fig:oss_YS_FFT_L2}  shows its spectral content via FFT. This figure also shows the FFTs of each rate, that of the argument of pericenter and that of the mean anomaly. Figure \ref{fig:oss_YS_FFT_L2_part} shows an enlargement of the previous figure around the annual periodicity. As we can see, there is no oscillating component with annual periodicity.
\begin{figure} [h]
	\centering
	\includegraphics[width=0.7\linewidth]{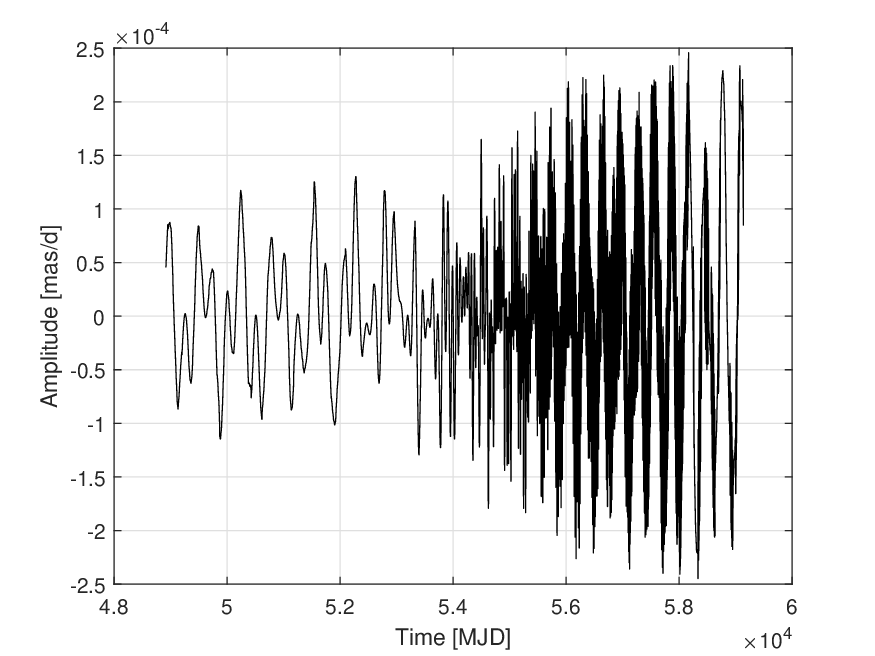}
	\caption{Time evolution of the LAGEOS II mean argument of latitude due to the Yarkovsky-Schach effect on a time interval of 28 years.}
	\label{fig:oss_YS_L2}
\end{figure}
\begin{figure} 
	\centering
	\includegraphics[width=0.7\linewidth]{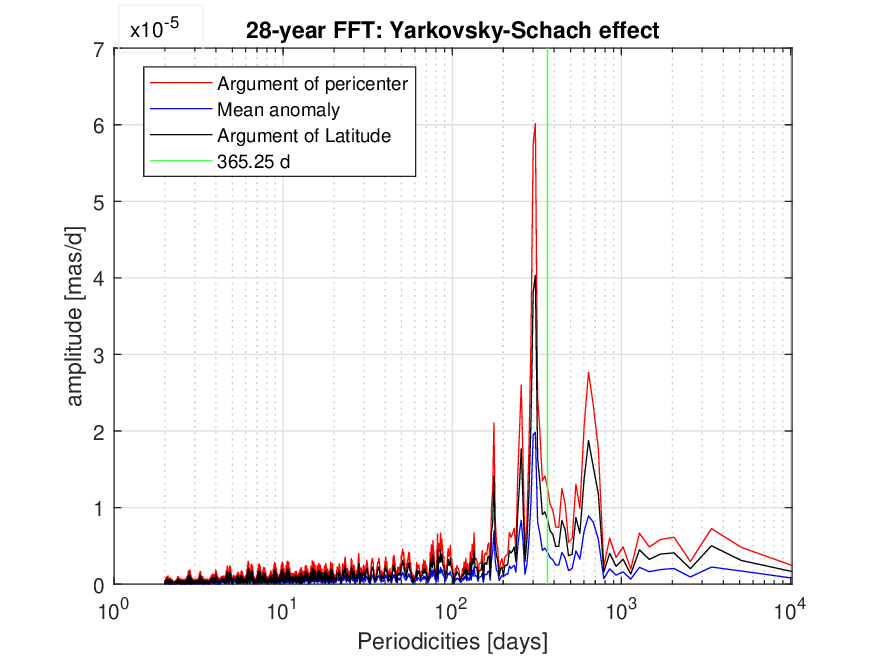}
	\caption{Yarkovsky-Schach effect. FFT of the LAGEOS II argument of pericenter (red) and mean anomaly (blue) over the 28 years period analyzed. The FFT of the argument of latitude (black) is also shown.}
	\label{fig:oss_YS_FFT_L2}
\end{figure}
\begin{figure} 
	\centering
	\includegraphics[width=0.7\linewidth]{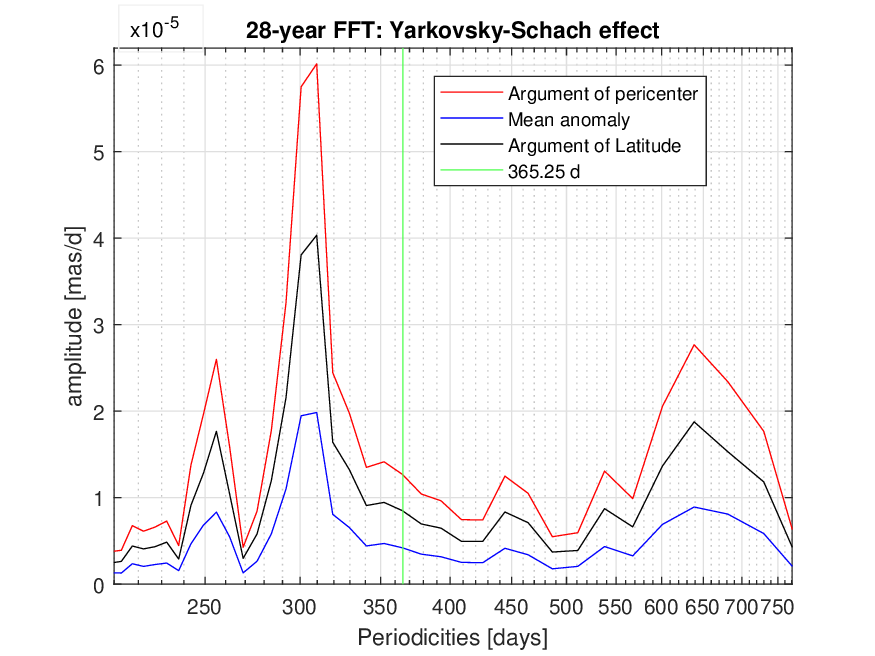}
	\caption{Zoom of Figure \ref{fig:oss_YS_FFT_L2} around the annual periodicity: no spectral component with annual periodicity is evident.}
	\label{fig:oss_YS_FFT_L2_part}
\end{figure}

Figures \ref{fig:oss_EY_L2}, \ref{fig:oss_EY_FFT_L2} and \ref{fig:oss_EY_FFT_L2_part} shows the results of our analysis in the case of the Earth-Yarkovsky effect.
The frequency analysis of the mean argument of latitude, see Figure \ref{fig:oss_EY_FFT_L2_part}, does not show evidence of an oscillating component with annual periodicity. On the other hand, it seems possible that an oscillating component with annual period is present in the satellite argument of pericenter, but not in its mean anomaly.

\begin{figure} [h]
	\centering
	\includegraphics[width=0.7\linewidth]{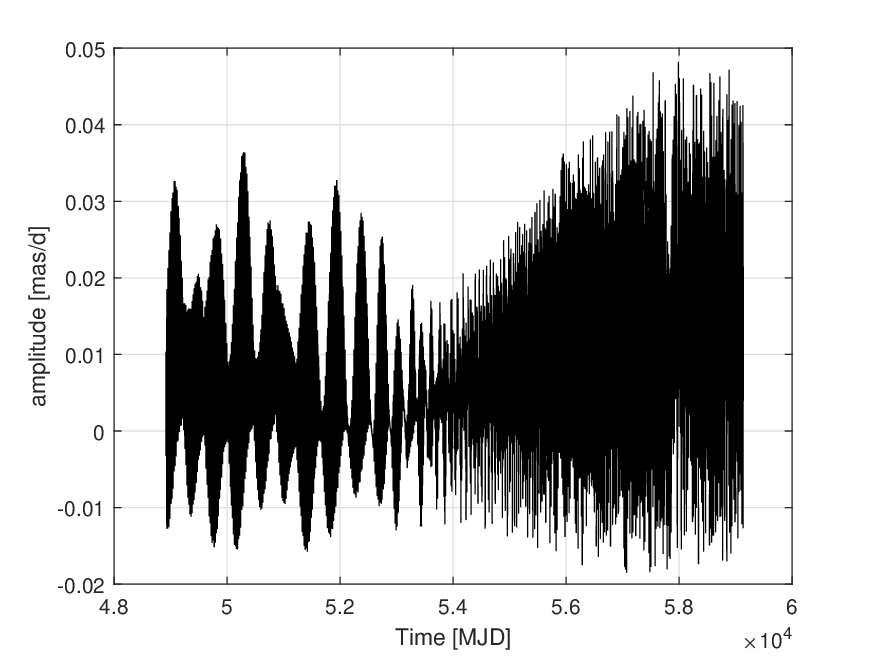}
	\caption{Time evolution of the LAGEOS II mean argument of latitude due to the Earth-Yarkovsky effect on a time interval of 28 years.}
	\label{fig:oss_EY_L2}
\end{figure}
\begin{figure} 
	\centering
	\includegraphics[width=0.7\linewidth]{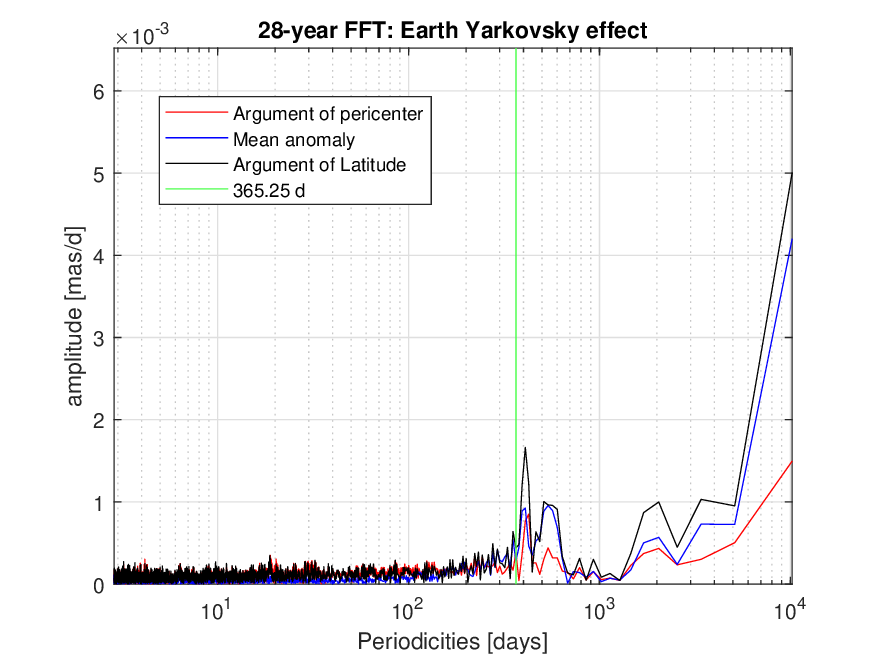}
	\caption{Earth-Yarkovsky effect. FFT of the LAGEOS II argument of pericenter (red) and mean anomaly (blue) over the 28 years period analyzed. The FFT of the argument of latitude (black) is also shown.}
	\label{fig:oss_EY_FFT_L2}
\end{figure}
\begin{figure} [h!]
	\centering
	\includegraphics[width=0.7\linewidth]{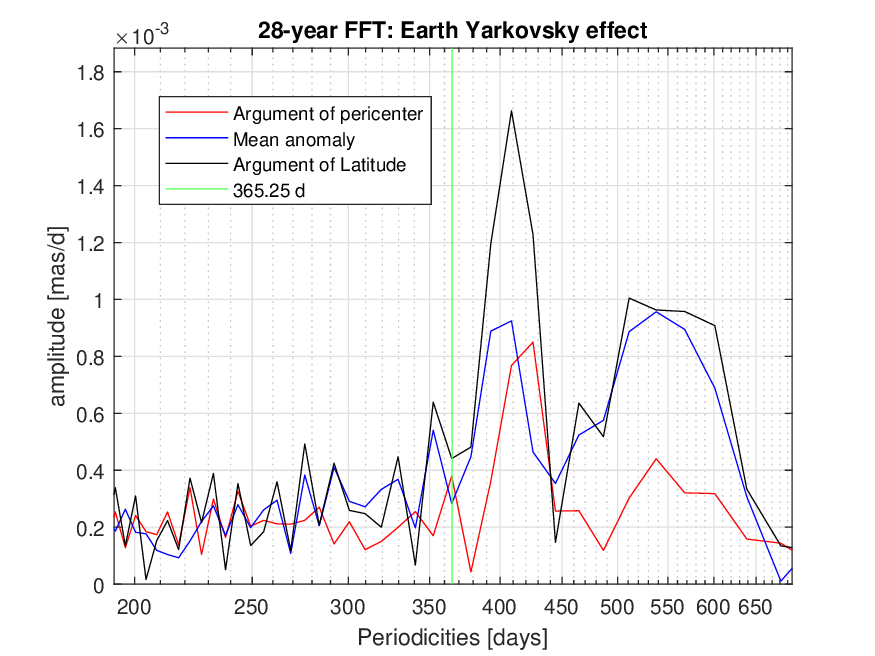}
	\caption{Zoom of Figure \ref{fig:oss_EY_FFT_L2} around the annual periodicity: no spectral component with annual periodicity is evident in the mean argument of latitude.}
	\label{fig:oss_EY_FFT_L2_part}
\end{figure}

The amplitude of the effect at the annual period in the rate of the argument of pericenter is very small. However, if we pessimistically assumed that the amplitude (about $4.4\times10^{-4}$ mas/day) of the FFT of the observable corresponding to the hypothetical annual periodicity was attributable to a spectral line, we would obtain a value for the error in $\alpha_1$ equal to $\sim2\times10^{-7}$. In fact, a very small value and in any case negligible compared to the systematic error previously estimated for the Earth's gravitational field.

As anticipated, an analysis of the impact of thermal effects was also conducted on LAGEOS and over a longer period of time, equal to 34 years from the launch of the satellite. No contribution of these perturbative effects to the annual periodicity is evident.

Therefore, based on our current best physical models, thermal thrust effects do not appear to be a significant source of systematic error that could mimic the signature of an LLI violation.

\subsection{End-to-End Analysis of Non-Gravitational Perturbations}\label{app:ngp_end_to_end}

In our second approach to error estimation, we performed a rigorous end-to-end quantitative analysis to evaluate the systematic bias originating from the main NGPs. This method provides a more consistent evaluation than a purely analytical estimate, as it processes high-fidelity NGPs models through the exact same measurement pipeline used for the final $\alpha_1$ estimation.

For each NGP effect considered---Solar Yarkovsky-Schach (YS), Earth-Yarkovsky (EY), and a conservative error in the Solar Radiation Pressure (SRP) model---we followed a consistent procedure:
\begin{enumerate}
    \item We generated the time series of the NGP-induced perturbations on the mean argument of latitude ($\dot{\ell}_0 = \dot{\omega} + \dot{M}$) for both LAGEOS and LAGEOS~II, based on our best available physical models.
    \item We processed these simulated perturbations through the two-satellite combination described by Eq.~(\ref{equ_comb}) to derive the corresponding time series for the observable $A_1(t)$.
    \item Finally, we fed this $A_1(t)$ series into our full Phase-Sensitive Detection (PSD) pipeline. The resulting DC output of the lock-In amplifier directly represents the systematic bias, $\alpha_1^{\text{NGP}}$, that the specific NGP would induce in our measurement.
\end{enumerate}

The results of this analysis are detailed below and summarized in Table \ref{tab:ngp_end_to_end}.

\subsubsection{Solar Yarkovsky-Schach (YS) Effect}
We first analyzed the signature of our YS thermal model. As anticipated from its known spectral properties, the effect does not project a significant DC component at the annual frequency. The output of the in-phase PSD analysis yields a systematic bias of $\alpha_1^{\text{YS}} = (+0.6 \pm 4) \times 10^{-8}$. This value is fully compatible with zero and is negligible compared to our final measurement uncertainty.

\subsubsection*{Earth-Yarkovsky (EY) Effect}
The same analysis was applied to the perturbations induced by the EY thermal effect. The magnitude of this effect is substantially smaller than that of the solar YS effect. The pipeline yields a systematic bias of $\alpha_1^{\text{EY}} = (+0.8 \pm 1) \times 10^{-11}$, which is entirely negligible.

\subsubsection*{Solar Radiation Pressure (SRP) Model Error}
To assess the impact of uncertainties in the SRP modeling, we simulated the effect of a conservative 0.5\% error in the radiation coefficients ($C_R$) for both LAGEOS satellites. The end-to-end analysis provides a potential systematic bias of $\alpha_1^{\text{SRP}} = (-0.4 \pm 4) \times 10^{-6}$. While this is the largest of the NGP contributions, its value is statistically consistent with zero and remains well below our final statistical error for $\alpha_1$.

 Notably, the SRP error model does not introduce a significant annual periodicity into our analysis. As shown in Figure \ref{fig:SRP_A1_t} (top), the FFT of the resulting $A_1^{SRP}(t)$ time series lacks any peak at the annual periodicity. This conclusion is further strengthened by the PSD analysis (bottom panel), where the application to $A_1^{SRP}(t)$ of a notch filter designed to remove signals around the annual period produces no discernible change in the output, confirming the absence of a relevant systematic signal from this source.
 \begin{figure}[h!]
 	\centering
 	\begin{subfigure}
 		\centering
 		\includegraphics[width=0.4\textwidth]{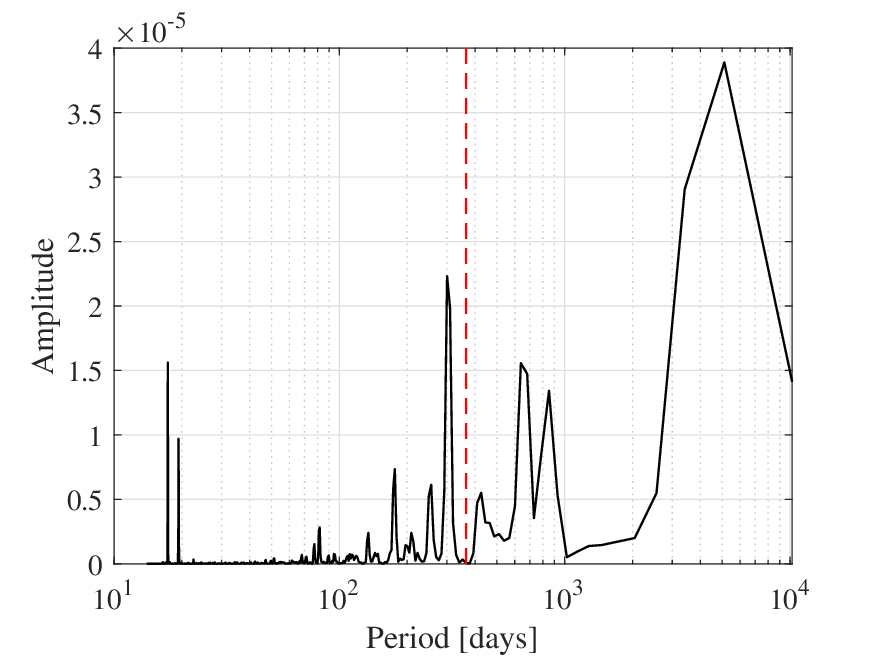}
 	
 	\end{subfigure}
 \begin{subfigure}
 	\centering
 	\includegraphics[width=0.4\textwidth]{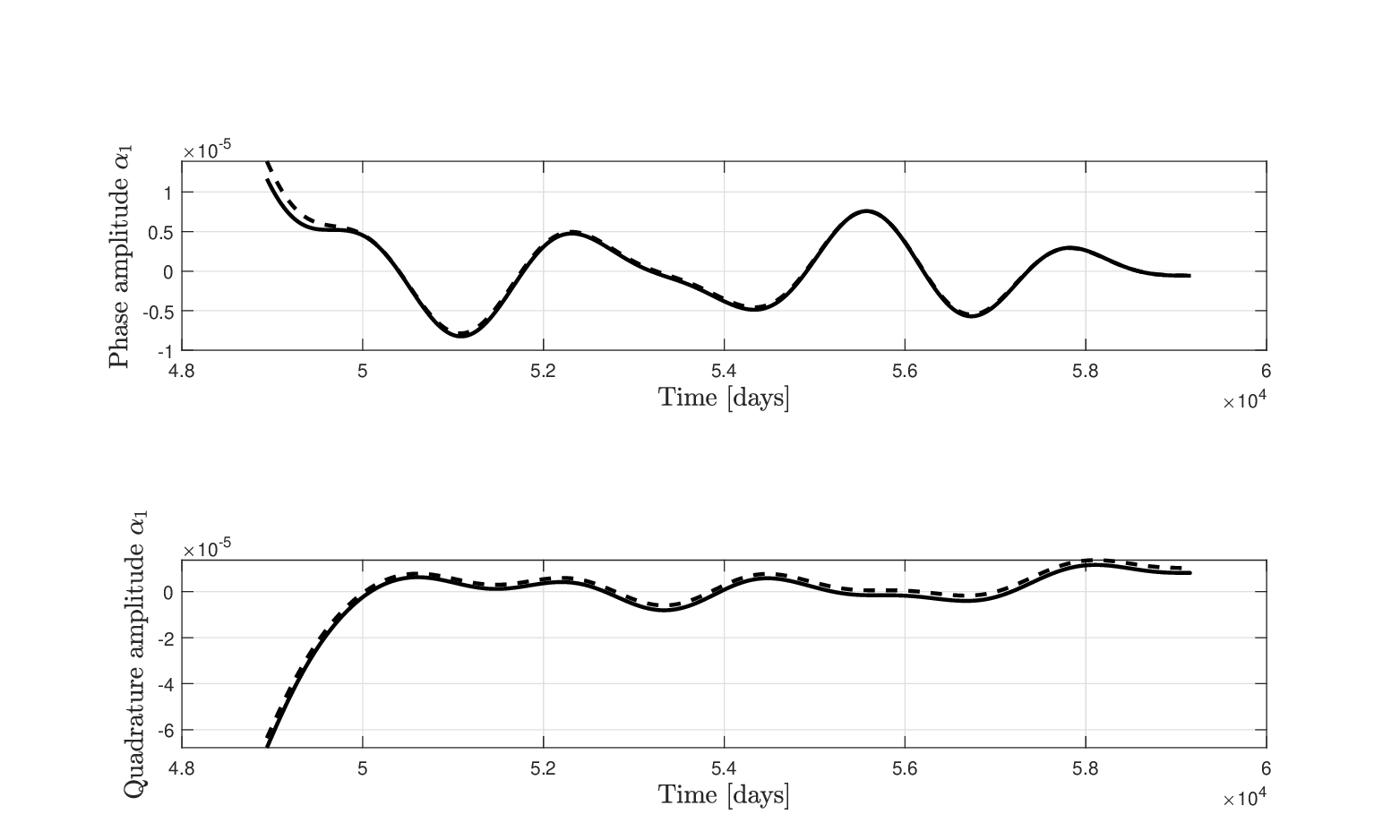}
 	
 \end{subfigure}
 \caption{FFT (top) of the time behavior for the variable $A_1(t)$ for the SRP model error and its PSD  (bottom). \label{fig:SRP_A1_t}}
 \end{figure}
 

\begin{table*}[htb!]
\caption{Systematic bias on $\alpha_1$ from the main NGPs, estimated via an end-to-end simulation. Each value represents the output of the full PSD pipeline when fed with the simulated signature of the corresponding NGP. \label{tab:ngp_end_to_end}}
\begin{ruledtabular} 
\begin{tabular}{lc}
Non-Gravitational Perturbation Source & Systematic Bias Estimate ($\alpha_1^{\text{NGP}}$) \\
\hline
Solar Yarkovsky-Schach (YS) Effect & $(+0.6 \pm 4.0) \times 10^{-8}$ \\
Earth-Yarkovsky (EY) Effect & $(+0.8 \pm 1.0) \times 10^{-11}$ \\
Solar Radiation Pressure (SRP) Error (0.5\%) & $(-0.4 \pm 4.0) \times 10^{-6}$ \\
\end{tabular}
\end{ruledtabular}
\end{table*}
\vskip 0.5cm
In conclusion, this quantitative end-to-end analysis confirms our previous assessments: the systematic biases introduced by the main non-gravitational perturbations are not a limiting factor for our measurement. The potential contribution from each effect is statistically consistent with zero and significantly smaller than both the leading gravitational uncertainties and our final statistical error. These results have been incorporated into the final error budget in Table \ref{tab:error-budget}.

\subsection{Sensitivity Analysis and Robustness Checks}\label{app:sensitivity}

In our third approach, we address the propagation of systematic errors through our complex, non-linear analysis pipeline. While a full Monte Carlo (MC) simulation is often considered the gold standard for such tasks, its application to our methodology is computationally prohibitive. A single POD run for one satellite over the ~30-year timespan requires approximately 8 hours with GEODYN~II and 16 hours with SATAN. An MC analysis with a statistically significant number of realizations (e.g., 1000) would thus demand several years of continuous computation, rendering it unfeasible.

Given these constraints, we have adopted a robust \emph{extreme sensitivity analysis} to rigorously investigate the stability of our $\alpha_1$ estimate against large, pessimistic variations in key model parameters. This approach addresses the core concern about uncaptured correlations and non-linear effects by testing the system's response under maximally stressed conditions. For each test, we performed a complete, independent POD and subsequent PSD analysis using the SATAN software, introducing a significant perturbation to a single component of the dynamical model.

The goal was twofold: to measure the induced shift in the central value of $\alpha_1$, and, critically, to verify if the estimated statistical uncertainty ($\sigma_{\alpha_1}$) remained stable. The main results are summarized below and in Table \ref{tab:sensitivity}.

\begin{enumerate}
    \item \textbf{Gravitational Field Coefficients ($\bar{C}_{2,0}$ and $\bar{C}_{4,0}$):} Recognizing the Earth's gravity field as a primary source of systematic error, we performed dedicated POD runs where the values of the quadrupole ($\bar{C}_{2,0}$) and hexadecapole ($\bar{C}_{4,0}$) coefficients were perturbed by $\pm5$ times their calibrated uncertainty. In both cases, the resulting shift in the central value of $\alpha_1$ was negligible. This result for $\bar{C}_{2,0}$ empirically confirms the effectiveness of our two-satellite combination (Eq.~\ref{equ_comb}) in canceling the error from this dominant coefficient.

    \item \textbf{Ocean Tides Model:} To explicitly test for potential spectral leakage from known periodic signals near the annual frequency, we performed a POD that completely excluded the annual ($S_a$) and semi-annual ($S_{sa}$) ocean tidal components from the FES2014 model. The subsequent analysis showed no significant alteration in the final $\alpha_1$ estimate.
    
    \item \textbf{Atmospheric Drag Model:} As an extreme test of non-gravitational force modeling, we performed a POD with the neutral atmospheric drag model (NRMLSISE) entirely removed. Even under this major simplification, no significant effect was observed on the final $\alpha_1$ result.
\end{enumerate}

\begin{table*}[hbt!]
\caption{Results of the \emph{extreme sensitivity analysis} performed with SATAN. Each row shows the value estimate for $\alpha_1$, the shift ($\Delta\alpha_1$) in the final $\alpha_1$ estimate and the change in its statistical uncertainty ($\Delta\sigma_{\alpha_1}$) resulting from the introduction of a pessimistic perturbation in a key model parameter (first column).  The reference value is the $\alpha_1$ estimate from the analysis performed with SATAN: $\alpha_1=(+1.2\pm1.5)\times10^{-5}$\label{tab:sensitivity}}
\begin{ruledtabular} 
\begin{tabular}{lccc}
Parameter / Model Perturbation & $\alpha_1$ value & Resulting Shift $\Delta\alpha_1$ & Change $\Delta\sigma_{\alpha_1}$ \\
\hline
$\bar{C}_{2,0}$ offset ($+ 5\sigma$) &   $(+1.1\pm1.5)\times10^{-5}$ & $-1 \times 10^{-6}$ & $0$ \\
$\bar{C}_{2,0}$ offset ($- 5\sigma$) &   $(+1.2\pm1.3)\times10^{-5}$ & $0$ & $-2 \times 10^{-6}$ \\
$\bar{C}_{4,0}$ offset ($+ 5\sigma$) &  $(+1.3\pm1.4)\times10^{-5}$  & $ +1 \times 10^{-6}$ & $-1 \times 10^{-6}$ \\
$\bar{C}_{4,0}$ offset ($- 5\sigma$) &  $(+1.1\pm1.5)\times10^{-5}$   & $-1 \times 10^{-6}$ & 0 \\
$S_a, S_{sa}$ tides removed & $(+1.2\pm1.4)\times10^{-5}$  & $0$ & $-1 \times 10^{-6}$  \\
$S_a, S_{sa}$ tides and atmospheric drag removed & $(+1.0\pm1.5)\times10^{-5}$  &  $-2 \times 10^{-6}$ & $0$\\
\end{tabular}
\end{ruledtabular}
\end{table*}

The central finding of this sensitivity analysis is the remarkable robustness of our result. As shown in Table \ref{tab:sensitivity}, the central value of $\alpha_1$ is largely insensitive to even extreme variations in the dominant model parameters. Crucially, the estimated statistical uncertainty, $\sigma_{\alpha_1} \sim 2 \times 10^{-5}$, remained stable across all tests. This indicates that our final error bar is not an artifact of a specific model choice but is a robust feature of the underlying data noise.

In conclusion, these extensive sensitivity analyses, while not a full MC simulation, provide strong evidence for the reliability of our systematic error assessment. They demonstrate that potential non-linearities and correlations related to these key parameters do not significantly bias our final $\alpha_1$ estimate or its overall uncertainty.

\subsection{Overall Error Budget and Discussion}\label{app:errori-discussione}

A cornerstone of this analysis is a comprehensive and conservative assessment of the systematic errors that could affect our measurement of $\alpha_1$. We have quantified the impact of potential mismodeling in all relevant gravitational and non-gravitational forces using three complementary approaches: spectral analysis of analytical models, end-to-end simulations, and extreme sensitivity tests.

A key strength of our methodology lies in the simultaneous analysis of the LAGEOS and LAGEOS~II satellites. This technique was specifically designed to mitigate the impact of the largest potential source of systematic error with an annual signature: the time-varying Earth's quadrupole moment, $\bar{C}_{2,0}$. As demonstrated in Section \ref{app:grav}, the carefully chosen linear combination of the satellites' observables, Eq.~(\ref{equ_comb}), effectively cancels the contribution of this dominant gravitational perturbation from the final solution for $\alpha_1$, significantly enhancing the robustness of our result.

Our complete systematic error budget is summarized in Table \ref{tab:error-budget}. The individual contributions were derived from the most conservative estimate among our different analysis approaches. For non-gravitational perturbations, we report the robust results from the end-to-end simulations described in Section \ref{app:ngp_end_to_end}, {which are consistent with the upper limits derived from our spectral analyses}.

\begin{table*}[htb!]
\caption{Systematic error budget for the measurement of $\alpha_1$. Each entry represents the most conservative estimate of the potential bias from different sources. The total systematic error is calculated as the root sum squares (RSS) of the individual contributions. \label{tab:error-budget}}
\begin{ruledtabular} 
\begin{tabular}{lcccccccc} 
{Error Source} & \multicolumn{2}{c}{{Gravitational}} & \multicolumn{2}{c}{{Tidal}} & \multicolumn{4}{c}{{Non-Gravitational}} \\
& $\bar{C}_{4,0}$ & $\bar{C}_{6,0}$ & {Ocean} & {Solid} & SRP  & {YS} & {EY} & \\
\hline
Systematic Error ($|\Delta\alpha_1|$) & $6.5 \times 10^{-6}$ & $6.0 \times 10^{-7}$ & $1.0 \times 10^{-7}$ &   $1.0 \times 10^{-9}$ & $4.0 \times 10^{-6}$ & $4.0 \times 10^{-8}$ & $1.0 \times 10^{-11}$ & \\
\hline
{Total Systematic Error (RSS)} & \multicolumn{7}{c}{$\sigma_{\text{sys}} \approx 7.7 \times 10^{-6}$} \\ 
\end{tabular}
\end{ruledtabular}
\end{table*}
The analysis of the error budget leads to two significant conclusions. 
First, the dominant source of systematic uncertainty is gravitational, stemming from the mismodeling of the Earth's hexadecapole coefficient, $\bar{C}_{4,0}$. All other contributions, particularly from non-gravitational and tidal forces, are at least an order of magnitude smaller. This underscores the effectiveness of our NGPs models and the robustness of the analysis against these challenging perturbations. {The largest error in these effects is related to the knowledge of the perturbation produced by direct solar radiation pressure. However, as shown by the end-to-end analysis presented in the previous Section \ref{app:ngp_end_to_end}, this error is not related to the DC conversion of an annual perturbation effect, but rather to the DC conversion of a broadband error in the model.}

Second, and most importantly, our measurement is demonstrably statistically limited, not systematically limited. The total systematic uncertainty, $\sigma_{\text{sys}} \approx 7.7 \times 10^{-6}$, is significantly smaller than the statistical errors obtained from our primary analyses:
\begin{itemize}
    \item For the GEODYN~II analysis, the statistical error ($\sigma_{\text{stat}} = 3 \times 10^{-5}$) is approximately {3.9 times larger} than the total systematic uncertainty.
    \item For the SATAN analysis, the statistical error ($\sigma_{\text{stat}} = 2 \times 10^{-5}$) is approximately {2.6 times larger} than the total systematic uncertainty.
\end{itemize}
This implies that the precision of our measurement is currently constrained by the stochastic noise in the data (likely related to the quantity and geometric distribution of the tracking data), rather than by deficiencies in our understanding of the underlying physics and orbital dynamics. The robustness of this conclusion was further confirmed by the sensitivity analyses presented in Section \ref{app:sensitivity}, which showed that even extreme errors in key model parameters do not significantly alter the final result.

Our final measurements for the PPN parameter are:
\begin{align*}
    \alpha_1 &= (+2 \pm 3) \times 10^{-5} \quad (\text{GEODYN II}) \\
    \alpha_1 &= (+1 \pm 2) \times 10^{-5} \quad (\text{SATAN})
\end{align*}
The non-zero central values, while larger than the systematic error budget alone, are statistically insignificant (well within one standard deviation) and can be confidently interpreted as random fluctuations around a true value of zero.
Finally, we showed that the reason why the statistical error is larger than the systematic error can be explained in terms of broadband noise, rather than by an underestimation of the systematic errors (see Section \ref{sec:measure}).

By combining our statistical and systematic errors in quadrature ($\sigma_{\text{total}} = \sqrt{\sigma_{\text{stat}}^2 + \sigma_{\text{sys}}^2}$), we obtain the total uncertainty for each analysis. Based on the results of the two conservative and independent analyses we performed with GEODYN~II and SATAN, our final constraint on a potential violation of Lorentz invariance in this context is:
\begin{equation}
    |\alpha_1| < 6.2 \times 10^{-5} \quad (95\%\, \text{C.L.})
\end{equation}
in the case of GEODYN~II, and:
\begin{equation}
    |\alpha_1| < 4.3 \times 10^{-5} \quad (95\%\, \text{C.L.})
\end{equation}
in the case of SATAN.
These results are consistent with General Relativity and {provide the tightest} constraints on this parameter from the analysis of satellite dynamics.
To confirm and potentially improve the bound on the $\alpha_1$ parameter obtained in this work, we intend to extend the analysis over a longer time horizon, also incorporating the eccentricity vector of the LAGEOS satellites as an observable sensitive to the presence of PFEs, i.e., to LLI violations.
Naturally, further refinement of the dynamical model, an ongoing effort, could help improve the POD and reduce the error budget, improving the overall quality of these measurements.

\end{document}